\documentclass[journal,10pt]{IEEEtran}

\usepackage{cite}

\ifCLASSINFOpdf
  \usepackage[pdftex]{graphicx}

\else

\fi

\usepackage{subfigure}
\usepackage{amsmath}

\usepackage{xcolor}

\usepackage{multicol}

\begin{document}

\title{Smoothed Phase-Coded FMCW: Waveform Properties and Transceiver Architecture}

\author{Utku~Kumbul,~\IEEEmembership{Graduate Student Member,~IEEE,}
        Nikita~Petrov,
        Cicero~S.~Vaucher,~\IEEEmembership{Senior Member,~IEEE,}
        and~Alexander~Yarovoy,~\IEEEmembership{Fellow Member,~IEEE}
\thanks{Utku Kumbul (u.kumbul@tudelft.nl) and Alexander Yarovoy are with the Microwave Sensing, Systems and Signals (MS3) group at the Faculty of Electrical Engineering, Mathematics and Computer Science, Delft University of Technology, Delft, Netherlands.}
\thanks{Nikita Petrov and Cicero S. Vaucher are with NXP Semiconductors, N.V., Eindhoven and Delft University of Technology, Delft, the Netherlands.}}

\markboth{Accepted for publication in IEEE Transactions on Aerospace and Electronic Systems. DOI 10.1109/TAES.2022.3206173}{Shell \MakeLowercase{\textit{et al.}}:}

\maketitle

\begin{abstract}
Smoothed phase-coded frequency modulated continuous waveform (SPC-FMCW), which is aimed to improve the coexistence of multiple radars operating within the same frequency bandwidth, is studied, and the receiving strategy with a low ADC sampling requirement is investigated. The Gaussian filter is applied to obtain smooth waveform phase transitions, and then quadratic phase lag compensation is performed before waveform transmission to enhance decoding. The proposed waveform is examined in different domains, and its waveform properties are analysed theoretically and demonstrated experimentally. Both simulation and experimental results show that the introduced waveform with the investigated processing steps helps combine all advantages of the FMCW waveform, including hardware simplicity and small operational bandwidth of the receiver, with the advantages of phase coding.

\end{abstract}


\IEEEpeerreviewmaketitle

\section{Introduction}

\IEEEPARstart{R}{adars} provide detection, tracking, and classification of targets under various weather conditions. As a consequence, radars are utilised in many areas such as surveillance,  meteorology,  defence and automotive systems. The dramatic increase in the number of radar sensors used for different applications has raised concerns about spectral congestion and the coexistence of radar sensors \cite{Brooker2007, Shannon2016Diversity, Griff2020, CananReview2020}. The mutual interference between multiple radar sensors downgrades the sensing performance of radar and needs to be mitigated \cite{Jianping2021, Neemat2019,Wildsmith2021,Jeroun2020}. Moreover, the radar systems used in civil applications (e.g. automotive radar, indoor monitoring) generally have limited processing power, preventing them from using computationally heavy techniques. In order to cope with these issues, designing a robust waveform with a low sampling requirement and improving the independent operation of multiple radars within the same frequency bandwidth is of interest.

Linear frequency modulated continuous waveform (FMCW) has been widely used in civil radar applications \cite{Bilik2019}. In the FMCW radar, the received signal is mixed with the complex conjugate of the transmitted signal for the stretch processing (also known as dechirping or deramping), which allows small analogue bandwidth of the receiver analogue-to-digital converter and a simple hardware structure \cite{Jankiraman}. Moreover, the FMCW radars can achieve good sensing performance with high resolution. However, the discrimination of FMCW between multiple radars is limited, and the FMCW radars suffer from radar-to-radar interference \cite{Benchter2016, Umehira2019, Roos2019}. For the purpose of unique waveform recognition, waveform coding has been widely used in radars \cite{Alland2019}. In particular, phase modulated continuous waveform (PMCW) provides high mutual orthogonality and thus improves the radar's robustness against interference \cite{Thillo2016, Xu2018, Bourdoux2020}. However, such coding spreads the waveform spectrum over a wide bandwidth and requires a dramatic increase in the receiver's analogue bandwidth.

Lately, phase-coded frequency modulated continuous waveform (PC-FMCW) has attracted much attention due to taking advantage of both FMCW and PMCW \cite{Reneau2014, Shannon2014, Cenk2019}. Applying coding to FMCW improves the waveform diversity and ensures the discrimination of self-transmitted signals from the waveforms transmitted by other radars \cite{FarukPhase}. Furthermore, phase-coded FMCW enables joint sensing and communication \cite{Lampel2019, FarukFrans, UtkuEuCAP}. In \cite{Reneau2014}, matched filtering is used as a receiver strategy to process PC-FMCW. For the matched filtering operation, the received signal is convolved with the complex conjugate of the transmitted signal. The conventional matched filtering operation in the digital domain demands the acquisition of the received signal with total bandwidth. Hence, this processing approach could not reduce the analogue receive bandwidth, and the key disadvantage of PMCW has been transferred to PC-FMCW. To decrease the waveform sampling requirement in the receiver, the dechirping based receiver structures have been studied \cite{Cenk2019,Lampel2019,FarukPhase}. In particular, compensated stretch processing has been suggested in \cite{Cenk2019}, where a filter bank has been applied to the sampled data after dechirping for all ranges of interest. Such a method, however, raises the computational complexity compared to the standard stretch processing as it obtains the range information via matrix multiplication instead of Fast Fourier Transform (FFT). To lower the computationally complexity, the dechirping and decoding receiver for PC-FMCW has been proposed in \cite{Lampel2019,FarukPhase}. There the dechirping is followed by the alignment of the coded beat signals for targets at different ranges using an ideal group delay filter. After alignment, all coded beat signals are decoded with the reference code, and the target range information is extracted from the beat signals via FFT. However, the group delay filter causes a quadratic phase shift (group delay dispersion effect) on the dechirped signal, resulting in the distortion of the code present in that signal. Consequently, the decoding becomes imperfect, which raises the sidelobes in the range profile.

The most popular phase coding scheme used for PC-FMCW in the preliminary studies \cite{FarukPhase, UtkuGeneralized} was binary phase shift keying (BPSK). The BPSK coding causes abrupt phase changes that lead to a large spectrum widening of the beat signal. As a consequence, the BPSK signal only with a small bandwidth compared to the sampling frequency (a few chips per chirps) can be used for sensing \cite{UtkuEuCAP}. In the case of BPSK code bandwidth being comparable to the sampling frequency of analog-to-digital converter (ADC), the sidelobe level significantly increases. Therefore, other phase modulation types with lower spectral broadening and thus better sensing performance are still of much interest.

In this paper, we have studied smoothed phase-coded frequency modulated continuous waveform (SPC-FMCW) to enhance the coexistence of multiple radars, and we have investigated the receiving strategy with a low ADC sampling requirement. The phase smoothing operation is proposed to obtain a smooth phase transition that addresses the bandwidth limitations of BPSK, and then the phase lag compensation is applied to the transmitted phase code to eliminate the undesired effect of the group delay filter. For analysis, we have used the Gaussian filter as a smoother and derived the waveform in different domains. Subsequently, we have investigated the waveform properties analytically. In addition, we have applied the proposed waveform to a real scenario and examined its sensing performance experimentally.

The rest of the paper is organised as follows. Section~\ref{sec:Signal_Model} describes the signal model for the generic PC-FMCW and gives the transceiver structure for the signal. Section~\ref{sec:TypeOfPhases} presents a smoothing operation to improve the phase transition of PC-FMCW. Section~\ref{sec:PhaseLagCompensation} provides the phase lag compensation, and Section~\ref{sec:BandwidthLimitation} investigates waveform properties of the resulting waveforms. Section~\ref{sec:Experiments} demonstrates the application of the waveforms to a real scenario. Finally, Section~\ref{sec:Conclusion} highlights the concluding remarks.

\section{Signal Model and Transceiver Structure}\label{sec:Signal_Model}

\begin{figure}[t]
    \centering
    \includegraphics[ width=1\linewidth]{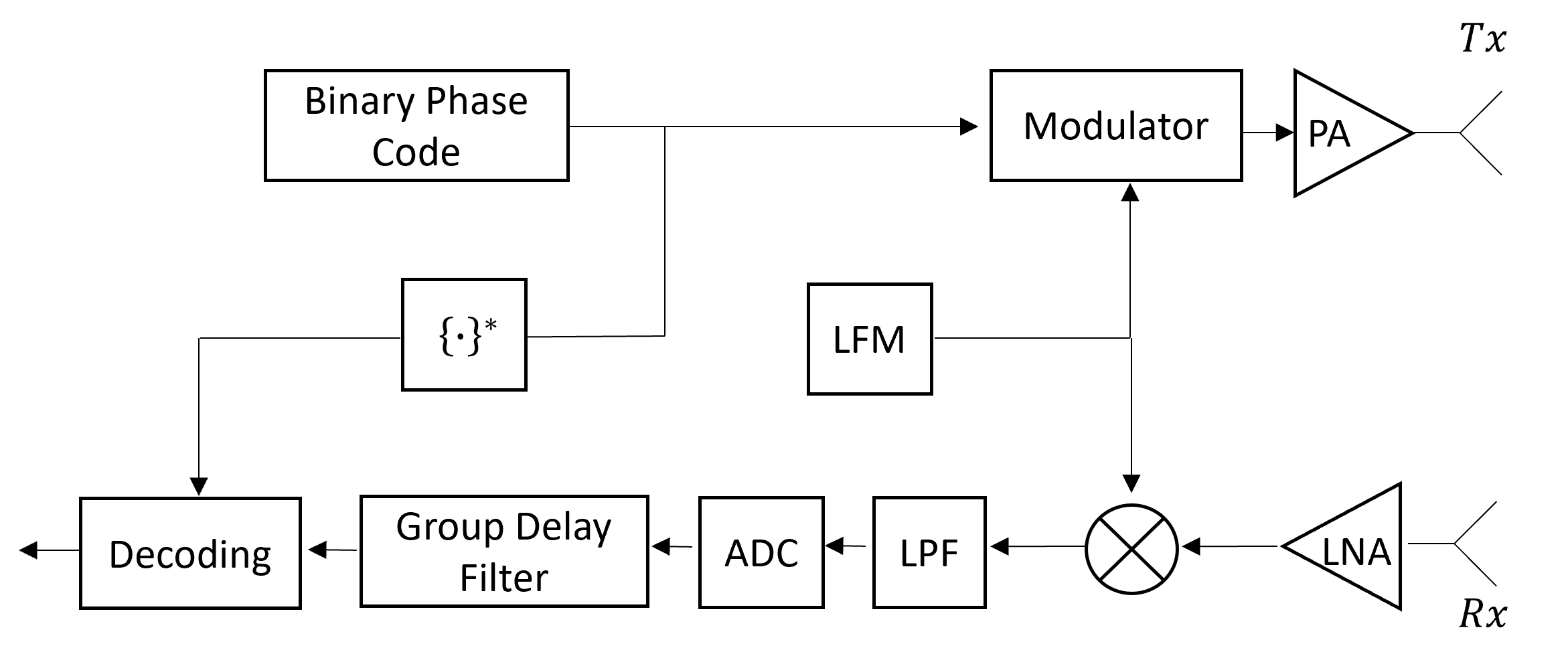}
    \caption{Block diagram of the state-of-the-art PC-FMCW transceiver structure}
    \label{fig:1}
\end{figure}

This section introduces the signal model and the state-of-the-art transceiver structure of phase-coded frequency modulated continuous waveform (PC-FMCW) \cite{FarukPhase} as illustrated in Figure~\ref{fig:1}.

The transmitted signal for frequency modulated continuous waveform (FMCW) can be written as:
\begin{equation}\label{LFMCW}
    x_{\text{FMCW}}(t)=e^{-j(2\pi f_c t+\pi kt^2)}, \, \, \quad t \in [0, \, T]
\end{equation}
where $f_c$ is the carrier frequency, $T$ is the sweep duration of the signal, $k=B/T$ is the slope of the linear frequency modulated waveform, and $B$ is the bandwidth. In PC-FMCW, the phase of FMCW changes according to the code sequence. The transmitted signal for the PC-FMCW radar can be represented as:
\begin{equation}
    x_{\text{T}}(t)=s(t)e^{-j(2\pi f_c t+\pi kt^2)},
\end{equation}
where $s(t)$ is a phase-coded\footnote{The phase code signal can also be kept inside the exponent as a phase term. We choose to write phase coding term as a separate signal component for more generic representation and ease of following.} signal. The received signal reflected from a moving point-like target can be written as:
\begin{equation}
    x_{\text{R}}(t)=\alpha_0\, s(t-\tau(t))e^{-j\left(2\pi f_c (t-\tau(t))+\pi k(t-\tau(t))^2\right)},
\end{equation}
where $\alpha_0$ is a complex amplitude proportional to the target back-scattering coefficient and propagation effects. Hereinafter, we substitute all the constant terms in $\alpha_0$ with no loss of generality. The round trip delay $\tau(t)$ for a target with constant velocity can be represented as:
\begin{equation}
    \tau (t)=\frac{2(R_0+v_0t)}{c}=\tau_0+\frac{2v_0 t}{c},
\end{equation}
where $R_0$ is the range, $v_0$ is the velocity, and $c$ is the speed of light. The range and velocity information of the target can be obtained by extracting the $\tau(t)$ from the received signal. In the state-of-the-art transceiver structure, the received signal is mixed with the complex conjugate of the uncoded transmit signal \eqref{LFMCW} for dechirping process \cite{FarukPhase}. The complex mixer output can be written as: 
\begin{equation}\label{equation5}
\begin{split}
x_{\text{M}}(t)=&x_{\text{R}}(t) {x^{*}_{\text{FMCW}}}(t)\\
=&\alpha_0\, s(t-\tau(t))e^{j(2\pi f_c\tau(t)+ 2\pi k \tau(t) t - \pi k {\tau(t)}^2)}\\
=&\alpha_0\, s(t-\tau_0-\frac{2v_0 t}{c})e^{j(2\pi f_c \tau_0 + 2\pi f_d t + 2\pi f_b t)},
\end{split}
\end{equation}
where $(\cdot)^*$ denotes the complex conjugate, $f_d=\frac{2v_0 f_c}{c}$ is the Doppler frequency and  $f_b$ is the beat frequency defined as:
\begin{equation}\label{beatFreqLabel}
    f_b=k\tau_0.
\end{equation}
Since the velocity of the target is much smaller than the speed of light as $v_0 \ll c$, the delayed code term can be approximated as $s(t-\tau_0-\frac{2v_0 t}{c})\approx s(t-\tau_0)$. Moreover, $e^{j(2\pi f_c \tau_0)}$ is constant phase term and thus incorporated into $\alpha_0$. 

In automotive radars, the Doppler frequency of the target is typically negligible compared to the frequency resolution of the beat signal, i.e. $f_d \ll f_s / N$, where $f_s$ is the sampling frequency of the beat signal and $N$ is the number of fast-time samples. Thus, we can approximate the $f_d + f_b \approx f_b$ without loss of generality. Note that we only neglected the Doppler frequency shift associated with fast-time, and there will be an additional term $\textrm{exp}(2\pi f_d \,mT)$ for velocity estimation in slow-time processing, where $m$ is the number of pulses. Since the group delay filter and decoding are only related to the fast-time processing part, we focus on signal analysis in fast-time. The slow-time processing is straightforward and the same as in the conventional FMCW automotive radars. Consequently, the mixer output in fast-time can be written as:
\begin{equation}\label{beat_signal}
    x_{\text{M}}(t)=\alpha_0\, s(t-\tau_0) e^{j(2\pi f_b t)} ,
\end{equation}
By taking the Fourier transform, the frequency-domain representation of the mixer output can be obtained as:
\begin{equation}
\begin{split}
    X_{\text{M}}(f)&=\alpha_0 \int_{-\infty}^{\infty} s(t-\tau_0) e^{j(2\pi f_b t)} e^{-j2\pi ft} \, dt \\
    &=\alpha_0 e^{-j(2\pi (f-f_b) \tau_0)} \int_{-\infty}^{\infty} s(t_1) e^{-j(2\pi (f-f_b) t_1)}  \, dt_1 \\
    &=\alpha_0S(f-f_b)e^{-j\left(2\pi (f-f_b) \tau_0\right)}\\
    &=\alpha_0S(f-f_b)e^{-j\left(\frac{2\pi f_b }{k}(f-f_b)\right)},
    \end{split}
\end{equation}
where for the final equality we used \eqref{beatFreqLabel}.

In the decoding process, the mixer output \eqref{beat_signal} is multiplied with the complex conjugate of the reference phase code for compensating phase changes initiated by the transmitted phase code. 
For a short-range radar application e.g. indoor monitoring, the delay can be neglected $s(t-\tau_0)\approx s(t)$, and the mixer output can be decoded by multiplying \eqref{beat_signal} with $s^*(t)$ directly \cite{UtkuEuCAP}. However, this assumption does not hold for the applications with $R\geq c/(2B_c)$, where $B_c$ is the bandwidth of phase-coded signal $s(t)$. For these applications, each coded beat signal (the response in all the range cells) is required to be aligned in fast-time to compensate the time delay before decoding. This alignment can be realized via the group delay filter either in time-domain \cite{FarukPhase}, or frequency-domain \cite{Lampel2019}. Assume we have a group delay filter with frequency response:
\begin{equation}
    H_{\text{g}}(f)=|H_{\text{g}}(f)| \angle{H_{\text{g}}(f)}=e^{j \theta_g(f)},
\end{equation}
and unity magnitude, $|H_{\text{g}}(f)|=1, \forall f$.

The Taylor series expansion of the phase response $\theta_g(f)$ around $f_b$ can be found as:
\begin{equation}
\begin{split}
    \theta_g(f)\big|_{f=f_b}=\theta(f_b)&+\frac{d\theta(f)}{df}\bigg|_{f=f_b} (f-f_b) \\
  & +  \sum\limits_{m = 2}^{{\infty}} \frac{1}{m!} \frac{d^m \theta(f)}{df^m}\bigg|_{f=f_b} (f-f_b)^m.
    \end{split}
\end{equation}
The resulting filter causes the group delay, $\tau_g(f)$, which is the first derivative of the phase response and shifts the envelope of the signal (The proof is given in Appendix C). To align coded beat signals, the group delay needs to eliminate $\tau_0$. Thus, the required group delay can be found as:
\begin{equation}
\tau_{\text{g}}(f)=-\frac{1}{2\pi} \frac{d\theta(f)}{df}\bigg|_{f=f_b}=-\tau_0=-f_b/k,    
\end{equation}
and consequently, the first derivative of the phase response can be written as:
\begin{equation}\label{derivative_Phase}
    \frac{d\theta(f)}{df}\bigg|_{f=f_b}=\frac{2\pi f_b}{k}.
\end{equation}
The group delay filter which gives the required group delay in \eqref{derivative_Phase} can be written as:
\begin{equation}
    H_{\text{g}}(f)=e^{j \theta_g(f)}=e^{j\frac{\pi f^2}{k}},
\end{equation}

\begin{figure}[t]

         \centerline{\includegraphics[width=0.9\linewidth]{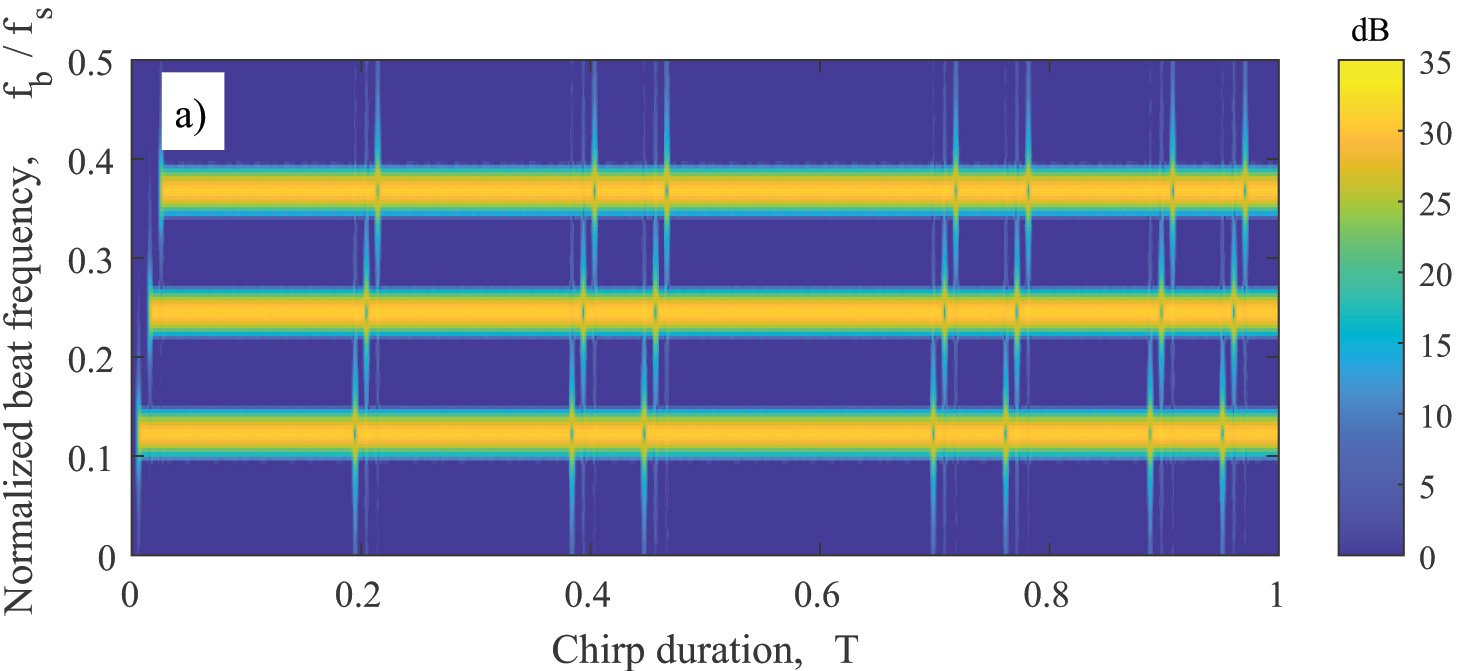}}
         \hfill
         \centerline{\includegraphics[width=0.9\linewidth]{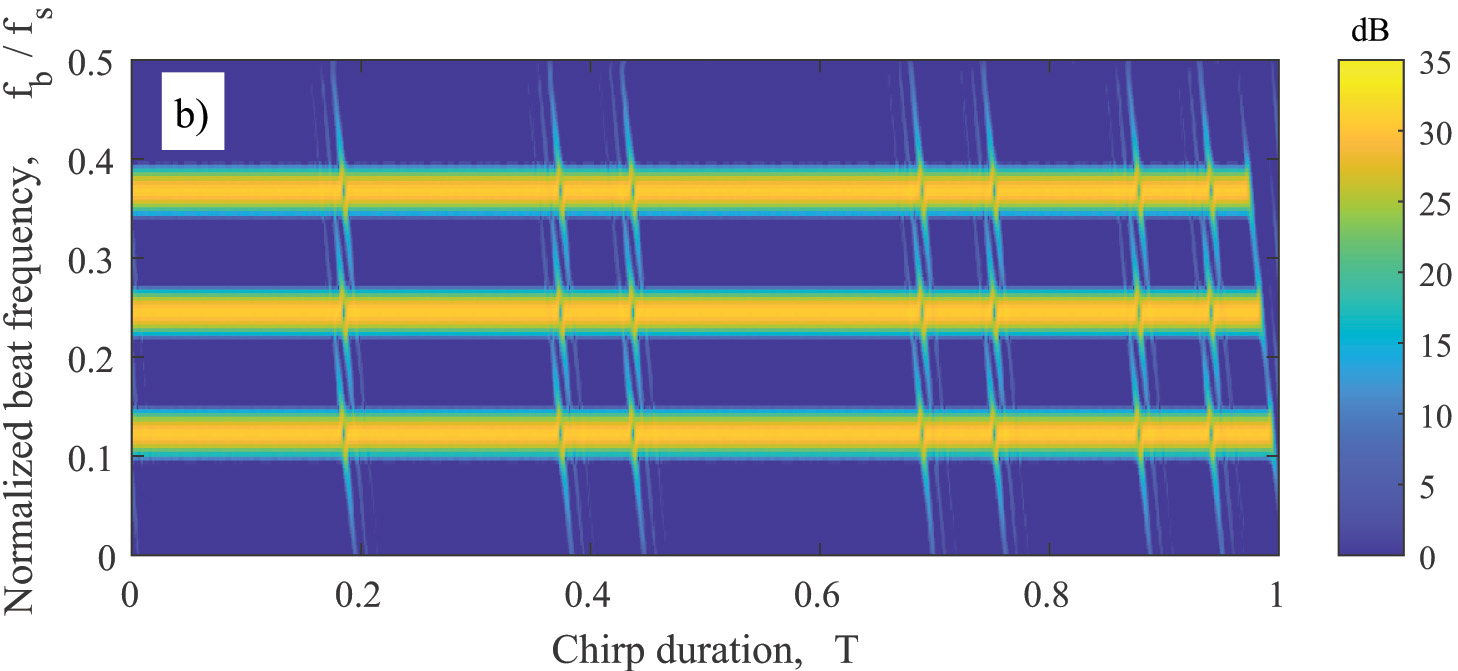}}
         \hfill
         \centerline{\includegraphics[width=0.9\linewidth]{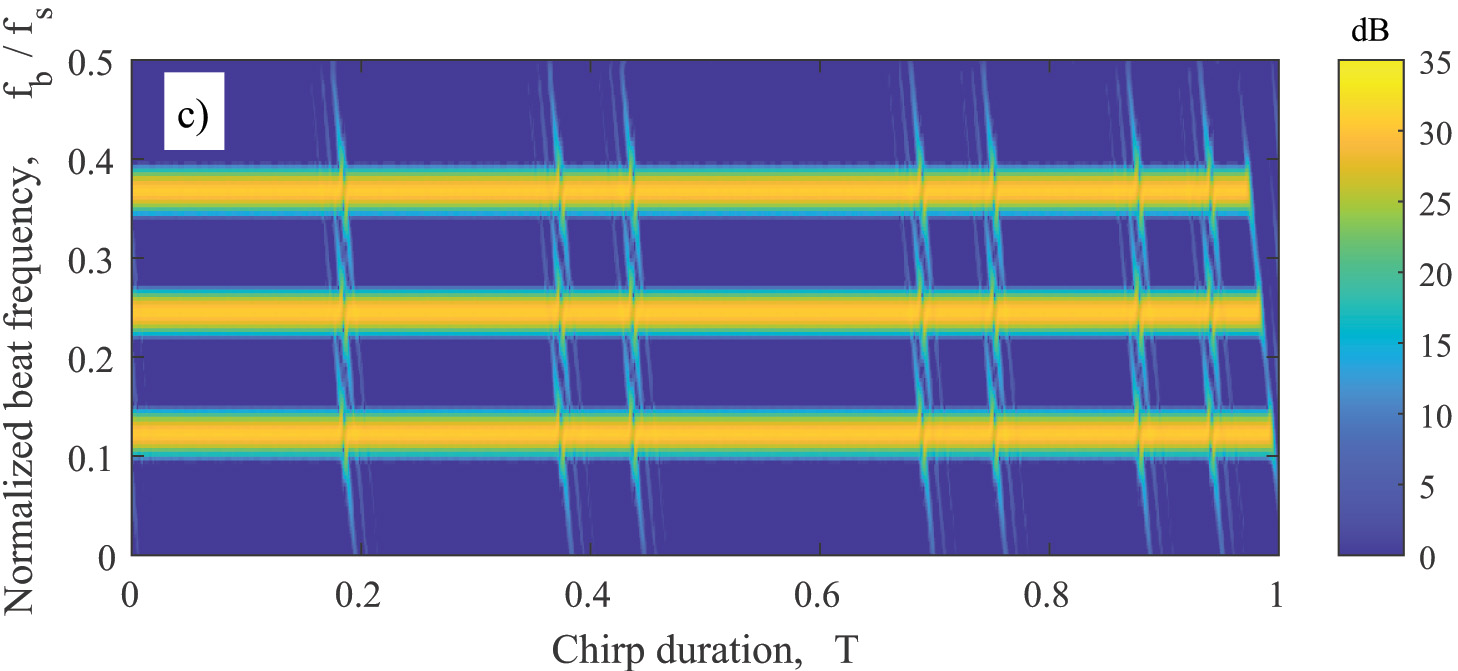}}
           
         \caption{Spectrogram of the BPSK phase-coded beat signals ($N_c=16$): a) Before the group delay filter b) After the group delay filter c) After decoding the group delay filter output }
\label{fig:2}
\end{figure}

We apply the group delay filter by multiplying the spectrum of mixer output as:
\begin{equation}
    Z_{\text{o}}(f)=X_{\text{M}}(f) H_{\text{g}}(f).
\end{equation}
For each beat frequency, the frequency characteristic of the group delay filter can be expressed by the Taylor series expansion of its phase response around $f_b$ as:
\begin{equation}\label{phaseTaylor}
     \theta_g(f)\big|_{f=f_b}=\frac{\pi {f_b}^2}{k}+\frac{2\pi f_b}{k}(f-f_b)+\frac{\pi}{k} (f-f_b)^2.
\end{equation}
Subsequently, multiplying the mixer output spectrum with the frequency characteristic of the filter gives:
\begin{equation}\label{group_out_frequ}
\begin{split}
     Z_{\text{o}}(f)&=X_{\text{M}}(f) e^{j\left(\frac{\pi {f_b}^2}{k}+\frac{2\pi f_b}{k}(f-f_b)+\frac{\pi}{k} (f-f_b)^2\right)}\\
   &=\alpha_0S(f-f_b)e^{-j\left(\frac{2\pi f_b }{k}(f-f_b)\right)} \\
   &\qquad \qquad \cdot e^{j\left(\frac{\pi {f_b}^2}{k}+\frac{2\pi f_b}{k}(f-f_b)+\frac{\pi}{k} (f-f_b)^2\right)}\\
   &=\alpha_0S(f-f_b) e^{j\left(\frac{\pi {f_b}^2}{k}+\frac{\pi}{k} (f-f_b)^2\right)}\\
   &=\alpha_0S(f-f_b) e^{j\left(\frac{\pi}{k} (f-f_b)^2\right)},
\end{split}
\end{equation}
where for the last equality we incorporated $\mathrm{exp}\left(\frac{\pi {f_b}^2}{k}\right)$ into $\alpha_0$ as it is a constant phase term (does not depend on frequency $f$). Then, taking the inverse Fourier transform of the group delay filter output \eqref{group_out_frequ} gives:
\begin{equation}\label{group_out_eq_1}
\begin{split}
    z_{\text{o}}&(t)=\mathcal{F}^{-1}\left\{\alpha_0S(f-f_b) e^{j\left(\frac{\pi}{k} (f-f_b)^2\right)}\right\}\\
    &=\mathcal{F}^{-1}\left\{\alpha_0S(f-f_b)\right\} \otimes  \mathcal{F}^{-1}\left\{e^{j\left(\frac{\pi}{k} (f-f_b)^2\right)}\right\}\\
   &=\left(\alpha_0s\left(t\right)  e^{j(2\pi f_b t)}\right) \otimes  \mathcal{F}^{-1}\left\{e^{j\left(\frac{\pi}{k} (f-f_b)^2\right)}\right\},
\end{split}
\end{equation}
where $\otimes$ denotes the convolution operation. Note that the delay $\tau_0$ is eliminated after the group delay filter for each coded beat signal. Moreover, the derived group delay filter has a quadratic frequency component within its phase response and applies different time delays to each frequency component. Consequently, the filter causes the so-called group delay dispersion effect shown as term $e^{j(\frac{\pi}{k}(f-f_b)^2)}$, which leads to a non-linear shift on the spectrum of the code signal. By substituting $\zeta=-j\frac{\pi}{k}$ and $f_1=f-f_b$, the dispersion effect can be written as:
\begin{equation}
\begin{split}
    h_{\text{dis}}(t)=&\mathcal{F}^{-1}\left\{e^{j\left(\frac{\pi}{k} (f-f_b)^2\right)}\right\} \\
     =&\int_{-\infty}^{\infty} e^{-\zeta{f_1}^2} \, e^{j2\pi f_1 t} \, df_1 \, e^{j2\pi f_b t}\\
     =&e^{j2\pi f_b t}\, e^{-\frac{{\pi}^2 t^2}{\zeta}}\int_{-\infty}^{\infty} e^{-\left(\sqrt{\zeta} f_1 -j\frac{\pi t}{\sqrt{\zeta}}\right)^2}  \, df_1 \\
      =&\sqrt{-\frac{k}{j}} \, e^{\pi \frac{k t^2}{j}} e^{j2\pi f_b t}.
    \end{split}
\end{equation}
Subsequently, the mixer output in time domain \eqref{group_out_eq_1} can be recast as:
\begin{equation}
    z_{\text{o}}(t)=\alpha_0 \, e^{j2\pi f_b t} \, \left(s(t) \otimes h_{\text{dis}}(t)\right).
\end{equation}

\begin{figure}[t]
         \centerline{\includegraphics[width=0.9\linewidth]{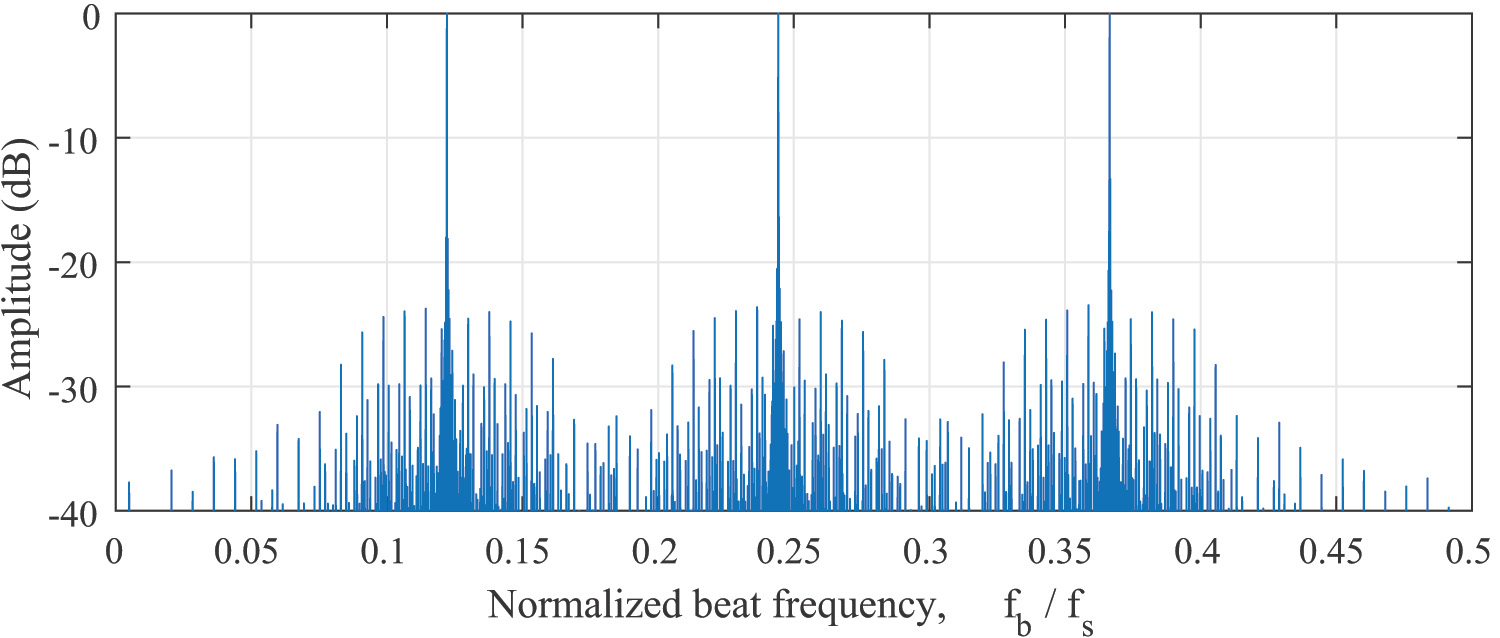}}
         \caption{The distorted range profile of decoded signal after the group delay filter ($N_c=64$) }
\label{fig:3}
\end{figure}

The spectrogram of the BPSK phase-coded beat signals are shown in Figure~\ref{fig:2}, where the system parameters are selected as $B=2$ GHz, $T=51.2$ $\mu$s and $N_c=16$. Subsequently. the code bandwidth $B_c=\frac{N_c}{T}=0.31$ MHz with these parameters ($N_c=16$). We use the same system parameters for the follow-up figures, if not mentioned otherwise. Moreover, we normalize the beat frequency with the maximum beat frequency, which is determined by the ADC sampling frequency as $f_{b_{\text{max}}}=f_s/2$. In Figure~\ref{fig:2}, we compare three cases: before the group delay filter, after the group delay filter and after decoding the group delay filter output. It can be seen that each coded beat signal has different time delays (associated with their corresponding range) before the group delay filter (Figure~\ref{fig:2} a). After using the group delay filter, we observe that each coded beat signal is aligned at the beginning (Figure~\ref{fig:2} b). Note that the signals with lower frequency are shifted less compared to signals with higher frequency. As a result of coded beat signal alignment, the decoding can be performed by multiplying the group delay filter output with the complex conjugate of the reference phase code. In an ideal decoding, the multiplication of codes gives $s(t)s^*(t)=e^{j\phi(t)}e^{-j\phi(t)}=1$. However, the spectrum of the code signal is shifted non-linearly ((Figure~\ref{fig:2} b) as it is convolved with $h_{\text{dis}}(t)$. Thus decoding becomes imperfect, and the code term is not removed properly (Figure~\ref{fig:2} c). The decoded beat signal can be written as \cite{Lampel2019}:
\begin{equation}
\begin{split}
z_{\text{d}}(t)&= z_{\text{o}}(t) s^*\left(t\right)\\
     &= \alpha_0 e^{j2\pi f_b t} \, \left(s(t) \otimes h_{\text{dis}}(t)\right) \,  s^*\left(t\right) \\
     &=\alpha_0 e^{j2\pi f_b t} \, \left(e^{j\phi(t)} \otimes h_{\text{dis}}(t)\right) \,  e^{-j\phi(t)}\\
      &= \alpha_0 e^{j2\pi f_b t} \, e^{j\epsilon(t)},
 \end{split}
\end{equation}
where $\epsilon(t)$ is the residual phase error due to the group delay filter dispersion that causes imperfection in decoding. The dispersion effect can be neglected for a narrow-band signal where the bandwidth of the phase-coded signal is very small compared to the sampling frequency $B_{c} \ll f_s$. However, the dispersion effect becomes crucial for a signal with a wide spectrum where $B_{c}$ is comparable to $f_s$. One example of such a signal is the BPSK phase-coded beat signal. In the time instance of phase shifts, the BPSK signal has a wide-spread spectrum due to abrupt phase changes. Applying non-linear phase shifts to its spectrum leads to huge imperfection in decoding. Consequently, the BPSK signal suffers from the distorted range profile after decoding, as demonstrated in Figure~\ref{fig:3}, where we use $N_c=64$. The distortion of the range profile raises for the long code sequences (the bandwidth of the chips increase). We will address the compensation of the group delay dispersion effect by applying quadratic phase lag to the waveform before transmission in Section~\ref{sec:PhaseLagCompensation}.

\section{Smooth Phase Transition Waveforms}\label{sec:TypeOfPhases}
This section presents the smoothing operation to improve the phase transition of the state-of-the-art waveform and reduces its spectral widening of the coded beat signal to obtain SPC-FMCW. 
\subsection{BPSK PC-FMCW}
\begin{figure}[t]
         \centerline{\includegraphics[width=0.9\linewidth]{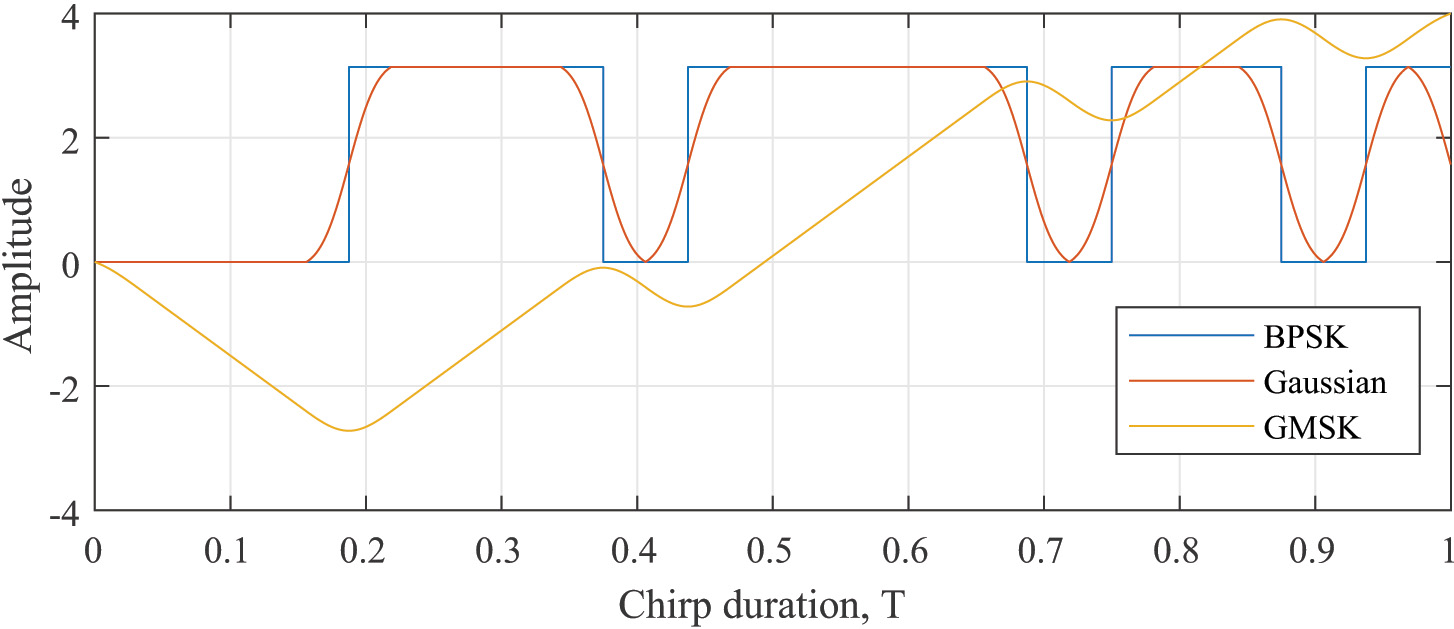}}
         \caption{Illustration of the phase types ($N_c=16$): BPSK, Gaussian, and GMSK}
\label{fig:4}
\end{figure}

In BPSK, the phase changes $\phi_{\text{bpsk}}(t)\in\{ 0,\pi\}$ as shown in Figure~\ref{fig:4} and the transmitted code term can be represented as:
\begin{equation}
c(t) =e^{j\phi_{\text{bpsk}}(t)} = \frac{1}{T} \frac{1}{T_c} \sum\limits_{n = 1}^{{N_c}} {e ^{j{\phi_n}}{\mathop{\rm rect}\nolimits} \left( {\frac{{t - (n-1/2){T_c}}}{{{T_c}}}} \right)},
\end{equation}
where $N_c$ is the number of chips within one chirp, $T_c=T/N_c$ is the chip duration, ${\rm rect}(t)=1, t\in[-T_c/2, \,\,\, T_c/2]$ and zero otherwise is the rectangle function, and $\phi_n$ denotes the phase corresponding to the $n^{\textrm{th}}$ bit of the  $N_c$ bits sequence.

\begin{figure}[t]

         \centerline{\includegraphics[width=0.9\linewidth]{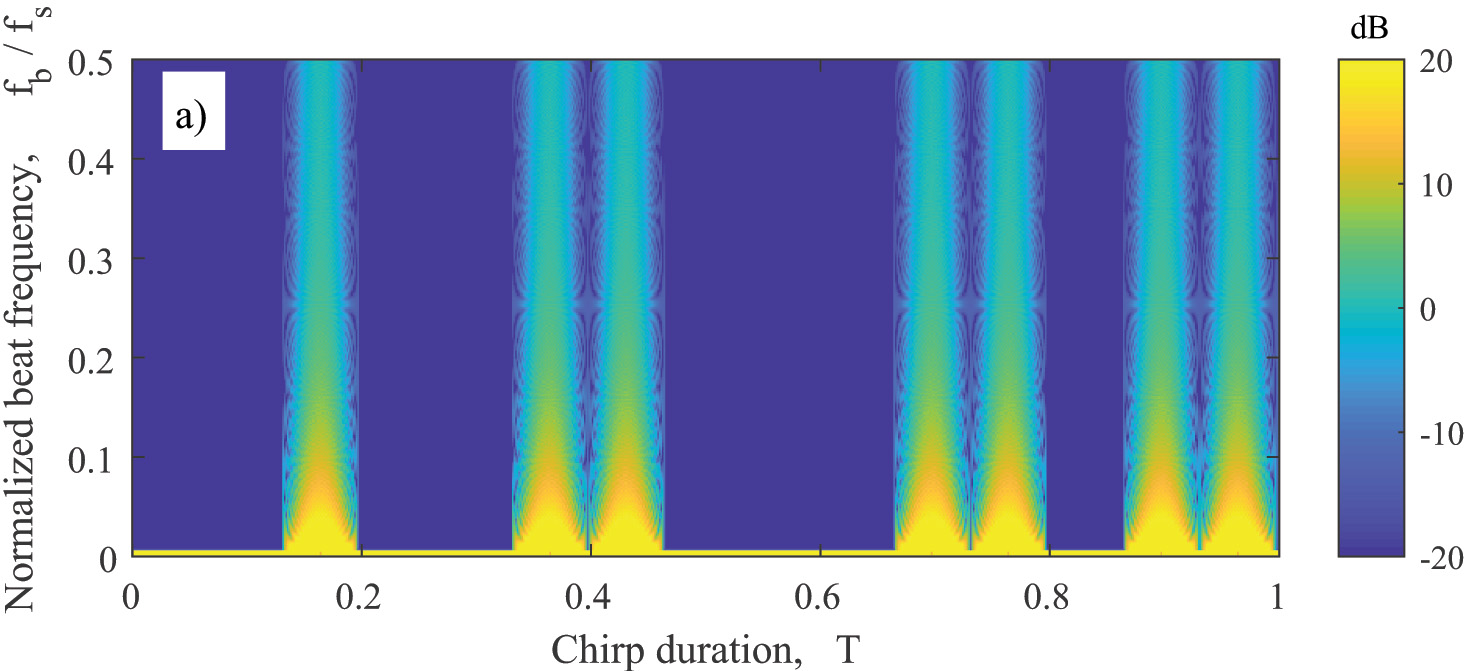}}
         \hfill
         \centerline{\includegraphics[width=0.9\linewidth]{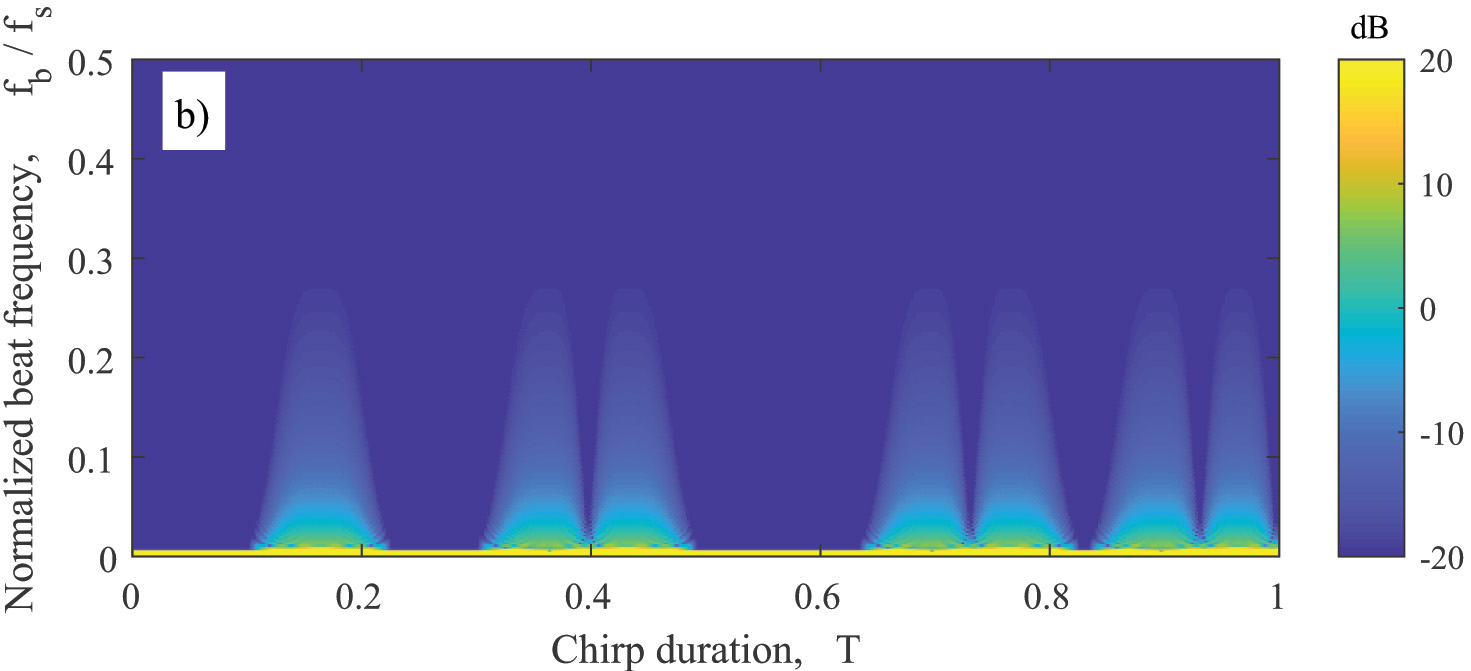}}
         \hfill
         \centerline{\includegraphics[width=0.9\linewidth]{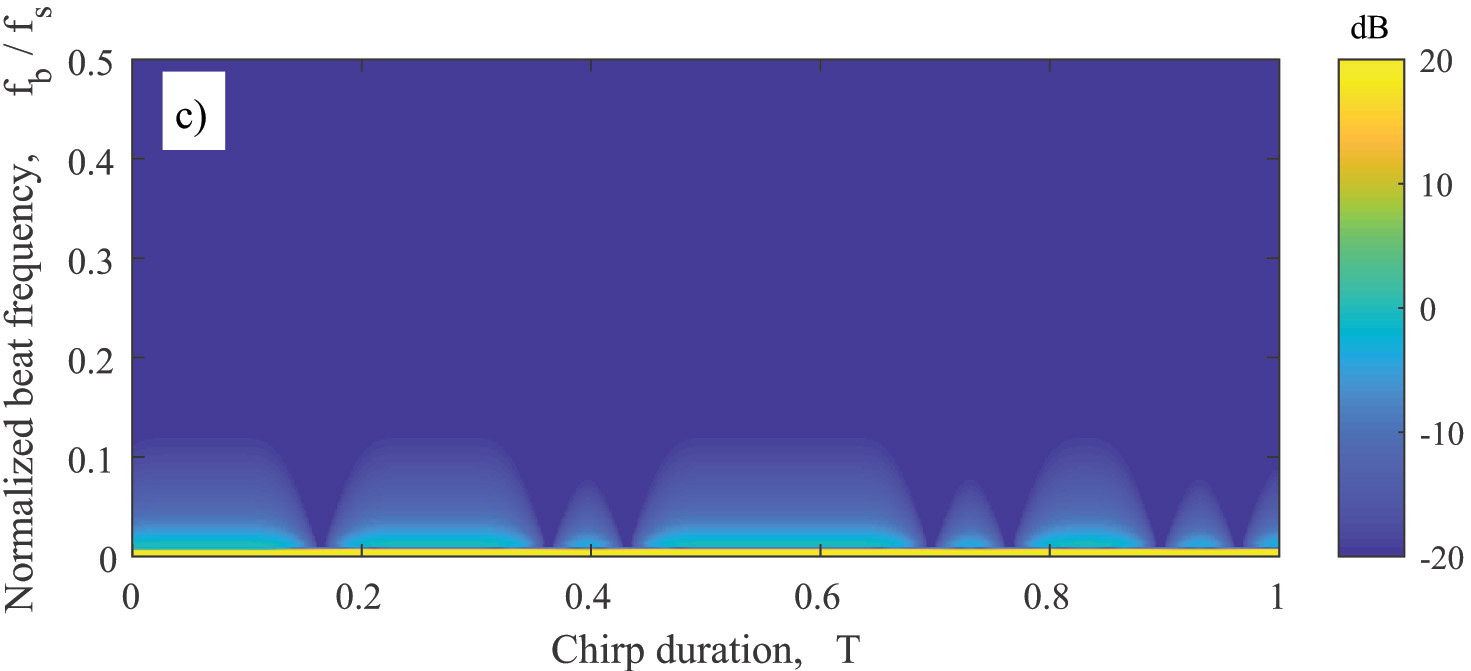}}
           
         \caption{Instantaneous frequency of the coded beat frequency signal associated with three PC-FMCW waveforms ($N_c=16$): a) BPSK b) Gaussian c) GMSK  }
\label{fig:5}
\end{figure}

In addition, analyzing the spectrogram of the phase-coded signal is complementary as it provides an additional perspective that may not easily be seen on the signal's time or frequency domain representation. The instantaneous frequency for the BPSK code sequence can be written as \cite{UtkuGeneralized}:
\begin{equation}\label{generalized_eq_1}
   \frac{1}{2\pi} \frac{d}{dt}\phi_{\text{bpsk}}(t)=\frac{1}{2\pi}\sum\limits_{n = 1}^{{N_c}} {({\phi_{n+1}}-\phi_n) \delta(t-n{T_c})}.
\end{equation}
The proof is given in Appendix A. The instantaneous frequency of BPSK is demonstrated in Figure~\ref{fig:5} a. It can be seen that the abrupt phase changes cause a short burst in the spectrum at the time instances of phase shifts, which mathematically comes from the derivatives of unit step functions and is represented as Dirac delta in \eqref{generalized_eq_1}.

For BPSK PC-FMCW, we can replace the $s(t-\tau_0)$ term with $c(t-\tau_0)$ and the mixer output becomes:
\begin{equation}
x_{\text{M}\textsubscript{bpsk}}(t)=\alpha_0c(t-\tau_0)e^{j(2\pi f_b t)},
\end{equation}
and the frequency-domain representation of the mixer output can be written as:
\begin{equation}\label{BPSK_freq}
\begin{split}
     &X_{\text{M}\textsubscript{bpsk}}(f)=\alpha_0C(f-f_b)e^{-j\left(2\pi (f-f_b) \tau_0\right)} \\
     &\quad=\frac{\alpha_0}{T} \sum\limits_{n = 1}^{{N_c}} {e ^{j{\phi_n}}} {\rm sinc}((f-f_b)T_c)e^{-j\left(2\pi (f-f_b) (\tau_0 + (n-\frac{1}{2})T_c)\right)} .
\end{split}
\end{equation}
The frequency spectrum of the mixer output \eqref{BPSK_freq} with $\tau_0=0$ is shown in Figure~\ref{fig:6} in blue color. The first null for the mixer output of BPSK PC-FMCW is defined by the ${\rm sinc}$ function as shown in \eqref{BPSK_freq}. Therefore, the first null location of the coded beat signal for BPSK PC-FMCW can be calculated as:
\begin{equation}
    f_b+B_c \, ,
\end{equation}
where $B_c=1/T_c$ is the bandwidth of chip. Assume we have an ideal Brick-wall low pass filter (LPF) as:
\begin{equation}
    L(f)={\rm rect}\left(\frac{f}{f_{cut}}\right).
\end{equation}
The output of the LPF can be represented as:
\begin{equation}
    O(f)=X_{\text{M}}(f)L(f).
\end{equation}
To include the $k^{\text{th}}$ null, the cut-off frequency of LPF can be written as:
\begin{equation}\label{eqLPFcut}
    \begin{split}
        f_{\text{cut}}&\geq f_{b_{\text{max}}}+\frac{1}{T_c}  k^{\text{th}}_{\text{null}} \\
        &\geq f_{b_{\text{max}}}+\frac{N_c}{T} k^{\text{th}}_{\text{null}}   \\
        &\geq k\tau_{\text{max}}+\frac{N_c}{T} k^{\text{th}}_{\text{null}} \\
        &\geq \frac{1}{T} \left(\frac{2B R_{max}}{c}+N_c k^{\text{th}}_{\text{null}} \right).
    \end{split}
\end{equation}
where $f_{b_{\text{max}}}=k \tau_{\text{max}}$ is the maximum beat frequency and $\tau_{\text{max}}=2R_{\text{max}}/c$ is the maximum round trip delay for the stationary target at the maximum range $R_{\text{max}}$. The cut-off frequency of LPF determines the minimum ADC sampling requirement which should be at least two times of $f_{\text{cut}}$. The BPSK coding results in substantial spectrum widening of the beat signal due to rapid phase shifts. Thus, the BPSK coding requires the sampling of a few multiples of code bandwidth. The spectrum width of a signal $x(t)$ can be calculated as \cite{Spectrum_Book}:
\begin{equation}
    \sigma_{\text{f}}=\sqrt{\frac{1}{{\textrm{P}}}\int_{-\infty}^{\infty} (f-\mu_f)^2 \left| X(f) \right|^2 \, df},
\end{equation}
where the total power of the spectrum can be defined as:
\begin{equation}
    \textrm{P}=\int_{-\infty}^{\infty} \left| X(f) \right|^2 \, df,
\end{equation}
and the mean frequency of the spectrum can be written as:
\begin{equation}
    \mu_f=\frac{1}{\textrm{P}}{\int_{-\infty}^{\infty} f \left| X(f)\right|^2 \, df}.
\end{equation}

The spectrum width of the coded beat signals associated with three PC-FMCW waveforms versus number of chips per chirp is shown in Figure~\ref{fig:71}. It can be seen that the spectrum widening of the coded beat signal increases as the number of chips per chirp raises (code bandwidth becomes larger, e.g. the normalized code bandwidth becomes $B_c/f_s=0.12$ for $N_c=1024$ and $f_s=160$ MHz). Using the BPSK code with large bandwidth comparable to the sampling frequency and filtering of the spectrum leads to increased sidelobe level. To cope with this problem, we can apply a smoother.

\subsection{Gaussian PC-FMCW}

A smoother can be applied to the phase of the code signal to reduce the spectrum widening of the coded beat signal. In this paper, we have used the Gaussian filter as a smoother for analysis. However, a different smoothing filter can be selected depending on the required spectral behavior of the application. The Gaussian filter can be represented as:
\begin{equation}
\begin{split}
    h(t)&=\sqrt{\frac{2\pi}{\ln{2}}} B_s e^{-\frac{2{\pi}^2 {B_s}^2}{\ln{2}}t^2}\\
    &=\frac{\eta}{\sqrt{\pi}} \, e^{- {\eta}^2 t^2},
\end{split}
\end{equation}
where $\eta=\sqrt{\frac{2{\pi}^2 {B_s^2}}{\ln{2}}}$ and $B_s$ is the $3$-dB bandwidth of the Gaussian filter. The Gaussian filter in frequency domain can be written as \cite{RouphaelBook2009}:
\begin{equation}
 H(f)=e^{-\frac{\ln(2)}{2} \left(\frac{f}{B_s}\right)^2} .
\end{equation}
Applying Gaussian filter to the binary code, we obtained the Gaussian binary code $\phi_{\text{gauss}}(t)=\phi_{\text{bpsk}}(t)\otimes h(t)$ as demonstrated in Figure~\ref{fig:4}. The instantaneous frequency for the Gaussian binary code can be written as:
\begin{equation}\label{instant_gauss}
    \frac{1}{2\pi}\frac{d}{dt}\phi_{\text{gauss}}(t)=\frac{\eta}{2\pi\sqrt{\pi}} \sum\limits_{n = 1}^{{N_c}} (\phi_{n+1} -\phi_n ) e^{-{\eta}^2(t-n{T_c})^2}.
\end{equation}
The proof is given in Appendix A. The equation \eqref{instant_gauss} shows that the abrupt phase changes are smoothed by the Gaussian filter, and the term $e^{-{\eta}^2(t-n{T_c})^2}$ is expected when the phase changes with respect to time. This can be seen in Figure~\ref{fig:5} b that the phase changes cause Gaussian shape in the instantaneous frequency of the Gaussian binary code.

\begin{figure}[t]

         \centerline{\includegraphics[width=0.9\linewidth]{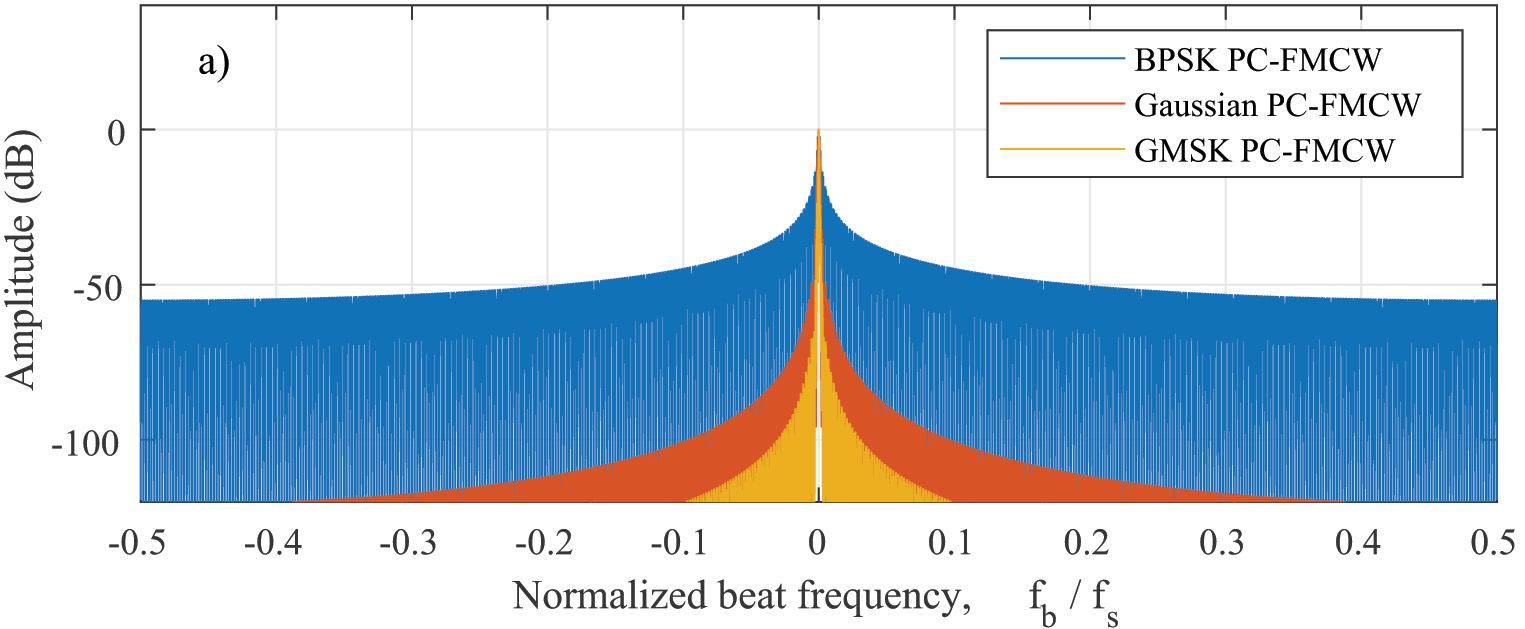}}
          \hfill\vfill
         \centerline{\includegraphics[width=0.9\linewidth]{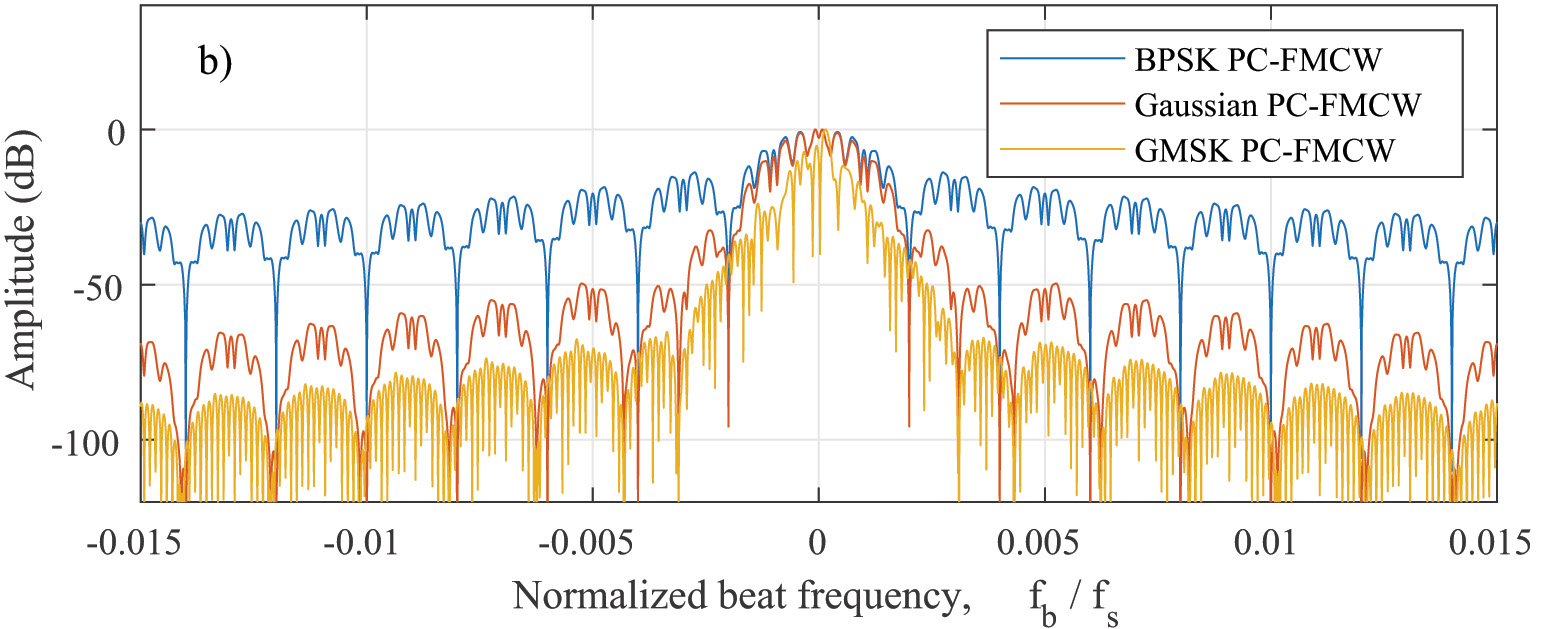}}
         
         \caption{Frequency spectrum comparison for the coded beat signals associated with three PC-FMCW waveforms with $N_c=16$: a) Full-band b) Zoomed}
\label{fig:6}
\end{figure}

In addition, the mixer output of Gaussian PC-FMCW can be represented as:
\begin{equation}
x_{\text{M}\textsubscript{gauss}}(t)=\alpha_0\left({c(t-\tau_0)\otimes h(t-\tau_0)}\right)e^{j(2\pi f_b t)},
\end{equation}
and its frequency spectrum can be written as:
\begin{equation}\label{eqGauss}
\begin{split}
    X_{\text{M}\textsubscript{gauss}}(f)=&\alpha_0C(f-f_b) \, \, H(f-f_b) \, e^{-j\left(2\pi (f-f_b) \tau_0\right)} \\
    =&\frac{\alpha_0}{T} \sum\limits_{n = 1}^{{N_c}} {e ^{j{\phi_n}}} {\rm sinc}((f-f_b)T_c) \\
    & \, e^{-j\left(2\pi (f-f_b) (\tau_0 + (n-\frac{1}{2})T_c)\right)} \, e^{-\frac{\ln(2)}{2} (\frac{f-f_b}{B_s})^2}.
\end{split}
\end{equation}
As seen in \eqref{eqGauss}, the first null of the mixer output for the Gaussian PC-FMCW is decided by the ${\rm sinc}$ function bounded by the Gaussian filter $H(f)$. Consequently, the first null location becomes $f_b + B_s$, and the required cut-off frequency to include the main lobe becomes:
\begin{equation}\label{minimumLPF}
    f_{\text{cut}}\geq f_{b_{\text{max}}}+ B_s.
\end{equation}
The frequency spectrum of the mixer output for Gaussian PC-FMCW \eqref{eqGauss} is demonstrated in Figure~\ref{fig:6} in red color. We observe that using a Gaussian filter reduces the spectrum widening of the coded beat signal. This can be seen in Figure~\ref{fig:71} where the spectrum width of the coded beat signal is lowered for Gaussian PC-FMCW compared to the BPSK coded signal.

\subsection{GMSK PC-FMCW}
Gaussian minimum shift keying (GMSK) is a popular modulation scheme in communication due to its low spectral spread. In GMSK, the binary code signal is filtered by a Gaussian filter, and the filtered code is integrated over time \cite{SimonBook2001,Svedek2009}. The resulting phase for GMSK becomes $\phi_{\text{gmsk}}(t)=\int_{-\infty}^{\infty}\phi_{\text{gauss}}(t)dt=\int_{-\infty}^{\infty} (\phi_{\text{bpsk}}(t)\otimes h(t)) dt$ as shown in Figure~\ref{fig:4}. The instantaneous frequency for the GMSK phase code can be obtained as:
%
\begin{equation}\label{instant_gmsk}
   \frac{1}{2\pi}\frac{d}{dt}\phi_{\text{gmsk}}(t)=\frac{1}{4\pi}\sum\limits_{n = 1}^{{N_c}} (\phi_{n+1} -\phi_n ) \, {\rm erf} \left(\eta (t-nT_c)\right).
\end{equation}
where $\rm{erf}(t)$ represents the error function. The proof is given in Appendix A. The \eqref{instant_gmsk} demonstrates that the Dirac delta term seen in BPSK coding due to abrupt phase change is replaced by the term ${\rm erf} \left(\eta (t-nT_c)\right)$ for the GMSK phase code. This behaviour can be observed in Figure~\ref{fig:5} c, where the phase changes lead to error functions (combination of left and right parts gives a smoothed rectangle shape) in the instantaneous frequency of the GMSK phase code.

\begin{figure}[t]
         \centerline{\includegraphics[width=0.9\linewidth]{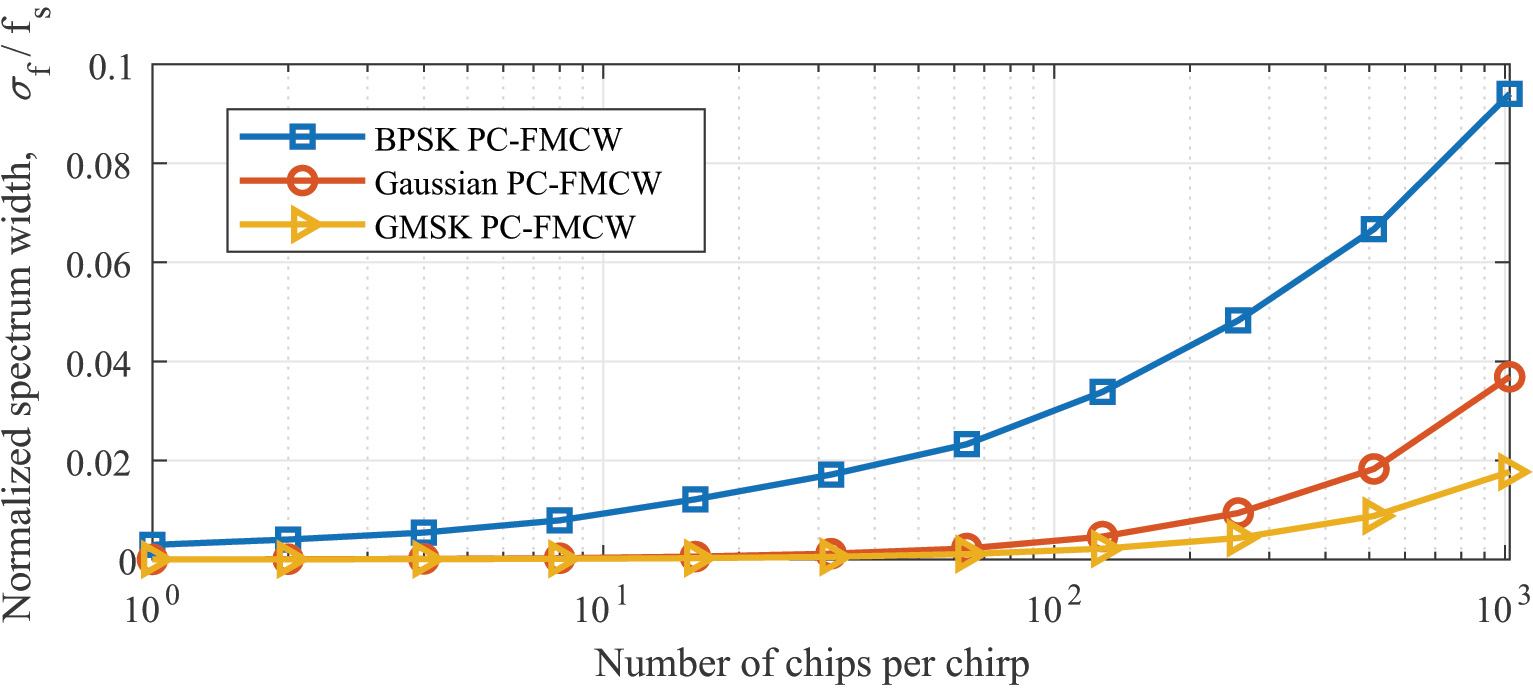}}
         \caption{Spectrum width of the coded beat signals associated with three PC-FMCW waveforms versus number of chips per chirp. Normalized code bandwidth $B_c/f_s=0.12$ for $N_c=1024$}
\label{fig:71}
\end{figure}

Subsequently, the mixer output of GMSK PC-FMCW can be represented as:
\begin{equation}
    x_{\text{M}\textsubscript{gmsk}}(t)=\alpha_0e^{j\phi_{\text{gmsk}}(t-\tau_0)} e^{j(2\pi f_b t)},
\end{equation}
and its frequency-domain representation can be written as \cite{RouphaelBook2009}:
\begin{equation}\label{eqGMSKFre}
\begin{split}
    X_{\text{M}\textsubscript{gmsk}}(f)\approx&\frac{\alpha_0}{T} \sum\limits_{n = 1}^{{N_c}} {e ^{j{\phi_n}}} {\rm sinc}^2((f-f_b)T_c) \\
    & e^{-j2\pi (f-f_b) (\tau(t)+(n-1/2)T_c)} \, e^{-\frac{\ln(2)}{2} (\frac{f-f_b}{B_s})^2}.
\end{split}
\end{equation}
Similar to the Gaussian case, the Gaussian filter $H(f)$ bounds the frequency components of the mixer output for GMSK PC-FMCW as shown in \eqref{eqGMSKFre}. Thus the first null location becomes $f_b + B_s$, and the required cut-off frequency to include the main lobe is the same as \eqref{minimumLPF}. In addition, GMSK PC-FMCW has a ${\rm sinc}^2$ term instead of a ${\rm sinc}$ function, which is seen in the BPSK code. This is because the GMSK phase code has a smoothed triangular shape while BPSK has rectangular, as illustrated in Figure~\ref{fig:4}. 

The frequency spectrum of the mixer output for GMSK PC-FMCW \eqref{eqGMSKFre} is shown in Figure~\ref{fig:6} in yellow color. It can be seen that taking the square of the ${\rm sinc}$ function and bounding it by the Gaussian filter further reduces the spectrum widening of the coded beat signal. We observe this in Figure~\ref{fig:71} as the spectrum width of the coded beat signal associated with GMSK PC-FMCW is lower (especially for large code bandwidth) compared to both BPSK PC-FMCW and Gaussian PC-FMCW. Consequently, better sensing performance (i.e. lower sidelobe level) is expected for a GMSK PC-FMCW when the bandwidth of the code increases and becomes comparable to ADC sampling.

\begin{figure}[t]
    \centering
    \includegraphics[ width=1\linewidth]{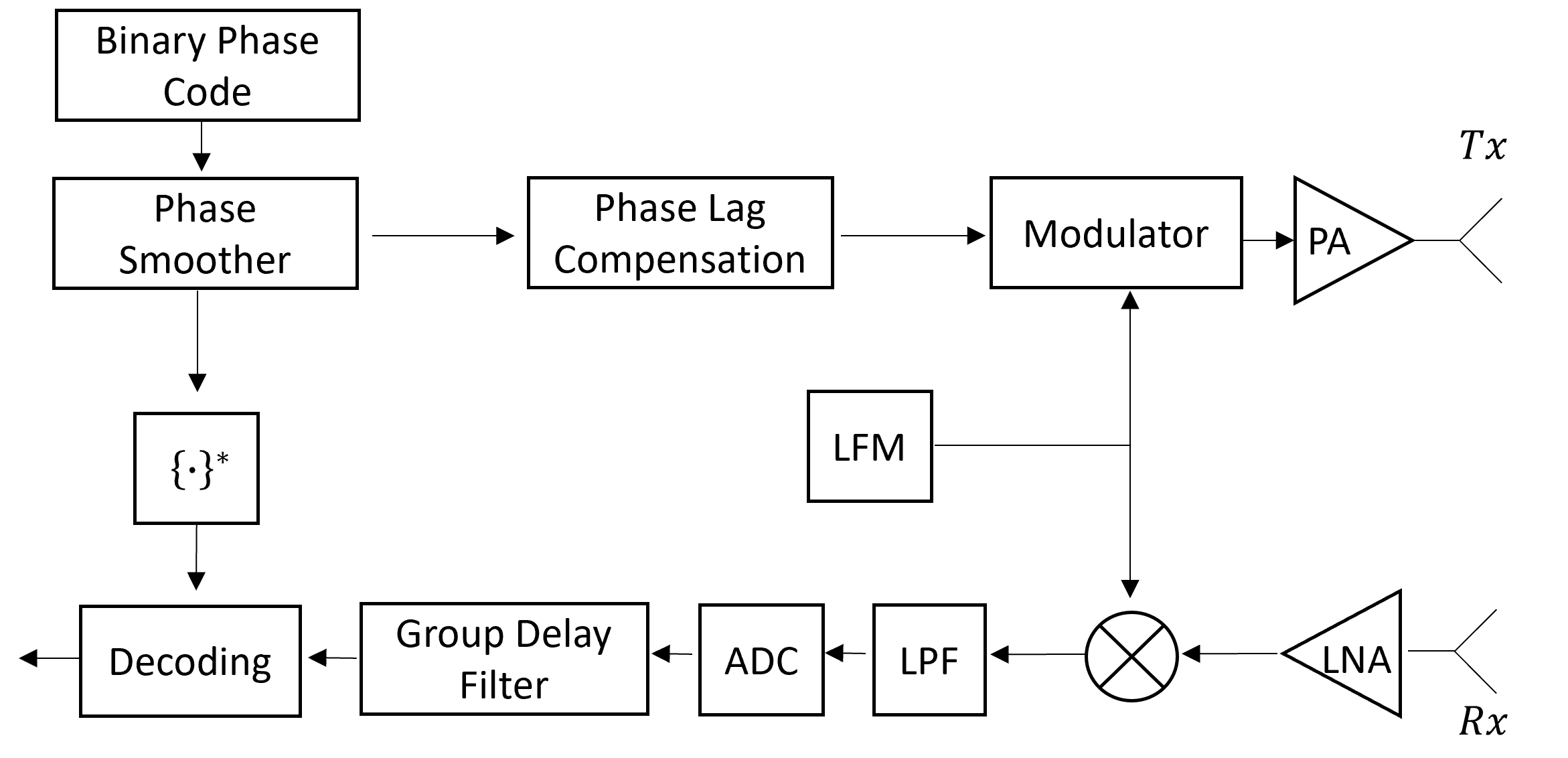}
    \caption{Block diagram of the proposed PC-FMCW transceiver structure}
    \label{fig:7}
\end{figure}

\begin{figure}[t]

         \centerline{\includegraphics[width=0.9\linewidth]{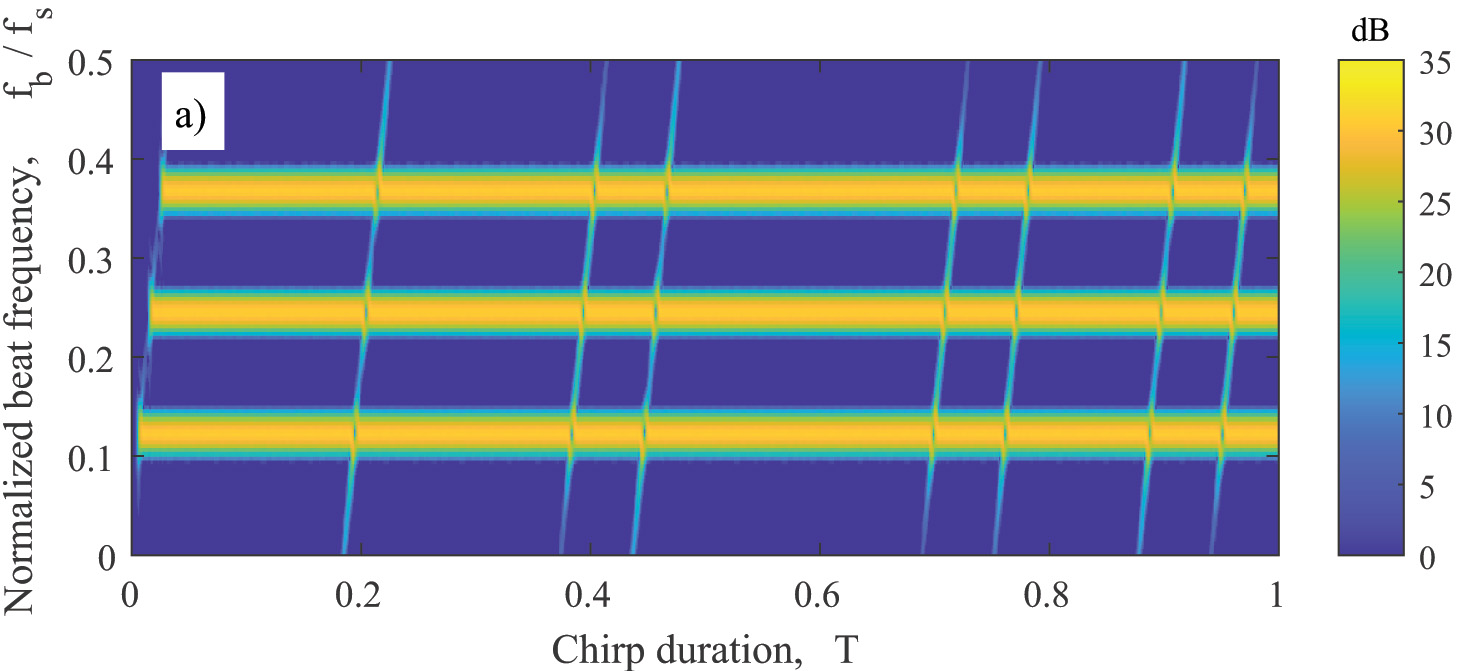}}
         \hfill
         \centerline{\includegraphics[width=0.9\linewidth]{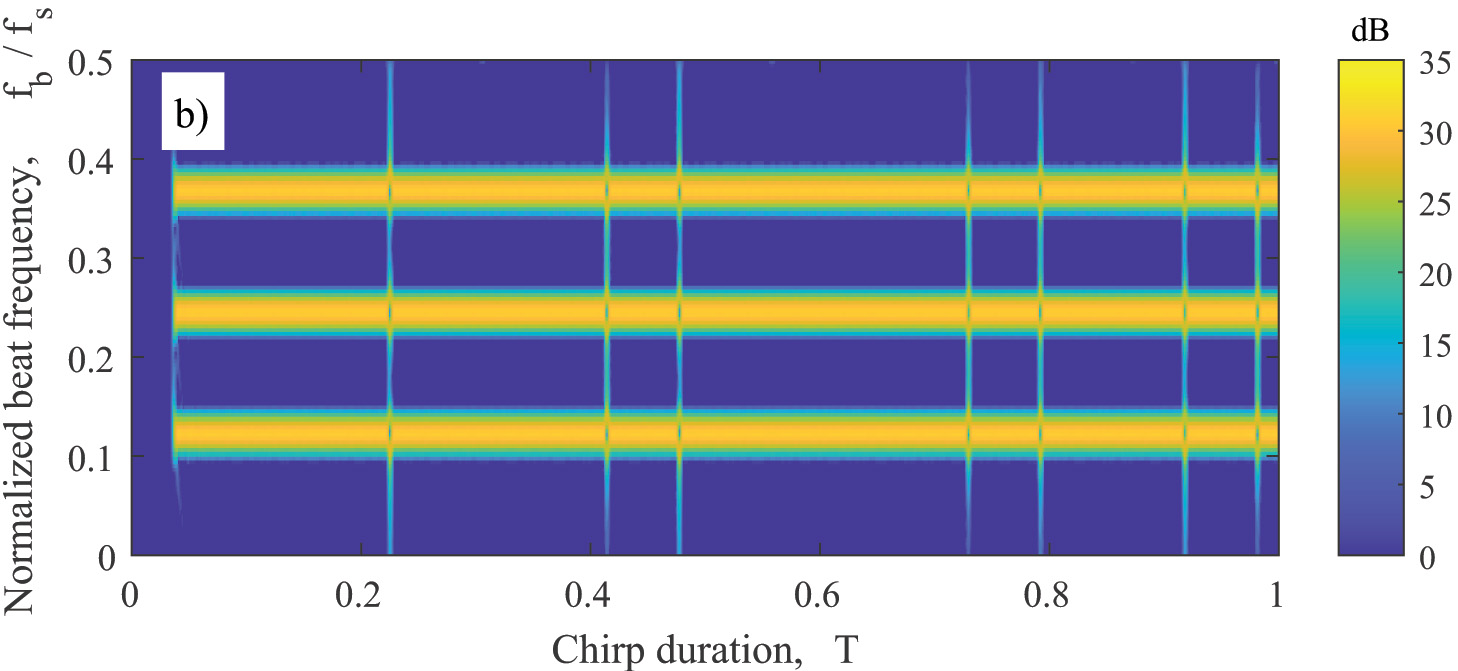}}
         \hfill
         \centerline{\includegraphics[width=0.9\linewidth]{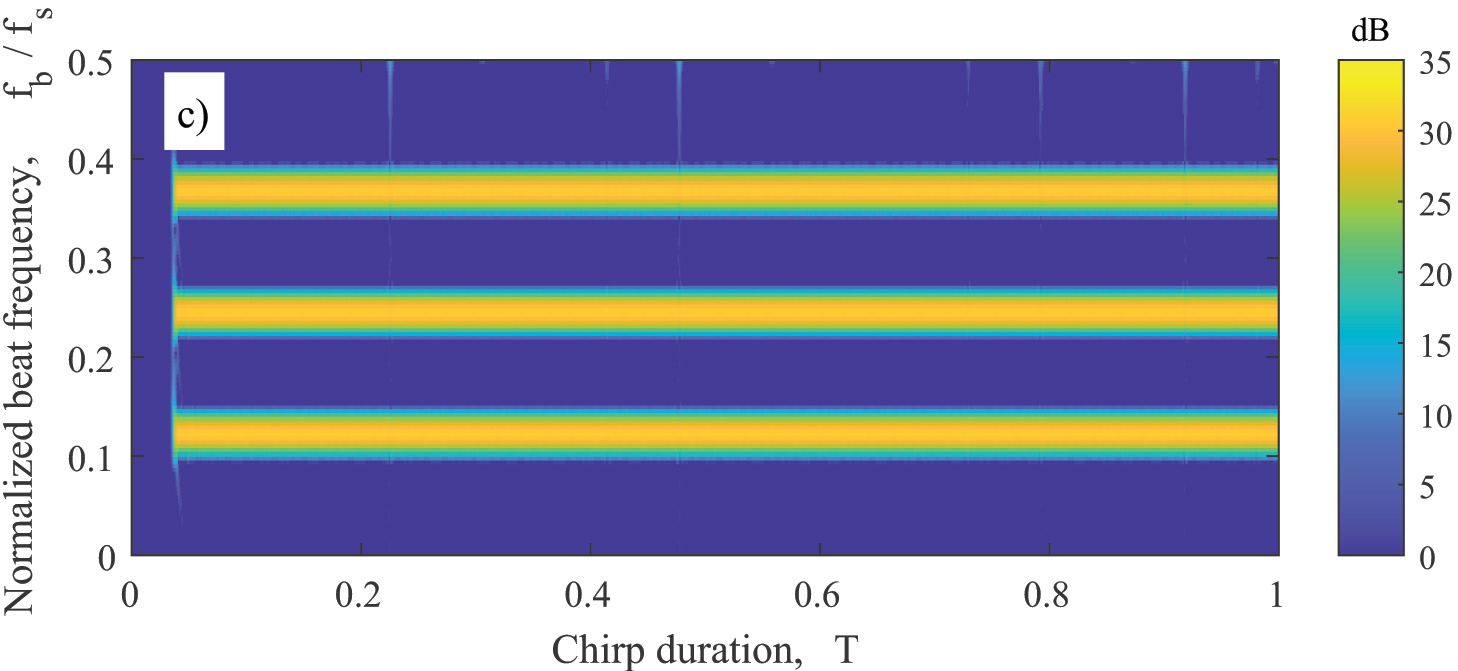}}
           
         \caption{Spectrogram of the BPSK phase-coded beat signals ($N_c=16$) with phase lag compensation applied: a) Before the group delay filter b) After the group delay filter c) After decoding the group delay filter output}
\label{fig:8}
\end{figure}

\section{Phase Lag Compensated Waveforms}\label{sec:PhaseLagCompensation}
This section introduces the phase lag compensated waveforms, and the proposed block diagram is shown in Figure~\ref{fig:7}. The signals that are modified due to the implementation of the phase lag compensation are denoted with a symbol ($\hat{.}$).

The group delay filter applies different time delays to each frequency component and causes a dispersion effect on the phase-coded signal, which leads to a distorted range profile, as explained in Section~\ref{sec:Signal_Model}. To eliminate the undesired effect of the group delay filter, we perform quadratic phase lag compensation on the transmitted code by multiplying its spectrum with the quadratic phase term as \cite{QuadraticPhase}:
 \begin{equation}
    \hat{S}(f)=S(f)e^{-j\frac{\pi f^2}{k}}.
\end{equation}
Then the mixer output \eqref{beat_signal} becomes:
\begin{equation}\label{PhaseLag_beat_signal}
    \hat{x}_{\text{M}}(t)=\alpha_0\hat{s}(t-\tau_0) e^{j(2\pi f_b t)},
\end{equation}
and
\begin{equation}
    \hat{X}_{\text{M}}(f)=\alpha_0S(f-f_b)e^{-j\left(\frac{2\pi f_b }{k}(f-f_b)+\frac{\pi }{k}(f-f_b)^2\right)},
\end{equation}
for time and frequency domain representation, respectively. Subsequently, the output of the group delay filter in the frequency domain becomes:
\begin{equation}\label{FreqWithPre}
\begin{split}
\hat{Z}_{\text{o}}(f)&=\hat{X}_{\text{M}}(f) e^{j\left(\frac{\pi {f_b}^2}{k}+\frac{2\pi f_b}{k}(f-f_b)+\frac{\pi}{k} (f-f_b)^2\right)}\\
       &=\alpha_0S(f-f_b) e^{j\left(\frac{\pi {f_b}^2}{k}\right)}\\
       &=\alpha_0S(f-f_b),
\end{split}
\end{equation}
where $\mathrm{exp}\left(\frac{\pi {f_b}^2}{k}\right)$ is a constant phase term (does not depend on frequency f), and thus it can be incorporated into $\alpha_0$. Note that the undesired term $\frac{\pi}{k}(f-f_b)^2$ caused by the phase response of the filter \eqref{phaseTaylor} is eliminated with the phase lag compensation. After taking the inverse Fourier transform, the time-domain representation of the new group delay filter output \eqref{FreqWithPre} becomes:
\begin{equation}\label{new_group_inTime}
    \hat{z}_{\text{o}}(t)=\alpha_0s\left(t\right)  e^{j(2\pi f_b t)}.
\end{equation}

In addition, we can shift the group delay filter output to the maximum delay (defined by maximum beat frequency, $\tau_{\text{max}}=\frac{f_{b_{\text{max}}}}{k}$) by multiplying its spectrum with linear phase delay $\mathrm{exp}\left(-j2\pi f \tau_{\text{max}}\right)$ for physical correctness and guarding the beginning of the next chirp. Consequently, the \eqref{new_group_inTime} is shifted to the maximum time delay as:

\begin{equation}\label{group45}
    z_{\text{a}}(t)=\alpha_0s\left(t-\tau_{\text{max}}\right)  e^{j(2\pi f_b t)}.
\end{equation}

The spectrogram of the BPSK phase-coded beat signals with phase lag compensation applied are shown in Figure~\ref{fig:8} ($B=2$ GHz, $T=51.2$ $\mu$s and $N_c=16$) for three cases: before the group delay filter, after the group delay filter and after decoding the group delay filter output. Note that by using phase lag compensation, the spectrum of the code signals is non-linearly shifted in the opposite direction before applying the group delay filter (Figure~\ref{fig:8} a). An example of the resulting signal \eqref{group45} can be seen in Figure~\ref{fig:8} b. It is observed that the group delay dispersion effect on the code signal is eliminated, and each coded beat signal is perfectly aligned after the group delay filter. Subsequently, we can apply the decoding signal, which is the complex conjugate of the reference phase code shifted to maximum beat frequency, and the decoded beat signal becomes:
\begin{equation}\label{eq_no_phase_error}
\begin{split}
         \hat{z}_{\text{d}}(t)&= z_{\text{a}}(t) s^*\left(t-\tau_{\text{max}}\right)\\
     &=  \alpha_0e^{j(2\pi f_b t )} .
\end{split}
\end{equation}

\begin{figure}[t]
         \centerline{\includegraphics[width=0.9\linewidth]{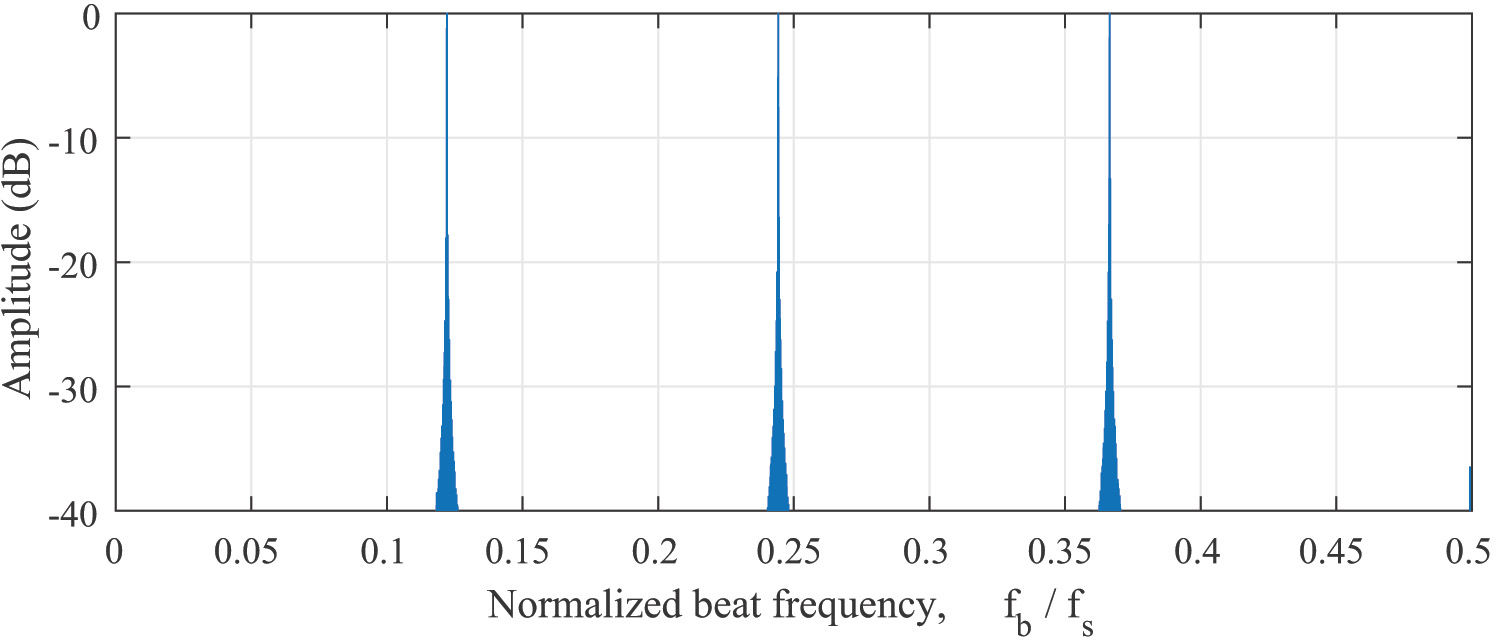}}
         \caption{The recovered range profile of decoded signal after using phase lag compensation and group delay filter ($N_c=64$)}
\label{fig:9}
\end{figure}

It can be seen in \eqref{eq_no_phase_error} that the code term is removed properly, and the residual phase error caused by the imperfection in decoding is eliminated by using the phase lag compensation (Figure~\ref{fig:8} c). As a consequence, the distorted range profile shown in Figure~\ref{fig:3} is recovered for a wide-band signal where $B_{c}$ is comparable to $f_s$ as illustrated in Figure~\ref{fig:9}. Moreover, the beat signals are obtained similar to the dechirped signal of conventional FMCW radar. This helps re-utilising all software algorithms previously developed for FMCW radar with the proposed waveform and transceiver structure.

\begin{figure}[t]

         \centerline{\includegraphics[width=0.9\linewidth]{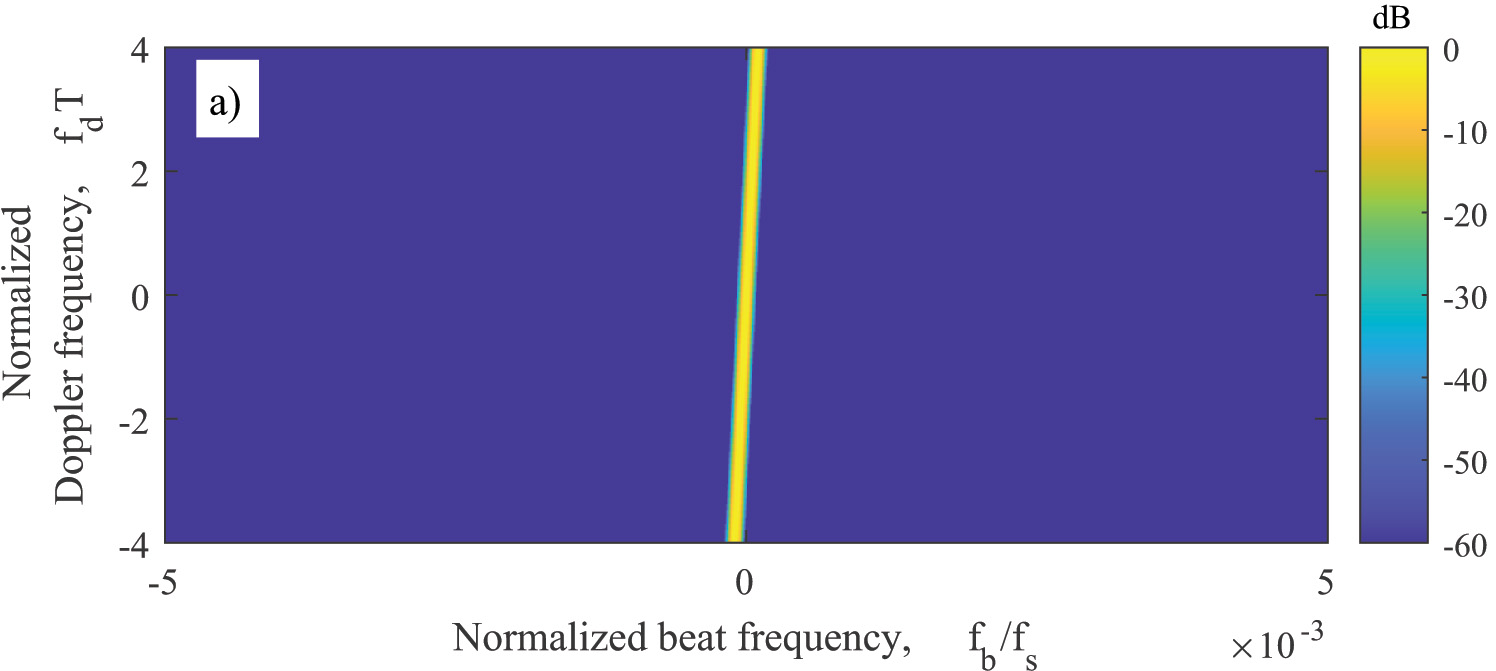}}
         \hfill\vfill
         \centerline{\includegraphics[width=0.9\linewidth]{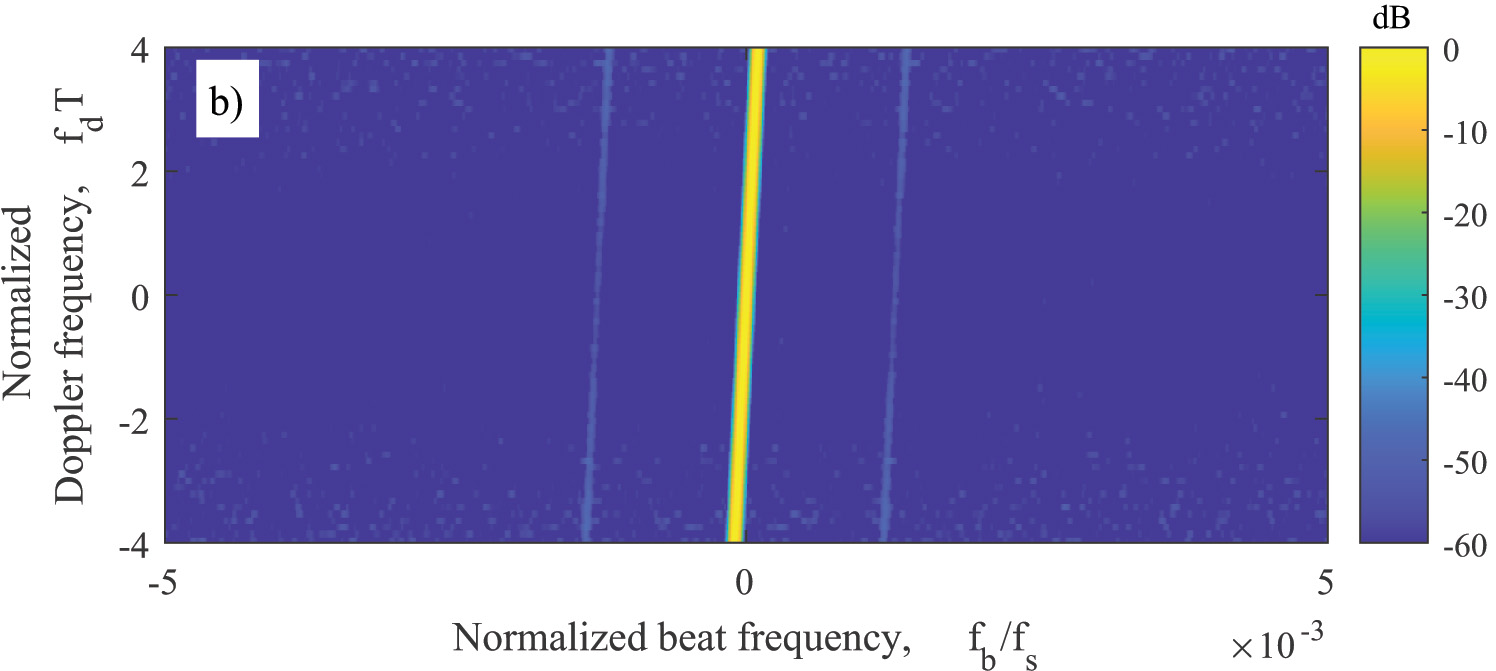}}
         \hfill\vfill
         \centerline{\includegraphics[width=0.9\linewidth]{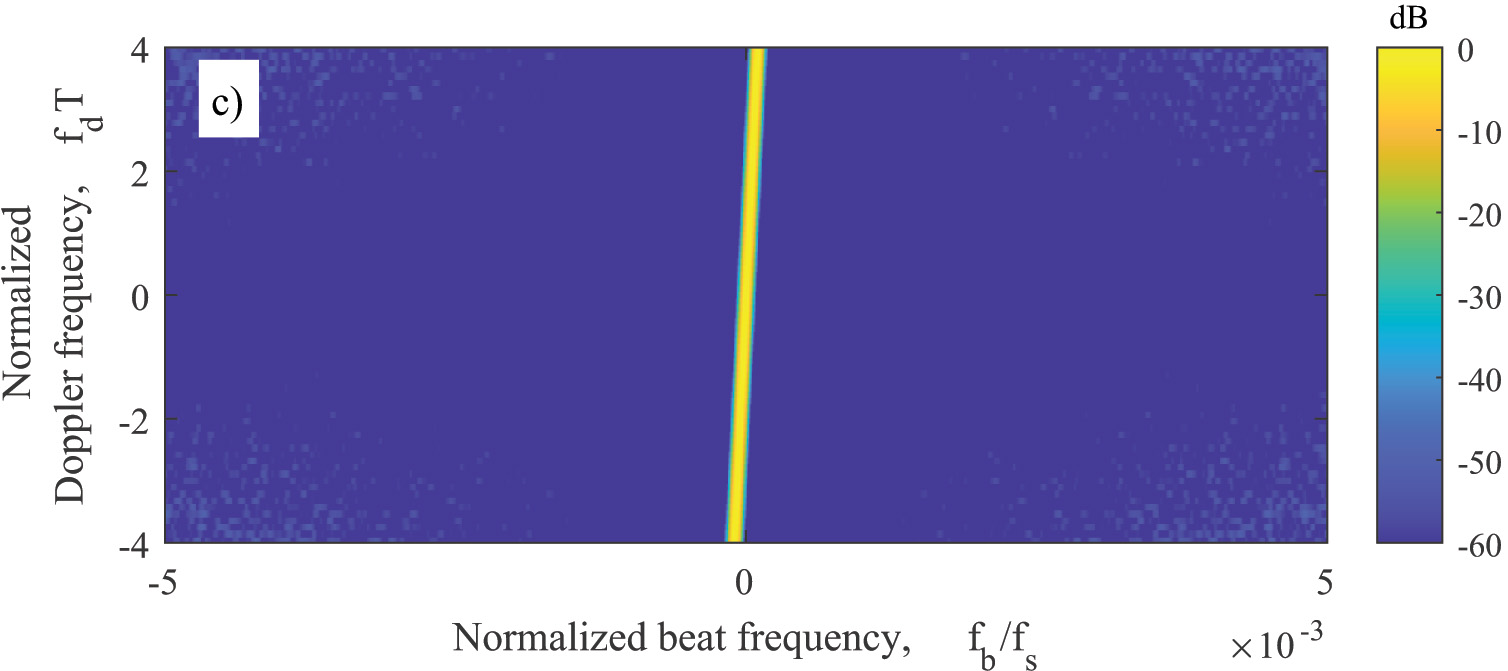}}
         \hfill\vfill
         \centerline{\includegraphics[width=0.9\linewidth]{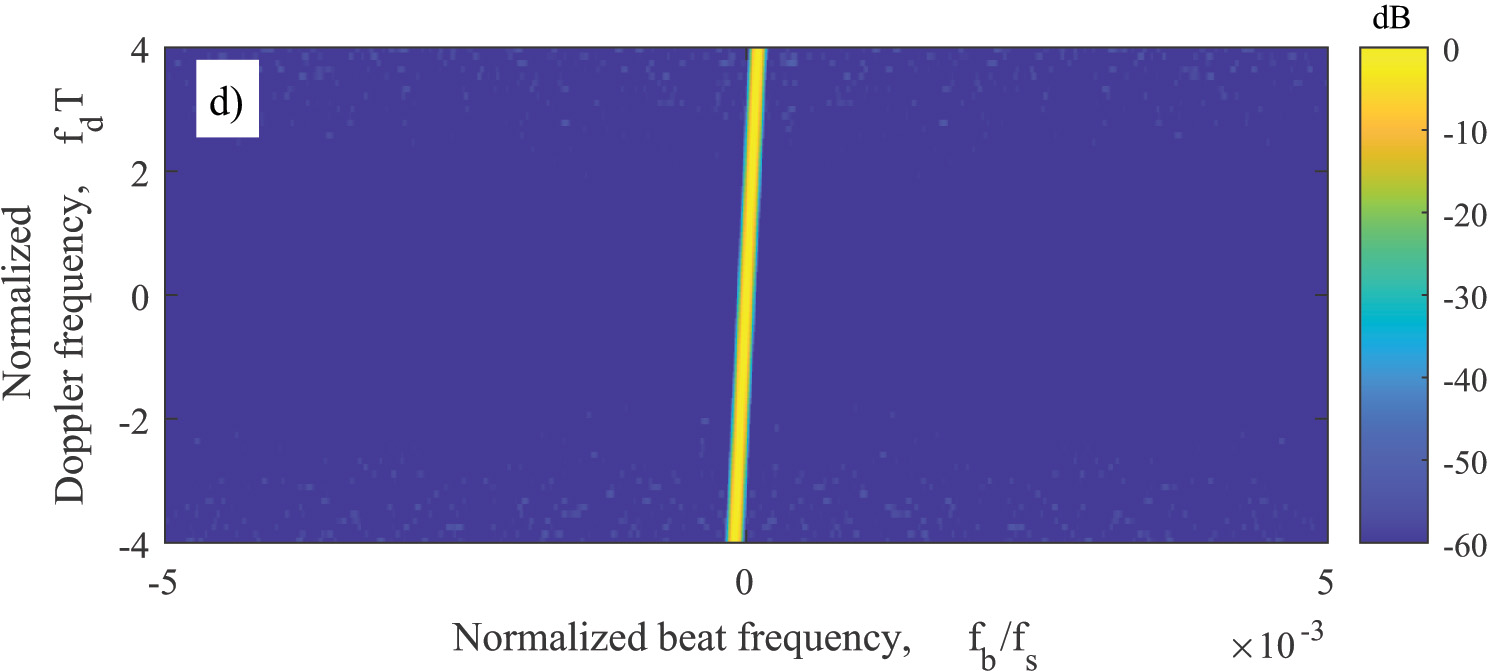}}
           
         \caption{Range profile with different Doppler frequency for FMCW and the phase lag compensated PC-FMCW waveforms with $N_c=1024$: a) FMCW b) BPSK c) Gaussian d) GMSK. The ridge is inclined in all cases but seems narrow due to zooming out the x-axis}
\label{fig:10}
\end{figure}

\section{Waveform Properties}\label{sec:BandwidthLimitation}

This section provides the properties of the phase lag compensated PC-FMCW. For the numerical simulations, we consider a radar operating with a carrier frequency $f_c=3.315$ GHz and transmitting the investigated waveforms with the chirp duration $T=1$ ms and the chirp bandwidth $B=200$ MHz. The phase lag compensated signal $\hat{s}(t)$ is used for phase coding, and we have used the random code sequence for all three PC-FMCW. The duration of the chip $T_c$ is controlled with the number of chips per chirp $N_c$ as $T_c=T/N_c$. To achieve a smoothed phase transition, the $3$-dB bandwidth of the Gaussian filter is set to two times the chip bandwidth $B_s=2B_c$. On the receiver side, \eqref{PhaseLag_beat_signal} is low-pass filtered with the cut-off frequency $f_{cut}=\pm20$ MHz and sampled with $f_s=40$ MHz. As a consequence, we have $N=40000$ range cells (fast-time samples) for this setting. The group delay filter is applied to the sampled signal to align the beat signals of different targets. Before decoding, the same LPF is applied to the reference phase-coded signal to prevent a signal mismatch. To focus on the waveform properties, we assume a noise-free scenario in the numerical simulations.

\subsection{Sensing}

The sensing performance of the phase lag compensated waveforms are assessed by using the investigated processing method and compared with FMCW. After proper decoding, the code term is removed, and the beat signal is recovered similar to the dechirped signal of traditional FMCW as explained in Section~\ref{sec:PhaseLagCompensation}.

To investigate the Doppler tolerance of the waveforms and proposed receiver strategy, we simulate the received signal after dechirping \eqref{equation5} as a function of Doppler frequency shift and plot the outcome of the introduced processing approach in a form similar to the ambiguity function in Figure~\ref{fig:10}. The presented plots show the behaviour of FMCW and three phase lag compensated PC-FMCW with $N_c=1024$ after processing. It can be seen that the inclined ridge associated with the chirped waveform ambiguity function is present and the same in a, b, c, and d. Thus, all considered waveforms have the Doppler tolerance of FMCW and exhibit the range-Doppler coupling, determined by the slope of the carrier chirp. Note that the x-axis is zoomed out to highlight the sidelobe differences between waveforms, and hence the inclined ridge seems like a narrow line. In the vicinity of the main lobe, they all have an identical response, determined by $100$ dB Chebyshev window, applied to the signals before range FFT. The sidelobes of three phase lag compensated PC-FMCW raises with the Doppler frequency shift; among them, the range profile degradation is minimal for GMSK.

\begin{figure}[t]

        \centerline{\includegraphics[width=0.9\linewidth]{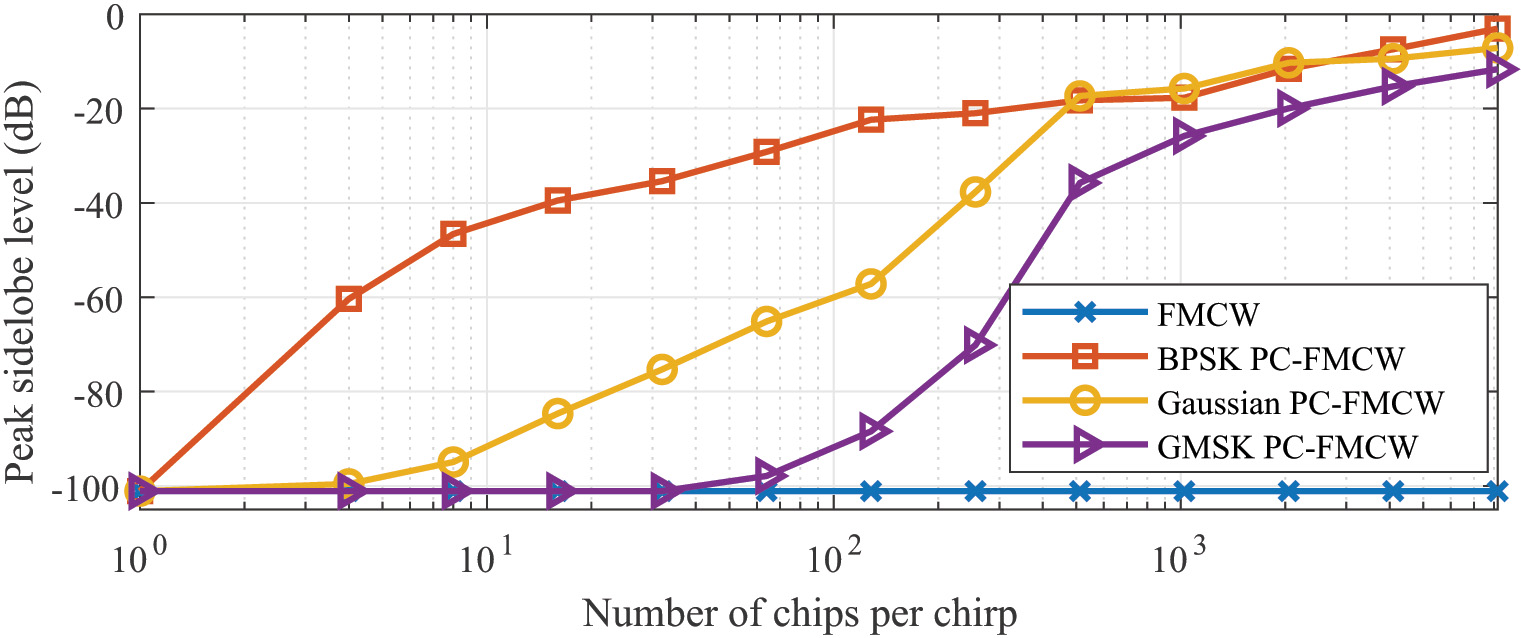}}
         
          \centering
         \textit{a)}
         
         \centerline{\includegraphics[width=0.9\linewidth]{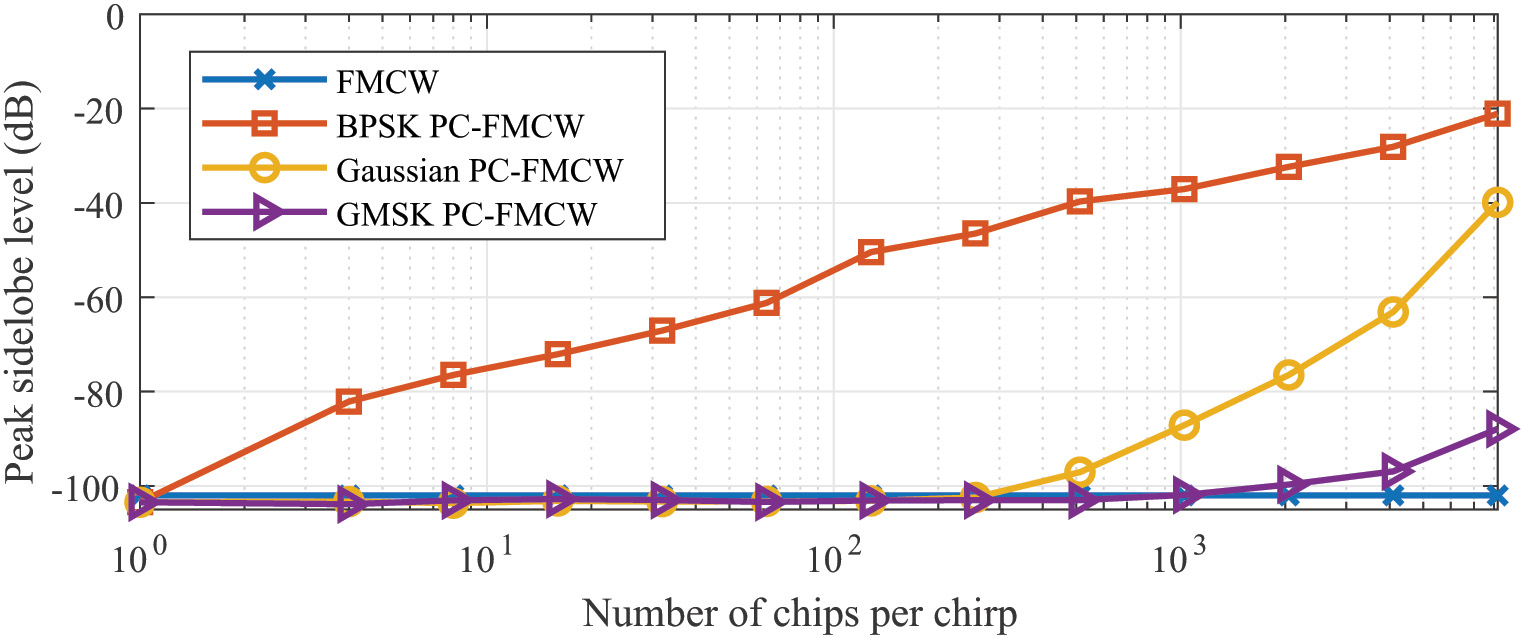}}
         
          \centering
         \textit{b)}
         
         \caption{PSL of the investigated waveforms at normalized target range $R/R_{\text{max}}=0.4$ versus the number of chips per chirp: a) No phase lag compensation b) With phase lag compensation}
\label{fig:11}
\end{figure}

The bandwidth of the chip $B_c$ raises as the number of chips per chirp $N_c$ increases. Consequently, the bandwidth of the chip becomes comparable to ADC sampling frequency, and the sidelobe level increases with the filtering of the spectrum. However, the spectrum widening of the coded beat signal is different for the three phase lag compensated PC-FMCW as explained in Section~\ref{sec:TypeOfPhases}. Therefore, they provide different peak sidelobe level (PSL). The beat signal PSL is defined by the maximum amplitude of the signal spectrum outside of the main lobe (first nulls) and can be written as:
\begin{equation}
    \text{PSL}= \max_{f \in {\mathcal{L} }} \left|\hat{Z}_{\text{d}}(f)\right|  \, \,\quad \mathcal{L}= (-\infty,-f_{l}) \cup (f_{r}, \infty),
\end{equation}
where $f_{l}$ and $f_{r}$ denote the frequency corresponding to the left and right parts of the first null, respectively, and $\mathcal{L}$ denotes the frequency interval.

\begin{figure}[t]

         \centerline{\includegraphics[width=0.9\linewidth]{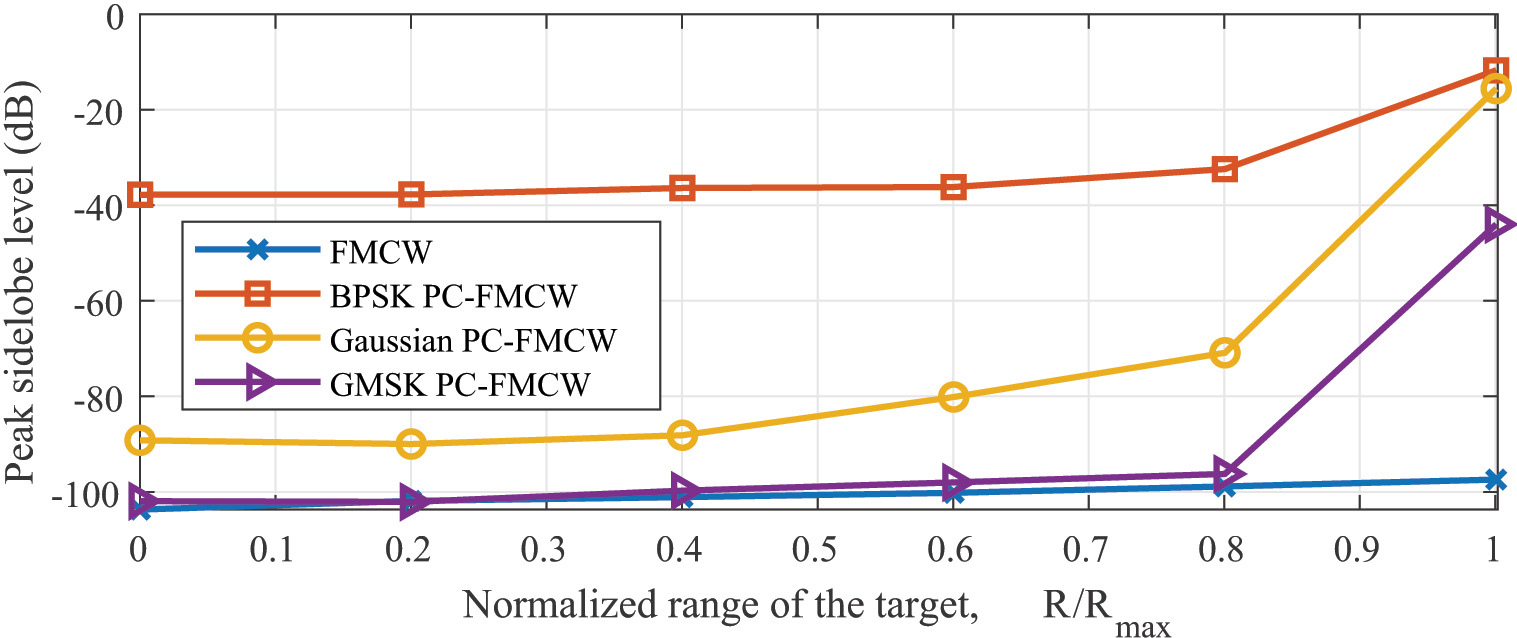}}
         
         \caption{PSL of the investigated waveforms (with phase lag compensation) with $N_c=1024$ versus the normalized range of the target with respect to the maximum range}
\label{fig:12}
\end{figure}

Next, we investigate the zero Doppler cuts of waveforms and compare their respective peak sidelobe levels. PSL of the investigated waveform at normalized target range $R/R_{\text{max}}=0.4$ as a function of the number of chips per chirp is demonstrated in Figure~\ref{fig:11} where the maximum range $R_{\text{max}}=\frac{c f_{b_{\text{max}}}}{2k}$ and the maximum beat signal is determined as $f_{b_{\text{max}}}=f_s/2$. To highlight the benefits of performing phase lag compensation, we also demonstrate the sensing performance of investigated waveforms without performing phase lag compensation in Figure~\ref{fig:11} a. It can be seen that applying phase lag compensation improves the PSL of three PC-FMCW waveforms (Figure~\ref{fig:11} b). Still, the PSL of BPSK PC-FMCW rapidly increases as the number of chips per chirp raises. On the other hand, we observe that the PSL of phase lag compensated GMSK PC-FMCW enhanced substantially, especially for long codes. Particularly, the PSL of GMSK PC-FMCW with $N_c=1024$ improved from $-25$ dB to $-100$ dB by using phase lag compensation. Consequently, GMSK PC-FMCW can provide PSL similar to FMCW up to $N_c=1024$. In addition, we illustrated the PSL of phase lag compensated waveforms with $N_c=1024$ as a function of the normalized target range in Figure~\ref{fig:12}. Note that the spectral widening and filtering of the spectrum become crucial for PC-FMCW as the target approaches the maximum range. GMSK PC-FMCW has favorable sensing performance among phase lag compensated waveforms and provides lower PSL.

\begin{figure}[t]

         \centerline{\includegraphics[width=0.9\linewidth]{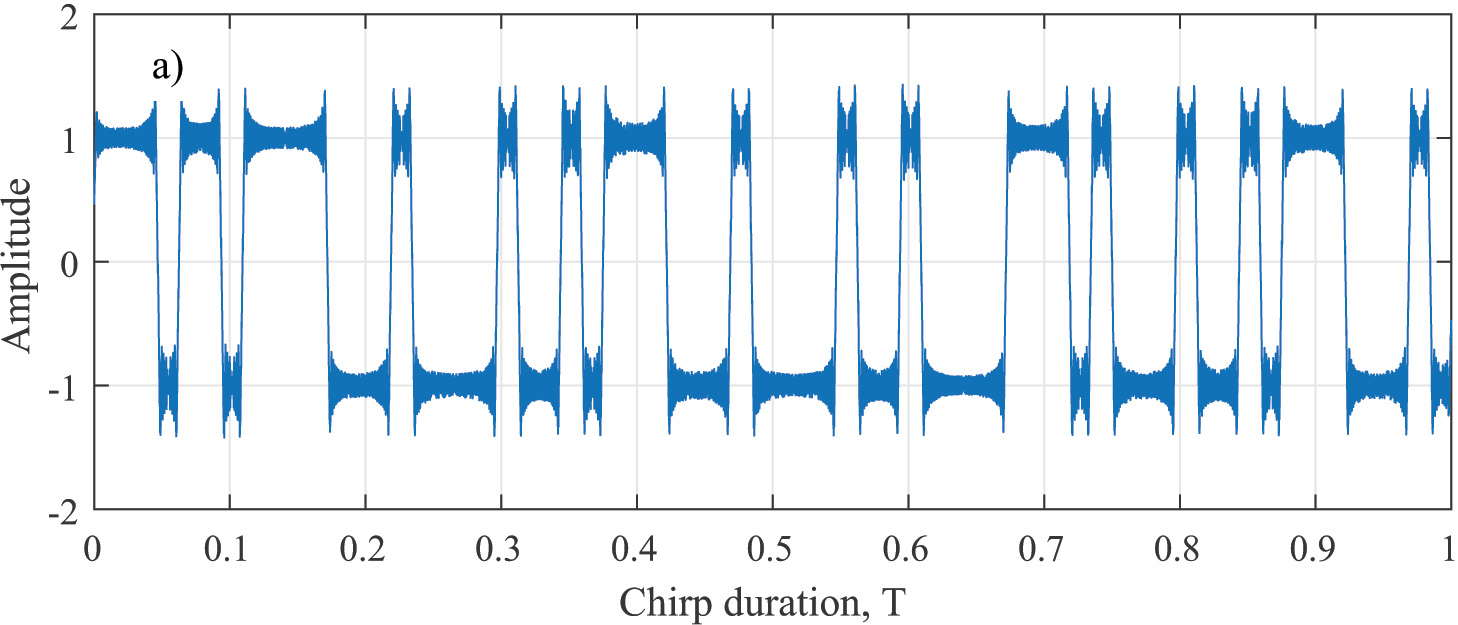}}
         \hfill
         \vfill
         \centerline{\includegraphics[width=0.9\linewidth]{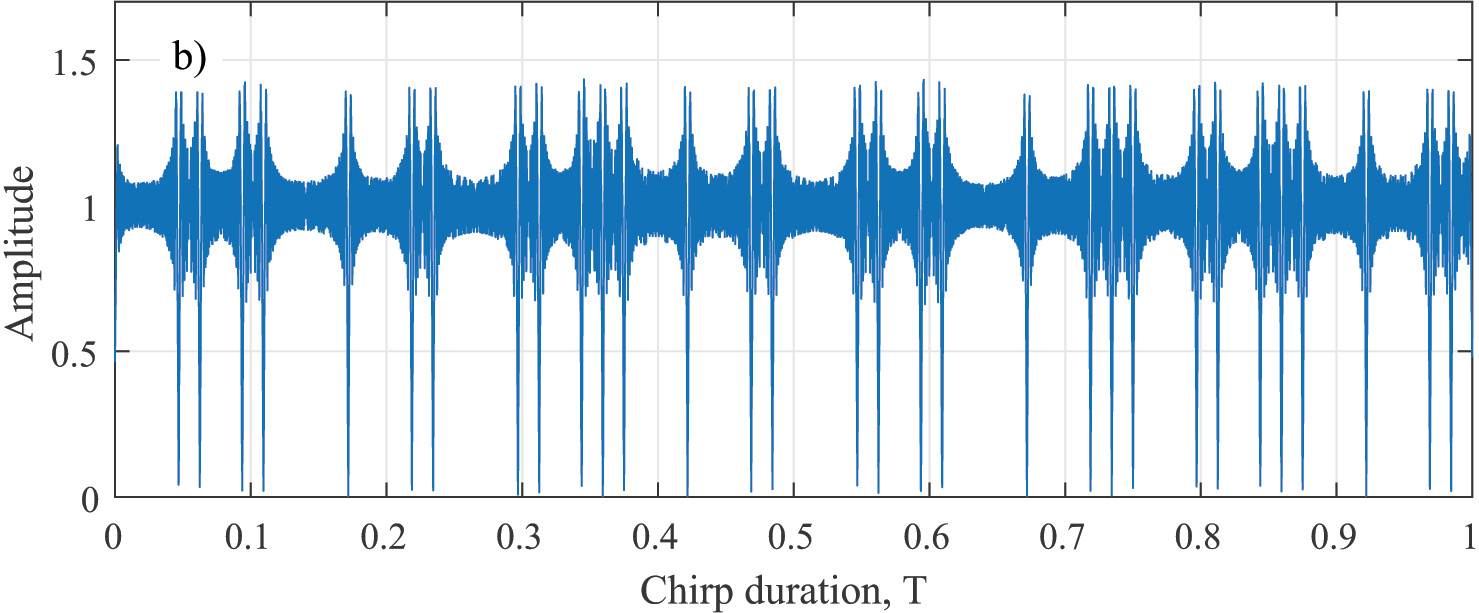}}
           
         \caption{Time-varying amplitude due to phase lag compensation: a) BPSK code $N_c=64$ b) Absolute value of the transmitted BPSK PC-FMCW }
\label{fig:88}
\end{figure}

\subsection{Peak to Average Power Ratio}
The quadratic phase lag compensation is applied to the spectrum of the transmitted code to eliminate the dispersion effect of the group delay filter. This quadratic phase lag compensation filter can be represented as:
\begin{equation}
    H_{\text{lag}}(f)=e^{-j\frac{\pi f^2}{k}},
\end{equation}
and the phase lag compensated code term in the time-domain can be written as:
\begin{equation}
    \hat{s}(t)=s(t) \otimes h_{\text{lag}}(t).
\end{equation}
To analyse the effect of quadratic phase lag compensation on phase-coded signal, let $\xi=j\frac{\pi}{k}$, then the quadratic phase lag compensation filter in the time-domain can be written as:
\begin{equation}\label{phaseLagEq}
\begin{split}
    h_{\text{lag}}(t)&=
    \int_{-\infty}^{\infty} e^{-\xi f^2} e^{j2\pi ft}\,df \\
    &=e^{-\frac{{\pi}^2 t^2}{\xi}}\int_{-\infty}^{\infty} e^{-\left(\sqrt{\xi}f-j\frac{\pi t}{\sqrt{\xi}}\right)^2} \,df \\
    &=\sqrt{\frac{k}{j}} e^{-\pi \frac{k t^2}{j}}.
\end{split}
\end{equation}
Subsequently, the result of the convolution for the BPSK code sequence becomes:
\begin{equation}\label{phase_lag_effect}
\begin{split}
    \hat{s}(t)&=c(t) \otimes h_{\text{lag}}(t)\\
    &=\frac{1}{T} \frac{1}{T_c} \frac{1}{2}\sum\limits_{n = 1}^{{N_c}}{e^{j({\phi_{n+1}}-\phi_n)}}{\rm erf}\left(\sqrt{\frac{\pi k}{j}}\left(t-n T_c\right)\right).
\end{split}
\end{equation}
The proof is given in Appendix B. The amplitude of the phase lag compensated BPSK code is shown in Figure~\ref{fig:88} a. The quadratic phase lag compensation applies different time delays to each frequency component of the transmitted phase code. During phase changes, the phase-coded signal has a wide spectrum and shifting the frequency components non-linearly creates ripples in the time-domain signal (Figure~\ref{fig:88} a). Moreover, the time interval between phase changes becomes shorter for a long code sequence, and ripples in the time-domain are collectively summed up as the adjacent phase shifts interfering with each other. Therefore, the amplitude of the transmitted waveform is not constant anymore (Figure~\ref{fig:88} b).

The time-varying amplitude initiated by the phase lag compensation leads to a high peak-to-average power ratio (PAPR). The PAPR of the signal can be represented as:
\begin{equation}
    \text{PAPR}=\frac{\text{max}{|x_{\text{T}}(t)|}^2}{\lim_{T\to\infty} \frac{1}{2T} \int_{-T}^{T}{|x_{\text{T}}(t)|}^2 \,dt}.
\end{equation}

\begin{figure}[t]

         \centerline{\includegraphics[width=0.9\linewidth]{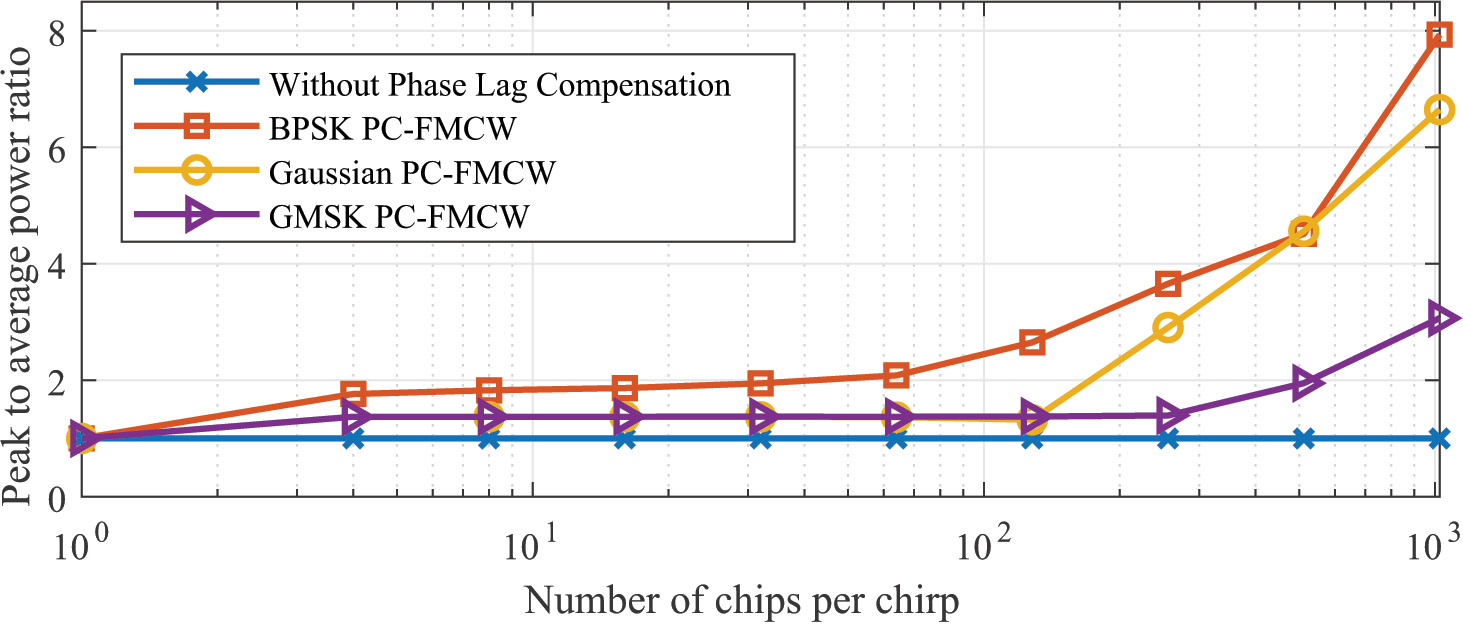}}
         
         \caption{Comparison of PAPR versus number of chips per chirp for phase lag compensated PC-FMCW waveforms}
\label{fig:13}
\end{figure}

The PAPR of the investigated waveforms is compared as a function of $N_c$ in Figure~\ref{fig:13}. It can be seen that the PAPR of the three PC-FMCW without phase lag compensation are constant and equal to 1. The PAPR increases for all three phase lag compensated PC-FMCW as $N_c$ raises. However, the effect of the phase lag compensation and the resulting amplitude variation decrease as the phase transition becomes smoother.  Note that the differences of PAPR between BPSK, Gaussian, and GMSK are comparable up to $N_c=64$ thereafter, PAPR varies notably. For long code sequences, GMSK PC-FMCW provides the lowest PAPR while BPSK PC-FMCW has the highest PAPR since the abrupt phase changes on the BPSK coding are affected more by the frequency-dependent shifts.

\begin{figure*}[!t]
\centering
\minipage{0.5\textwidth}
\centering
\includegraphics[width=0.9\textwidth]{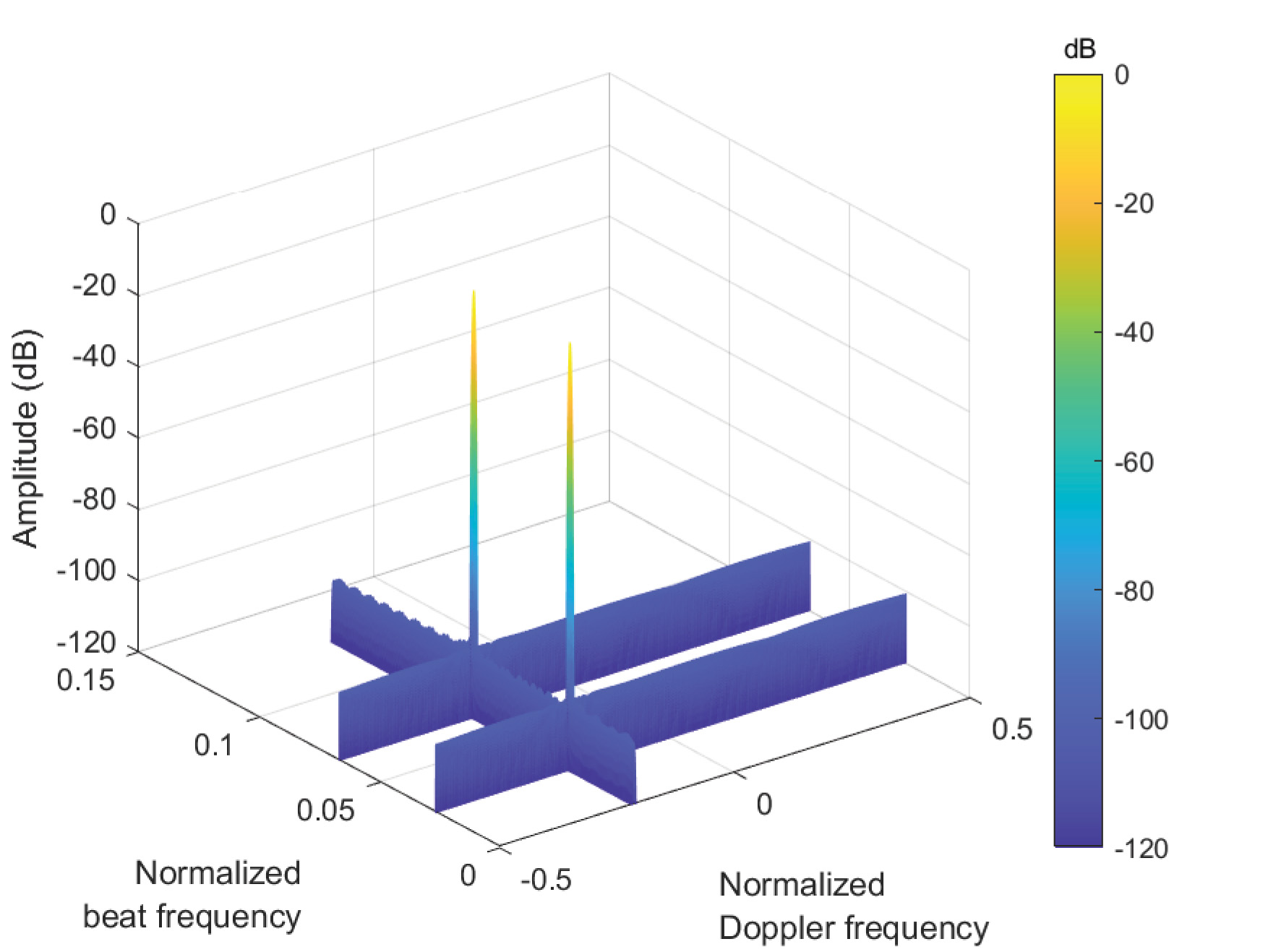}\\
\textit{a)}%
\endminipage  \hfill
\minipage{0.5\textwidth}
\centering
\includegraphics[width=0.9\textwidth]{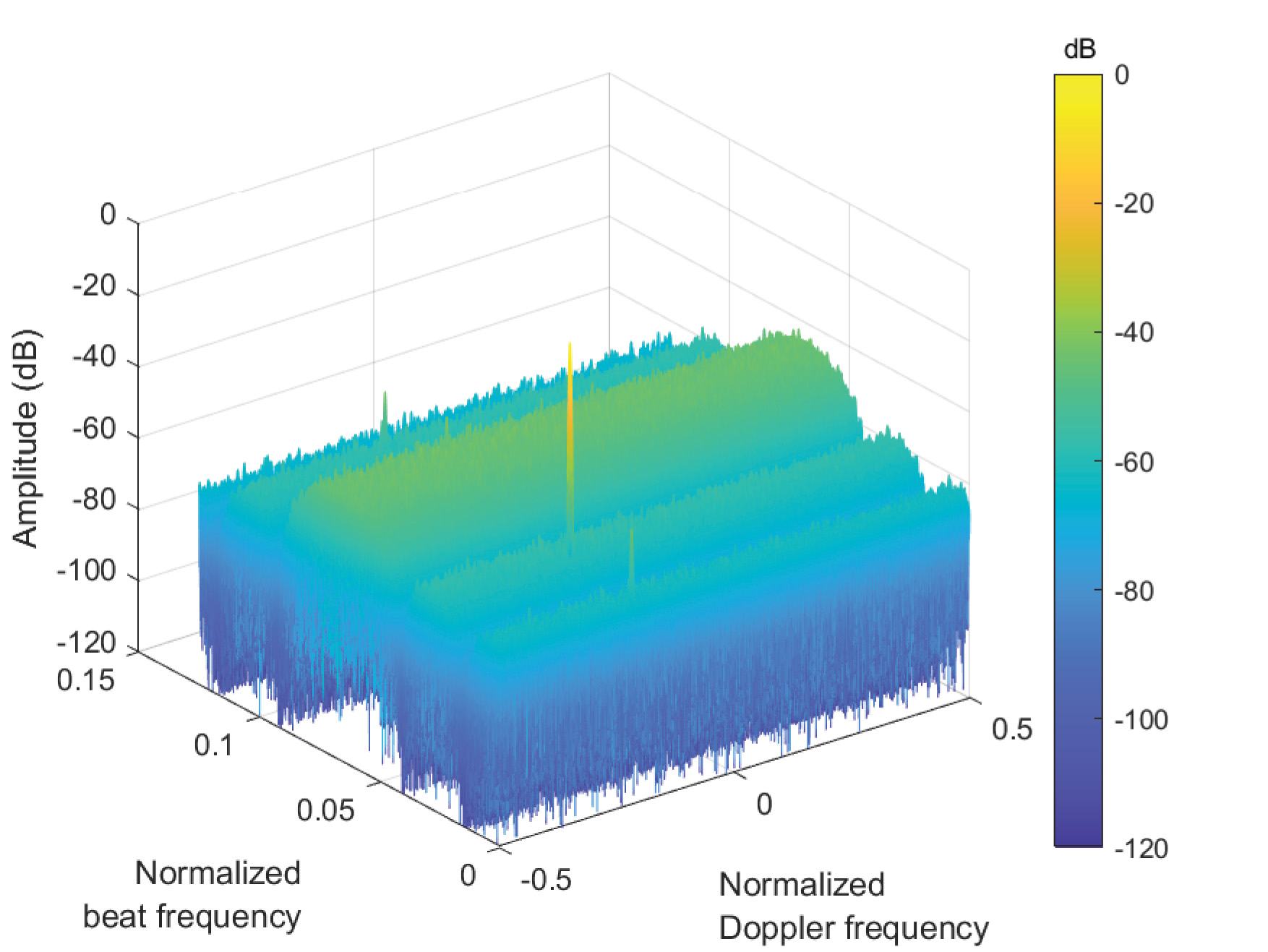}\\
\textit{b)}%
\endminipage  \hfill
\vspace{1mm}
\minipage{0.5\textwidth}
\centering
\includegraphics[width=0.9\textwidth]{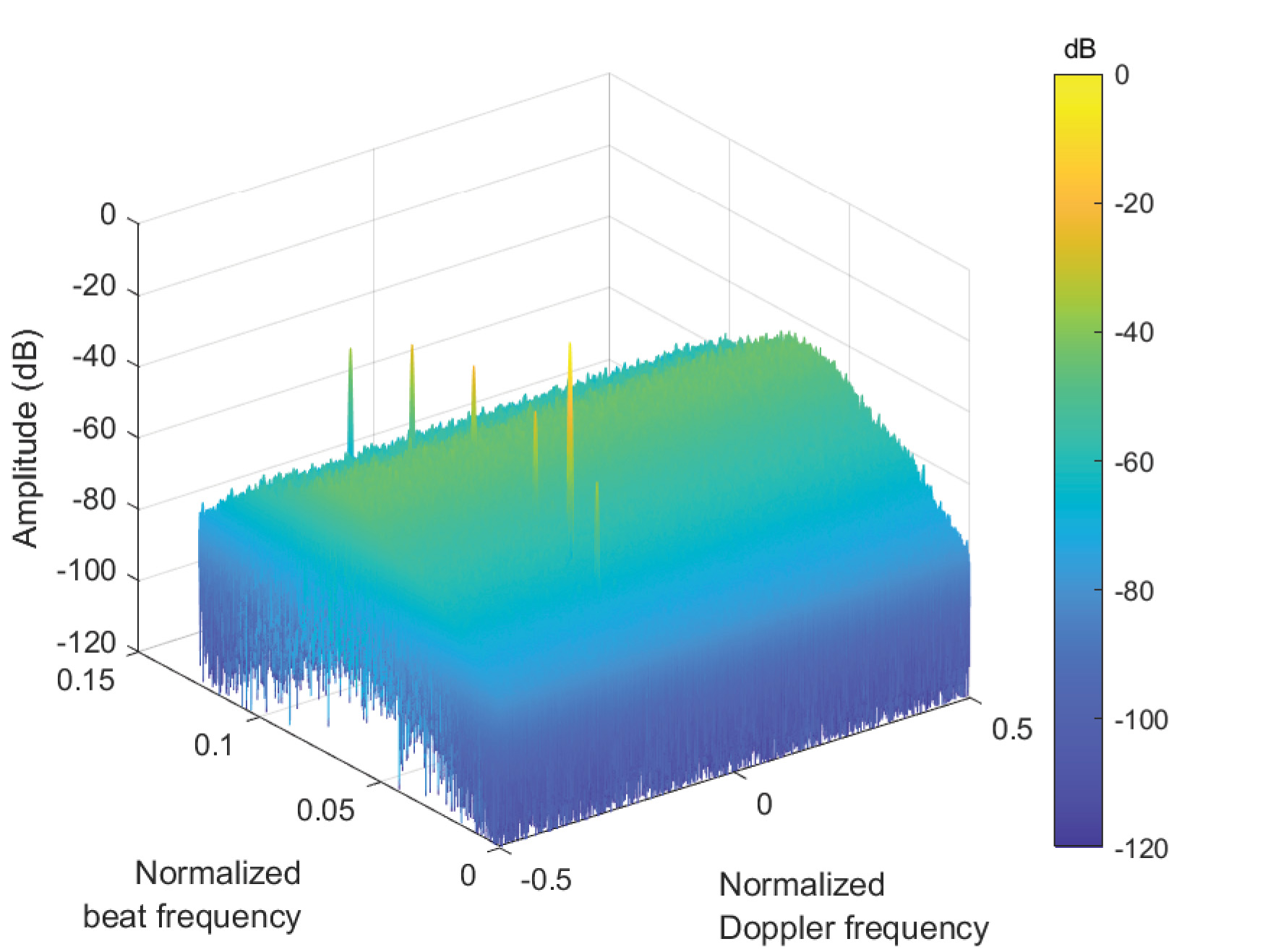}\\
\textit{c)}%
\endminipage  \hfill
\minipage{0.5\textwidth}
\centering
\includegraphics[width=0.9\textwidth]{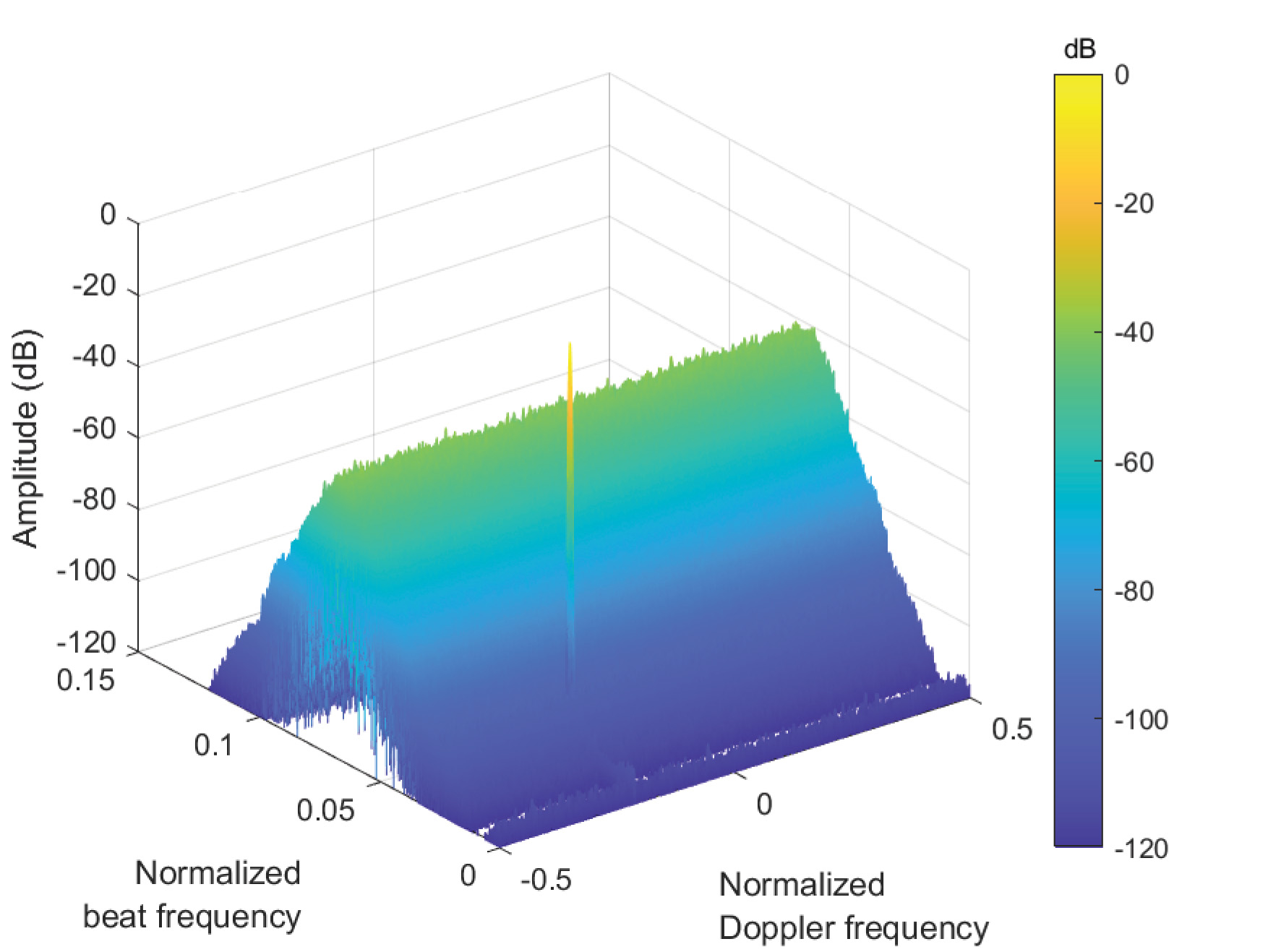}\\
\textit{d)}%
\endminipage  \hfill
\caption{Comparison of cross-isolation between two beat signals associated with three phase lag compensated PC-FMCW waveforms with different random codes: a) FMCW (no code) b) BPSK c) Gaussian d) GMSK }
\label{fig:14}
\end{figure*}

\subsection{Mutual Orthogonality}\label{sec:MutualOrthogonality}

The coding spreads the power of the signals in the beat frequency domain. As each transmitted PC-FMCW uses its phase-coded signal, only the correct signal passes through the received signal, which is matched to this code. The signals with other code sequences are not matched to this code, leading to the spread of the power over range. The theoretical limits of the suppression are equal to the spreading factor and can be written as \cite{Jankiraman}:
\begin{equation}
    SP\equiv10\log\left(\frac{BT}{BT_c}\right)=10\log_{10}(N_c).
\end{equation}
Assume the first radar (victim) transmits PC-FMCW with the phase lag compensated code $\hat{s}_1(t)$ to detect a target. The received signal reflected from the target with complex coefficient $\alpha_1$ can be written as:
\begin{equation}
  x_{\text{R}_1}(t)=\alpha_1\hat{s}_1(t-{\tau}_1)e^{-j\left(2\pi f_c (t-{\tau}_1)+\pi k(t-{\tau}_1)^2\right)}
\end{equation}
To illustrate the mutual orthogonality assessment, consider the worst-case scenario when a second radar is perfectly synchronized with the first radar and transmits PC-FMCW with the phase lag compensated code $\hat{s}_2(t)$. The signal transmitted from the second radar is delayed in time and captured by the first radar with complex coefficient $\alpha_2$ as:
\begin{equation}
  x_{\text{R}_2}(t)=\alpha_2\hat{s}_2(t-{\tau}_2)e^{-j\left(2\pi f_c (t-{\tau}_2)+\pi k(t-{\tau}_2)^2\right)},
\end{equation}
where ${\tau}_2$ is the round trip delay between the first and second radars. Subsequently, the total received signal on the first radar is the combination of received signals and can be written as:
\begin{equation}
  x_{\text{R}}(t)=x_{\text{R}_1}(t)+x_{\text{R}_2}(t).
\end{equation}

The total received signal is mixed and dechirped with the uncoded transmit signal of the first radar. The mixer output gives the summation of two coded beat signals. Subsequently, the group delay filter is applied to the mixer output and aligns coded beat signals at the maximum delay as discussed in Section~\ref{sec:PhaseLagCompensation}. The output of the group delay filter can be represented as:
\begin{equation}
    g_{\text{o}}(t)=\alpha_1 s_1(t-\tau_{\text{max}})e^{j(2\pi k \tau_1 t)}+\alpha_2 s_2(t-\tau_{\text{max}})e^{j( 2\pi k \tau_2 t)}.
\end{equation}
%
During decoding, the group delay filter output is decoded with the complex conjugate of the first code shifted to the maximum delay $s_1(t-\tau_{\text{max}})$. After decoding, the beat signal reflected from the target is obtained similar to the dechirped signal of conventional FMCW, while the beat signal initiated by the second radar remains coded as:
\begin{equation}
\begin{split}
    d_{\text{o}}(t)=&d_{1}(t) + d_{2}(t)\\
    =& \alpha_1 e^{j(2\pi k \tau_1 t)}+\alpha_2{s}^*_1(t-\tau_{\text{max}})s_2(t-\tau_{\text{max}})e^{j( 2\pi k \tau_2 t)},
\end{split}    
\end{equation}
where $d_{1}(t)$ and $d_{2}(t)$ are the decoded signals. Subsequently, we investigate the cross-isolation between two beat signals in the spectrum of the decoded signal output. The cross-isolation can be defined as:
\begin{equation}
    \text{Cross-isolation}=  \frac{\max_{f \in \forall} \left|D_1(f)\right|}{\max_{f \in \forall}\left|D_2(f)\right|} ,
\end{equation}
where $D_1(f)$ and $D_2(f)$ are the spectrum of decoded signals associated with $d_1(t)$ and $d_2(t)$, respectively.

\begin{figure}[t]

         \centerline{\includegraphics[width=0.9\linewidth]{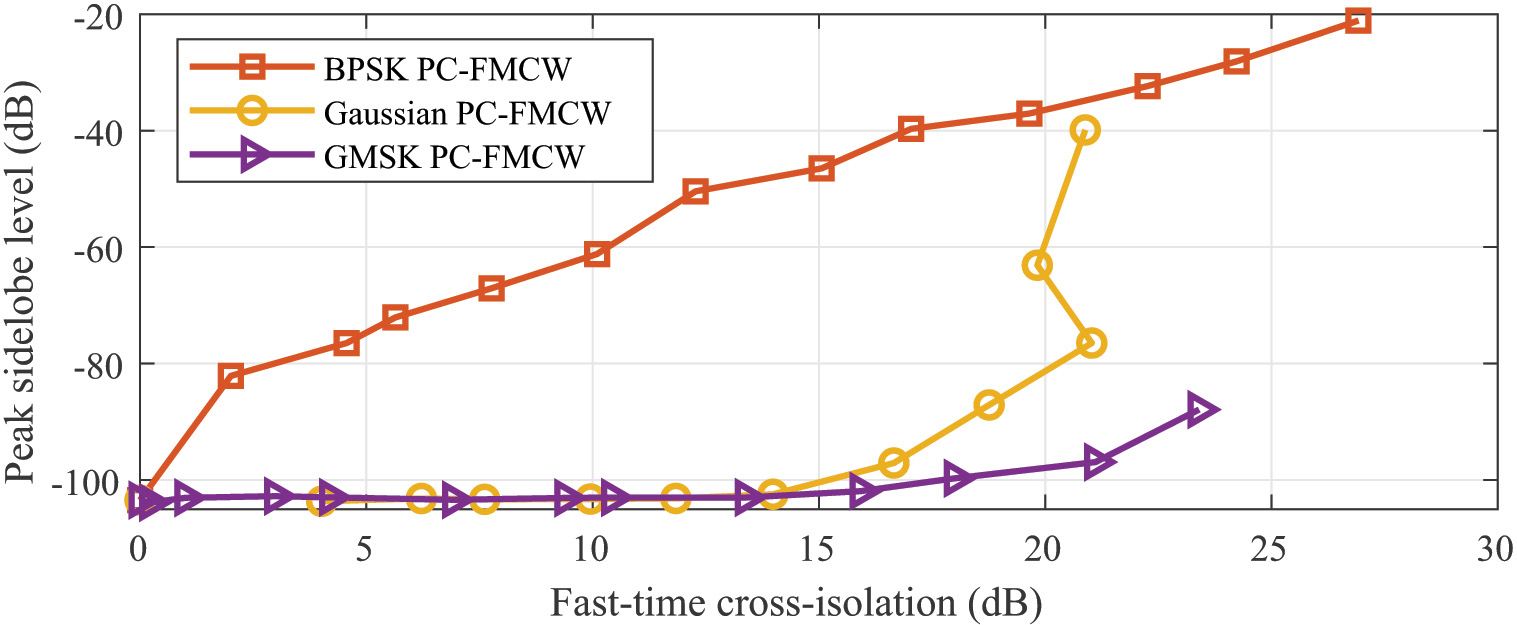}}
         
         \caption{Comparison of PSL versus cross-isolation in fast-time for phase lag compensated PC-FMCW waveforms}
\label{fig:142}
\end{figure}

In Figure~\ref{fig:14}, we compare the cross-isolation between the two beat signals associated with PC-FMCW waveforms with different random code sequences. We consider the number of chips per chirp $N_c=1024$ and the number of chirp pulses $N_p=512$. It is shown in Figure~\ref{fig:14} that the second radar causes a beat signal according to $f_{b_2}=k\tau_2$ which can be seen as a ghost target for a perfectly synchronized case (which is very difficult to generate in a real-life scenario and is just used for the proof of the mutual orthogonality concept), and it can not be distinguished from the target in the traditional FMCW (Figure~\ref{fig:14} a). However, in the phase-coded FMCW cases, the beat signal initiated by the second radar $f_{b_2}$ remains coded, and thus its power is spread over both fast-time and slow-time. This cross-isolation between two beat signals associated with BPSK PC-FMCW, Gaussian PC-FMCW, and GMSK PC-FMCW are given in Figure~\ref{fig:14} b, c, and d, respectively. The theoretical upper-boundary limit regarding the suppression of the beat signal $f_{b_2}$ is $10\log_{10}(512)+10\log_{10}(1024)=57$ dB for a perfectly orthogonal code (combined with the suppression in both slow-time and fast-time). However, the three phase lag compensated PC-FMCW are not perfectly orthogonal after applying the phase lag compensation and filtering. Their resulting suppression behaviours in the fast-time are different according to their phase modulation type, as demonstrated in Figure~\ref{fig:14}. In particular, Gaussian PC-FMCW has the local peaks between phase-coded signals, and it gives the worst suppression performance. BPSK PC-FMCW spreads the power of $f_{b_2}$ to all range cells as the spectrum of BPSK has a significant spectrum broadening of the beat signal. GMSK PC-FMCW spreads the power of the  $f_{b_2}$ as a Gaussian shape (Triangular in dB scale) over the range cells defined by the $3$-dB bandwidth of the Gaussian filter (smoother bandwidth) $B_s$. Consequently, GMSK PC-FMCW has a narrower spreading characteristic than BPSK, which might help to avoid masking of targets with weak radar cross section (RCS) outside of main lobe. Moreover, it can be seen in Figure~\ref{fig:142} that GMSK PC-FMCW can provide high cross-isolation while achieving low PSL. These facts favour the usage of GMSK PC-FMCW over BPSK PC-FMCW.

\section{Experiments}\label{sec:Experiments}

This section demonstrates the experimental results related to the sensing and cross-isolation performance of the phase lag compensated PC-FMCW waveforms. The experimental investigation of the waveforms has been done using PARSAX radar \cite{Parsax2008}. We use the proposed transceiver structure for each PC-FMCW, and we apply the traditional dechipring transceiver structure for the FMCW waveform, which is used as a benchmark. We use random code sequences with $N_C=1024$ for the three phase lag compensated PC-FMCW and choose the system parameters as given in Table~\ref{tab1}. To emphasize the advantage of GMSK, we choose ADC sampling frequency as $f_{\rm{ADC}}=2$ MHz so that the code bandwidth becomes comparable to ADC sampling. Moreover, we applied Chebyshev windowing with $80$ dB suppression and compared it with a rectangle windowing case to highlight the sensing performance of the waveforms. In addition, we normalized all the range profiles by the maximum of the range profile.

\begin{table}[t]
\caption{System Parameters}
\begin{center}
\begin{tabular}{|c|c|c|}
\hline
Chirp bandwidth& $B$ & $40$ MHz \\
\hline
Chirp duration & $T$& $1 $ ms \\
\hline
Intermediate frequency & $f_{\rm{IF}}$ & $125$ MHz \\
\hline
IF sampling frequency  & $f_s$ & $400$ MHz \\
\hline
Carrier frequency & $f_c$& $3.315$ GHz \\
\hline
ADC sampling frequency & $f_{\rm{ADC}}$& $2$ MHz \\
\hline
Number of chips & $N_c$& $1024$ \\
\hline
Chip duration & $T_c$& $0.97 \mu$ s \\
\hline
Chip bandwidth & $B_c$& $1.024$ MHz \\
\hline
Smoother bandwidth & $B_s$& $2.048$ MHz \\
\hline
\end{tabular}
\label{tab1}
\end{center}
\end{table}

\begin{figure}[b]
\centerline{\includegraphics[width=0.6\linewidth]{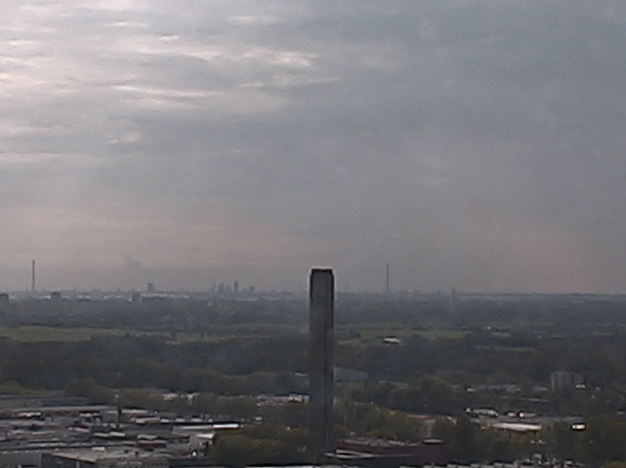}}
\caption{Illustration of the stationary target}
\label{fig:15}
\end{figure}

\begin{figure*}[!t]
\centering
\minipage{0.5\textwidth}
\centering
\includegraphics[width=0.9\textwidth]{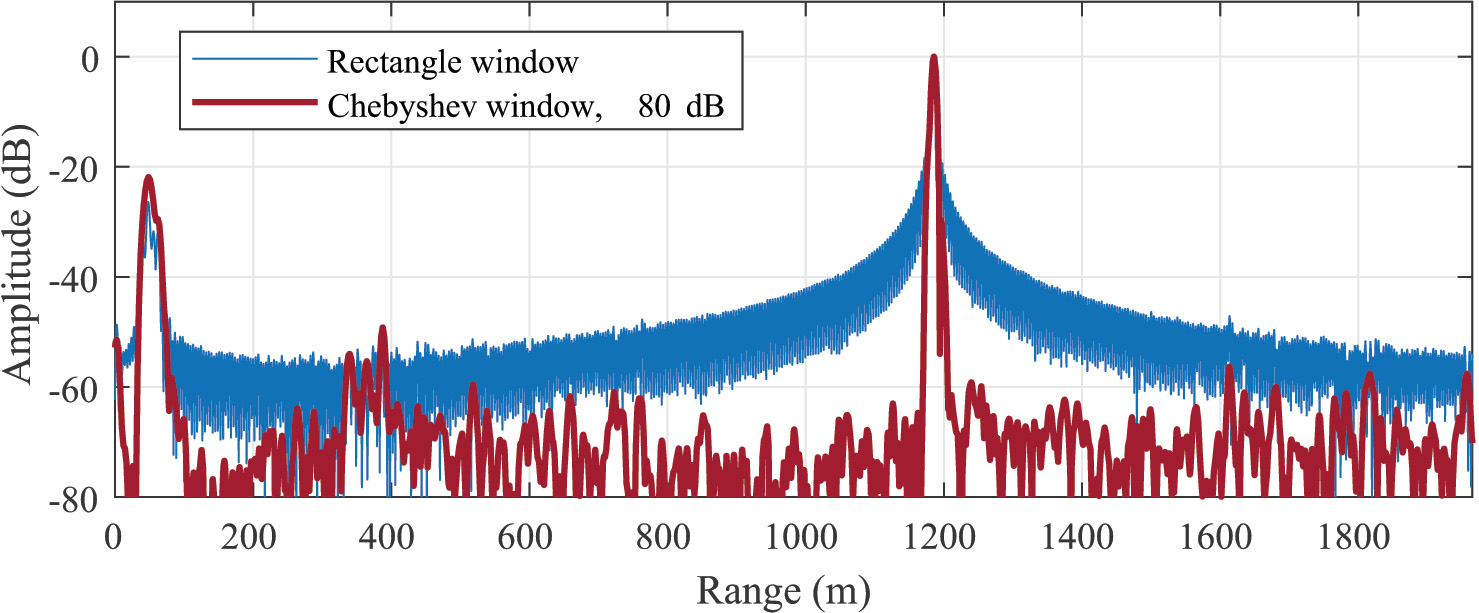}\\
\textit{a)}%
\endminipage  \hfill
\minipage{0.5\textwidth}
\centering
\includegraphics[width=0.9\textwidth]{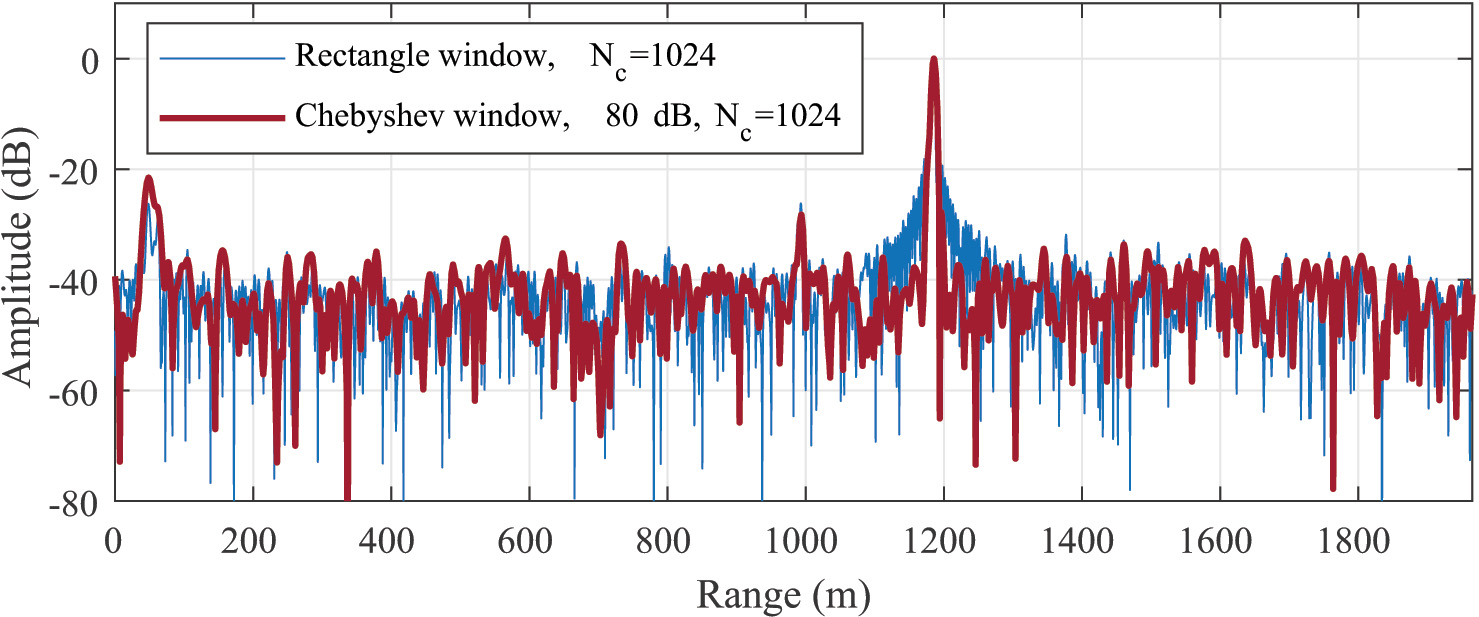}\\
\textit{b)}%
\endminipage  \hfill
\vspace{1mm}
\minipage{0.5\textwidth}
\centering
\includegraphics[width=0.9\textwidth]{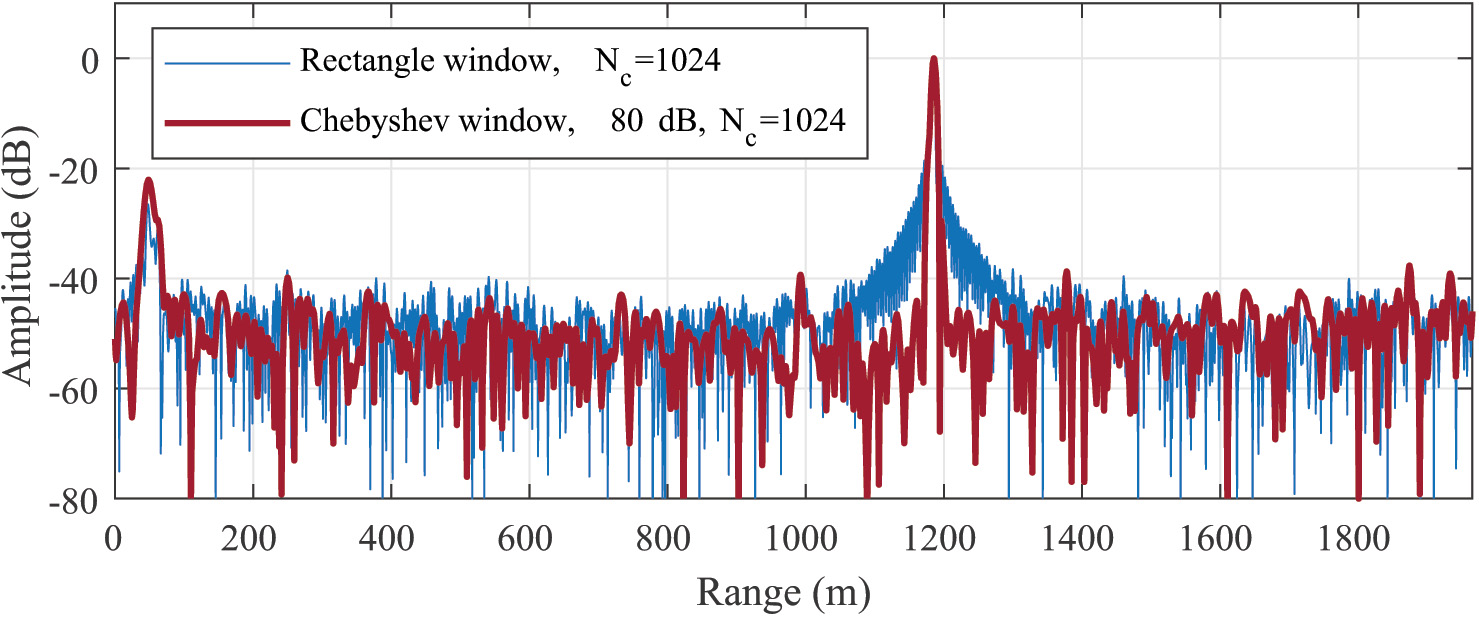}\\
\textit{c)}%
\endminipage  \hfill
\minipage{0.5\textwidth}
\centering
\includegraphics[width=0.9\textwidth]{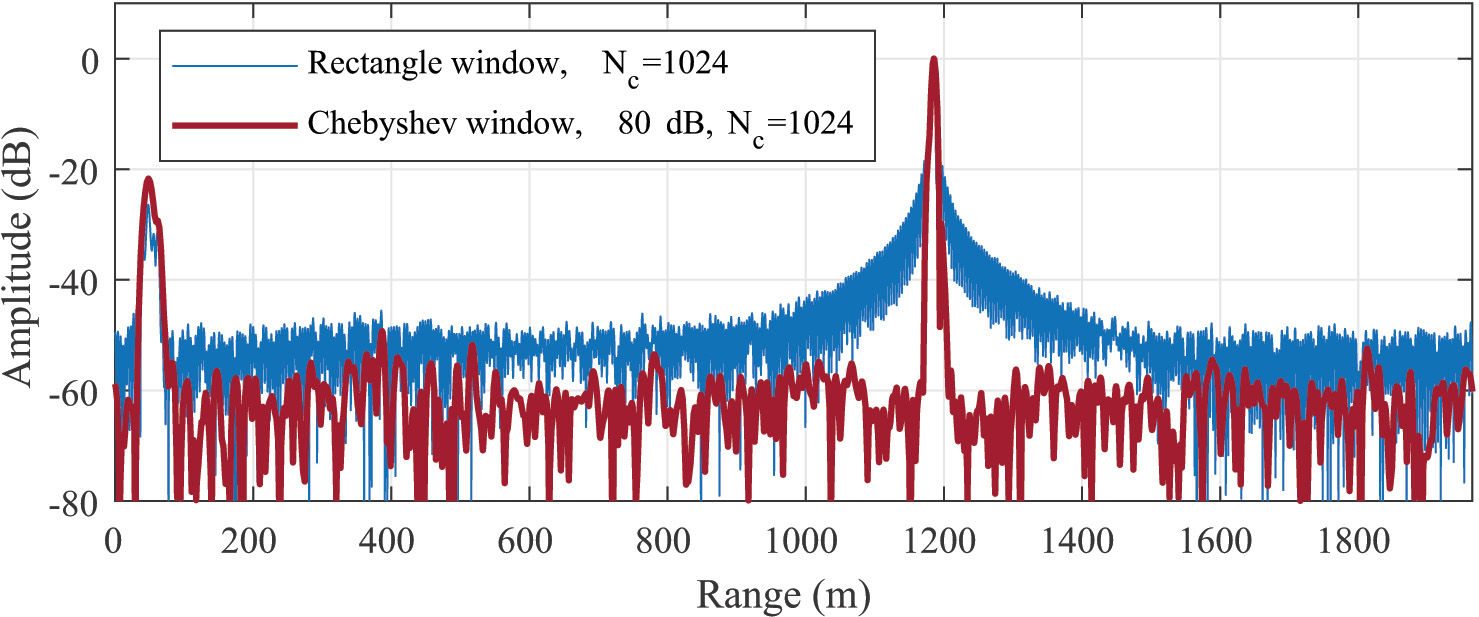}\\
\textit{d)}%
\endminipage  \hfill
\caption{Stationary target range profiles for phase lag compensated PC-FMCW waveforms: a) FMCW b) BPSK c) Gaussian d) GMSK }
\label{fig:16}
\end{figure*}

\begin{figure}[b]
\centerline{\includegraphics[width=0.9\linewidth]{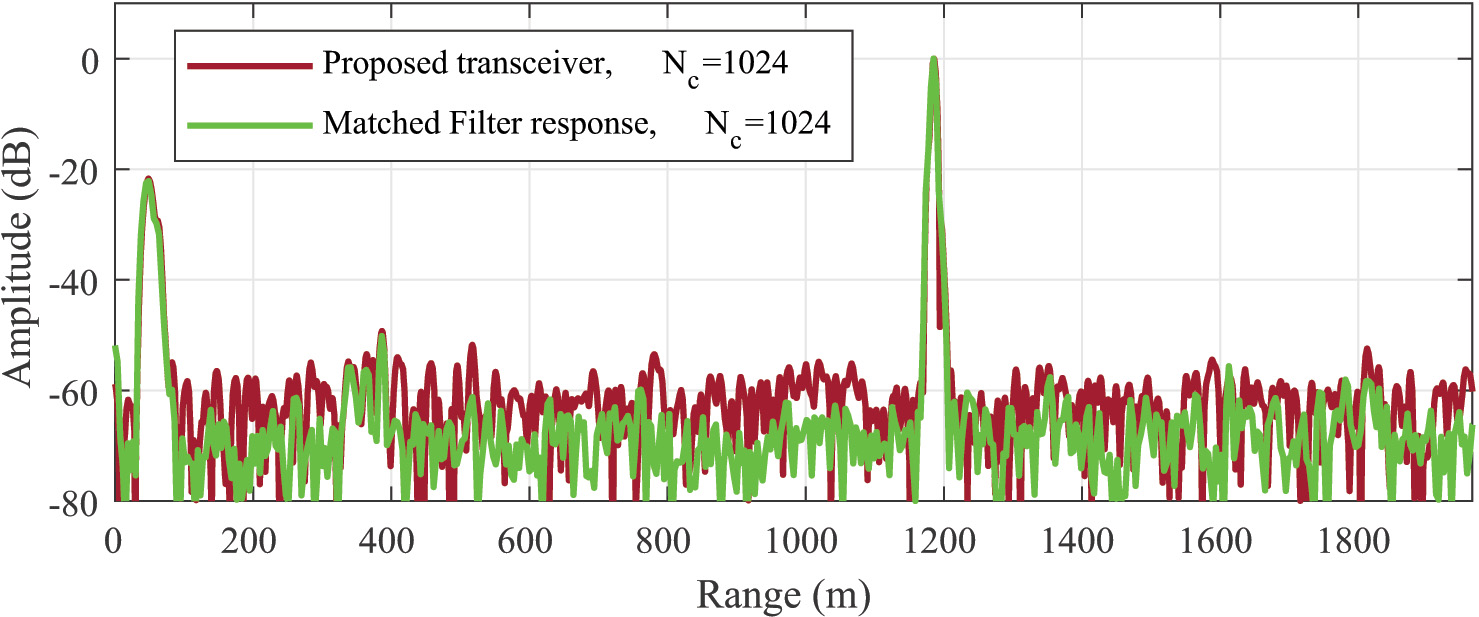}}
\caption{Range profile for phase lag compensated GMSK PC-FMCW in case of the proposed transceiver and Matched filter}
\label{fig:17}
\end{figure}

\subsection{Sensing Performance of One Waveform}

In this subsection, we transmit only one waveform at a time to validate the sensing performance of the waveforms. The resulting waveforms are performed in the real scenario to detect both stationary and moving targets. Note that the experimental environment is dynamic for the moving target experiment. To detect the same car and compare the sensing performance of the waveforms, we transmit four waveforms sequentially with $128$ chirp pulses in each waveform.

\subsubsection{Stationary Target Experiment}
For the stationary target experiment, we look at the chimney located at $1185$ m away from the radar, as shown in Figure~\ref{fig:15}. The range profiles of the four different waveforms are demonstrated in Figure~\ref{fig:16}. At the chimney location, the noise-clutter level of the range profile is around $\sim-60$ dB for FMCW, and it provide $\sim60$ dB dynamic range after applying Chebyshev windowing. It can be seen that BPSK and Gaussian have increased sidelobes and provides $\sim30$ dB and $\sim40$ dB dynamic ranges, respectively. This is due to the fact that BPSK and Gaussian have substantial broadening in the beat frequency and the coded beat signals have a wide spectrum. As a result, the sensing performance of BPSK and Gaussian suffers from limited ADC sampling. On the other hand, widening of the coded beat signal spectrum is reduced by using GMSK as explained in Section~\ref{sec:TypeOfPhases}. Thus, GMSK is expected to provide better sensing performance while the code bandwidth becomes comparable to ADC sampling. We observe this behaviour as GMSK provides the best and closest performance to FMCW by providing $\sim60$ dB dynamic range in the vicinity of the chimney. In addition, we demonstrate the range profile of GMSK PC-FMCW by using Matched Filter receiving strategy and compared it with the response of the proposed transceiver structure in Figure~\ref{fig:17}. It can be seen that the proposed transceiver structure gives a very similar result to the matched filter response for GMSK PC-FMCW.

\subsubsection{Moving Target Experiment}
\begin{figure}[b!]
\centerline{\includegraphics[width=0.6\linewidth]{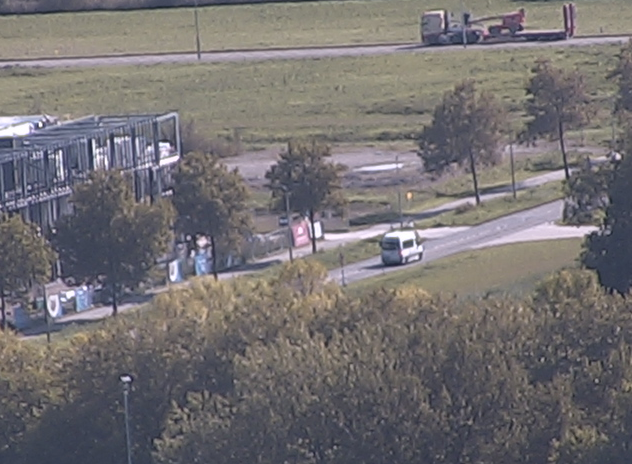}}
\caption{Illustration of the moving target}
\label{fig:18}
\end{figure}

\begin{figure*}[!t]
\centering
\minipage{0.5\textwidth}
\centering
\includegraphics[width=0.9\textwidth]{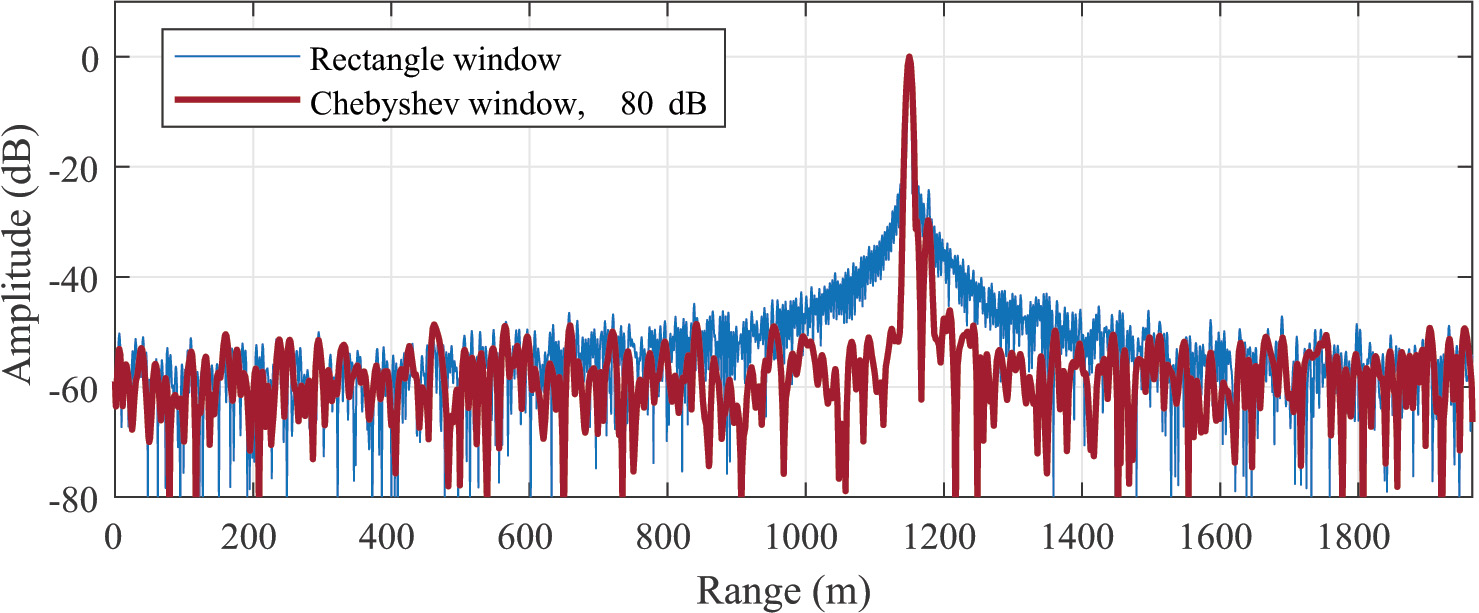}
\includegraphics[width=0.9\textwidth]{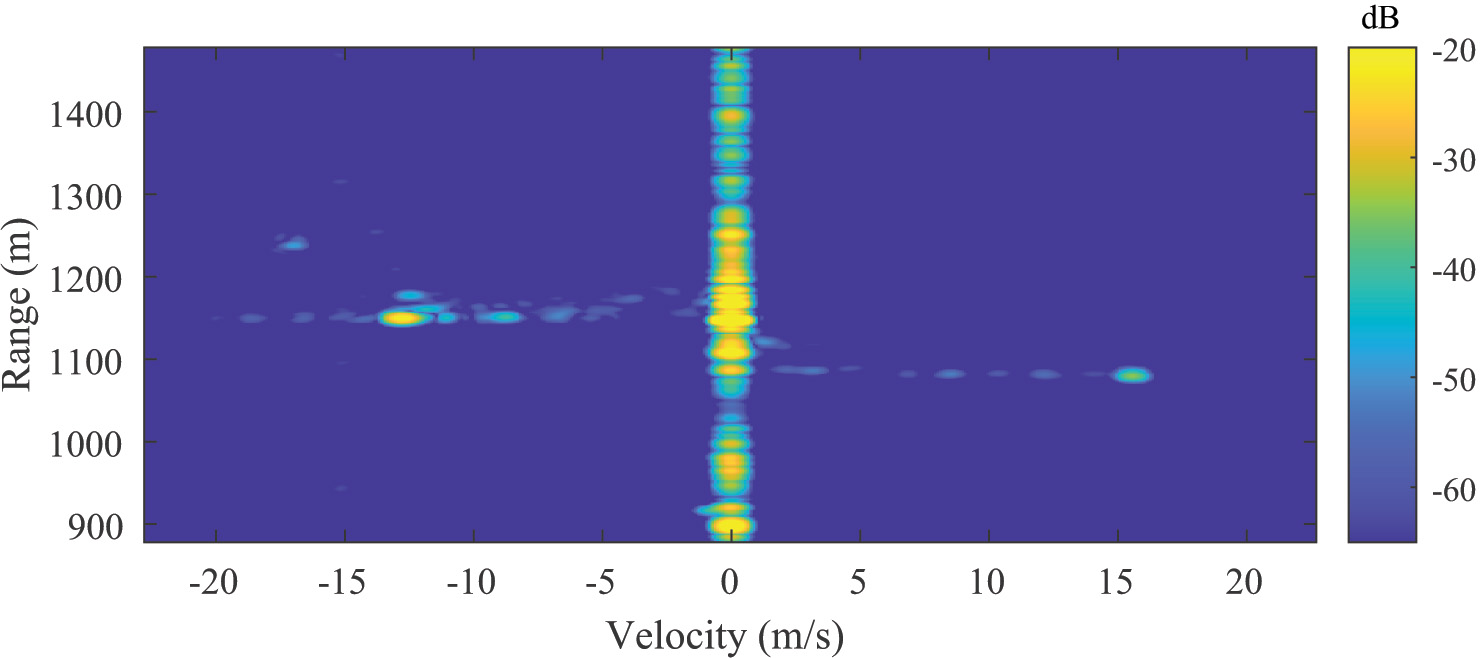}\\
\textit{a)}%
\endminipage  \hfill
\minipage{0.5\textwidth}
\centering
\includegraphics[width=0.9\textwidth]{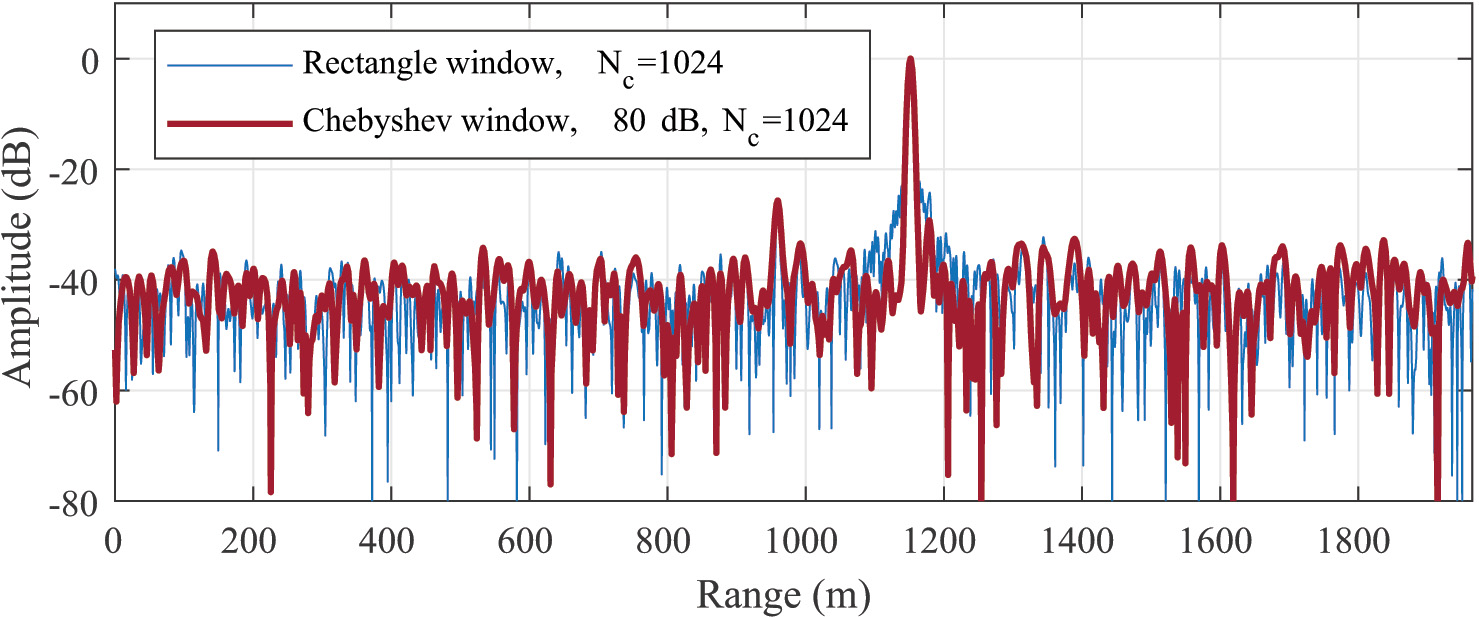}
\includegraphics[width=0.9\textwidth]{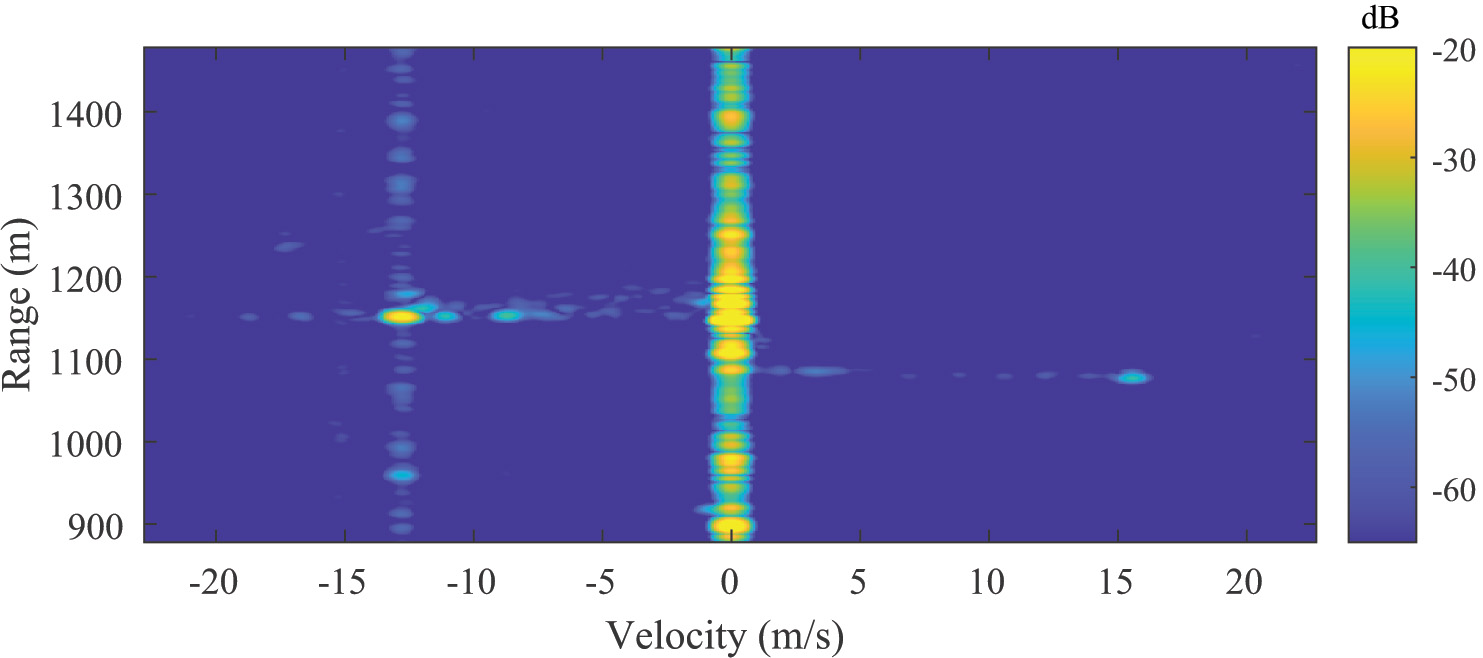}\\
\textit{b)}%
\endminipage  \hfill
\vspace{1mm}
\minipage{0.5\textwidth}
\centering
\includegraphics[width=0.9\textwidth]{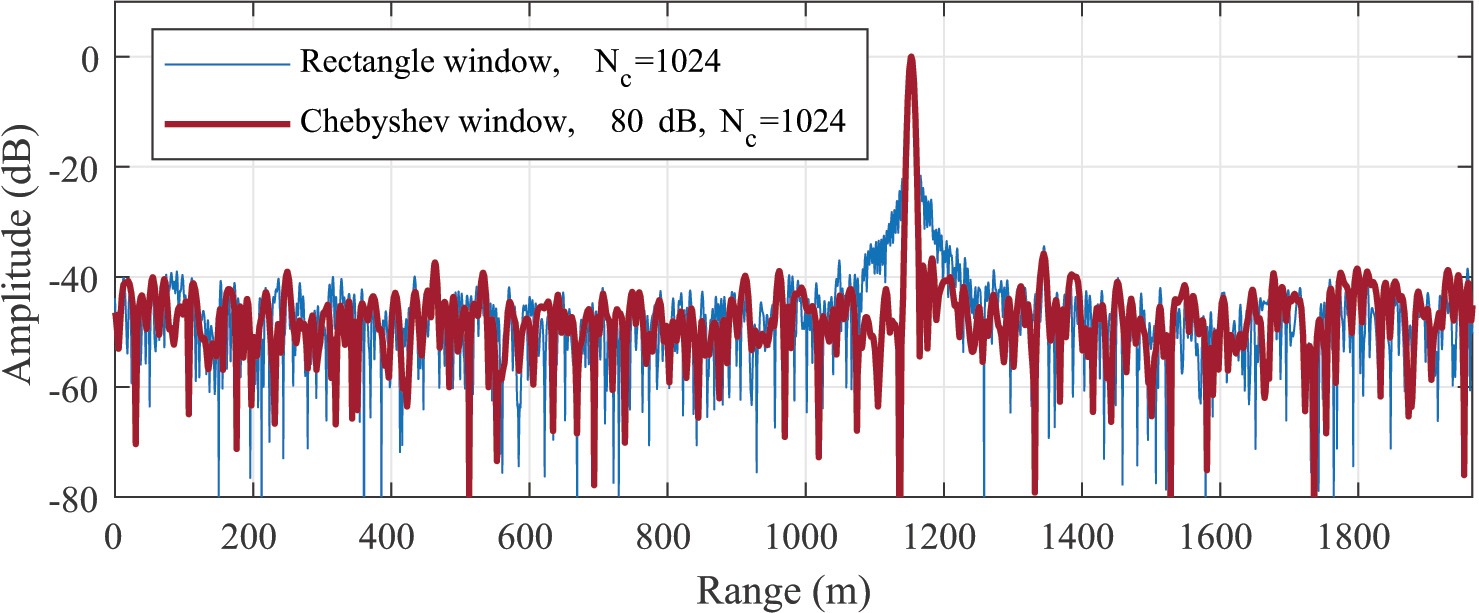}
\includegraphics[width=0.9\textwidth]{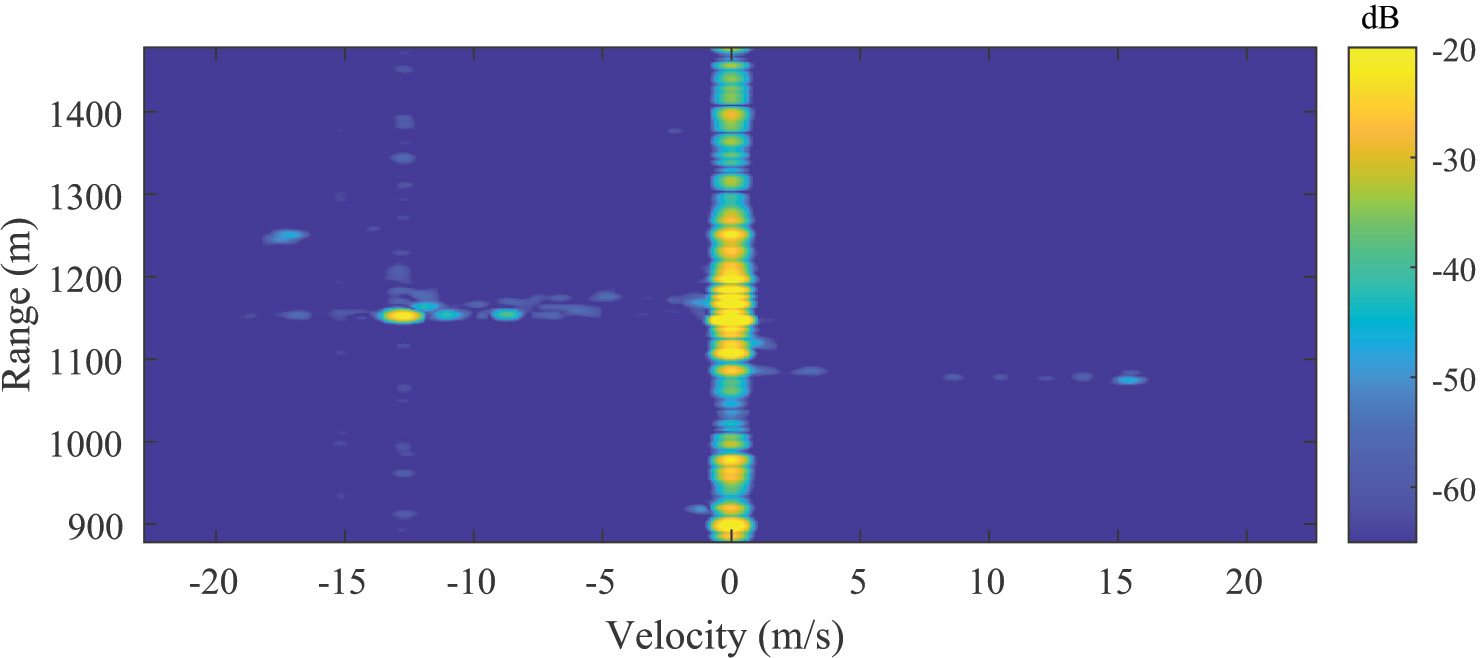}\\
\textit{c)}%
\endminipage  \hfill
\minipage{0.5\textwidth}
\centering
\includegraphics[width=0.9\textwidth]{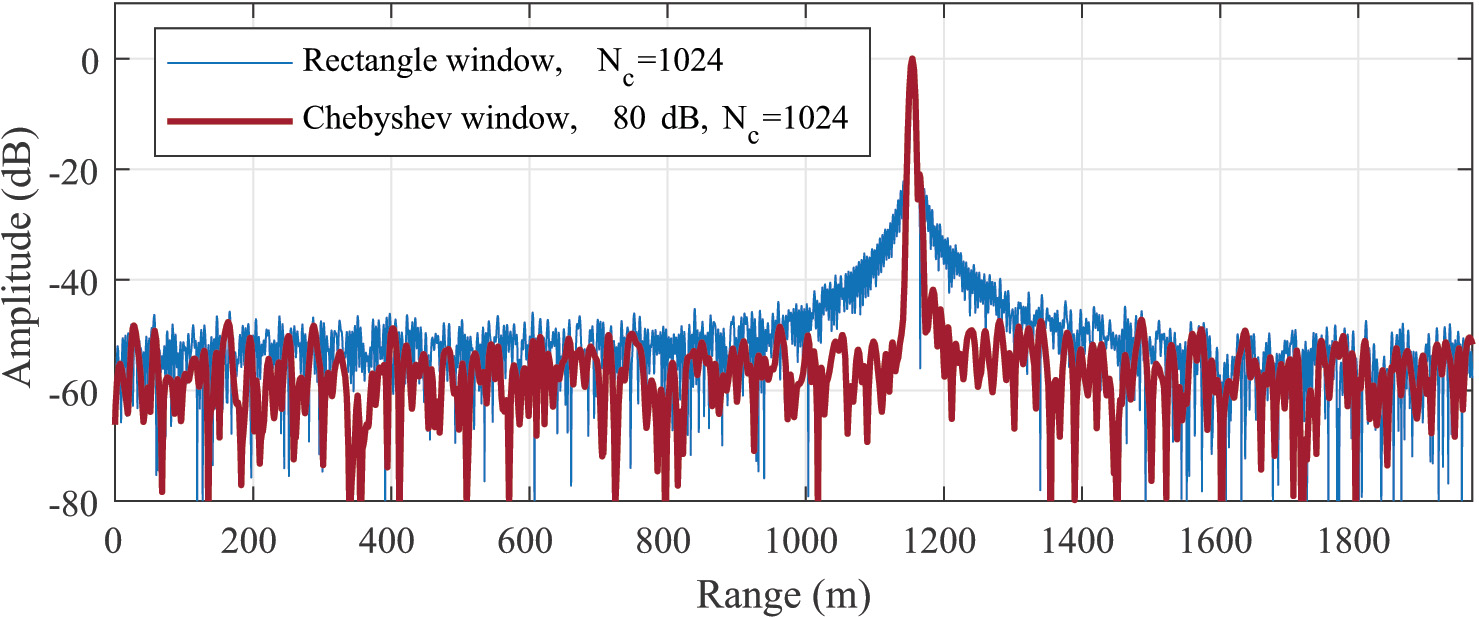}
\includegraphics[width=0.9\textwidth]{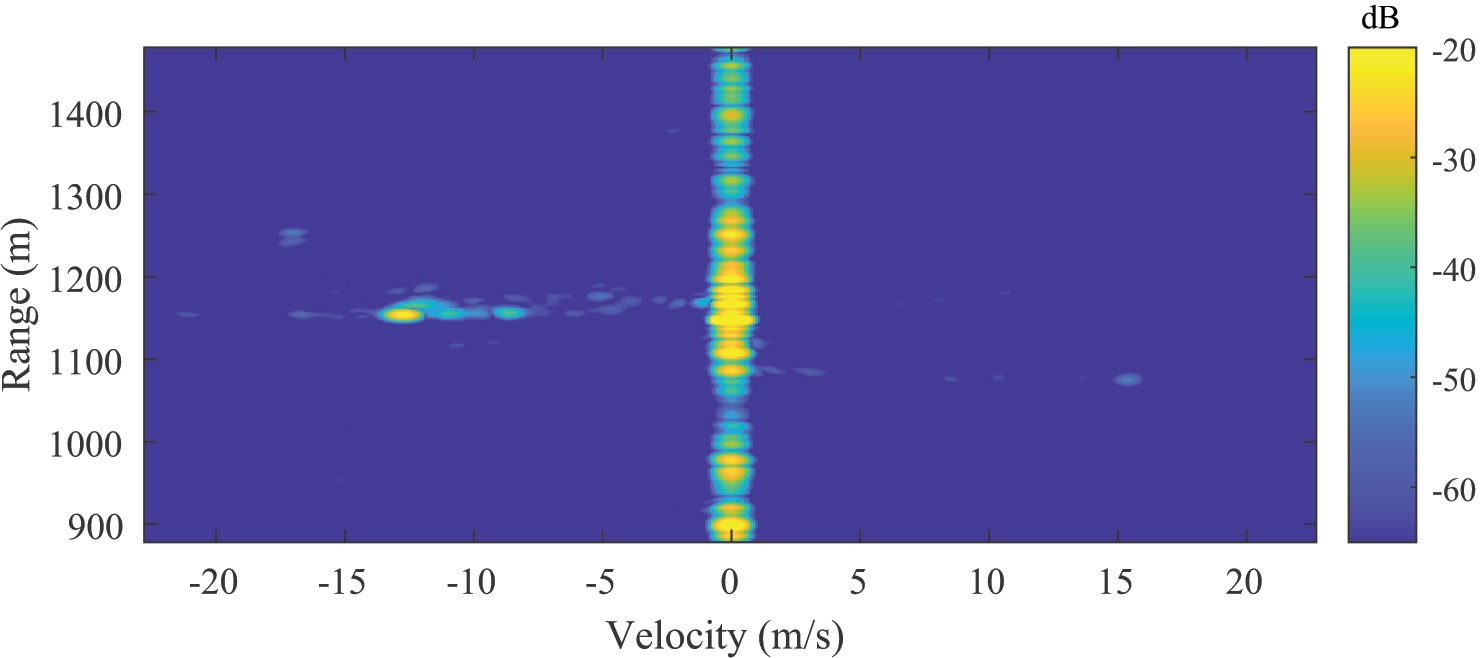}\\
\textit{d)}%
\endminipage  \hfill
\caption{Moving target range and range-Doppler profiles for phase lag compensated PC-FMCW waveforms: a) FMCW b) BPSK c) Gaussian d) GMSK}
\label{fig:19}
\end{figure*}

The Doppler tolerance of the investigated waveforms is validated by the moving target experiment where we observe the road and detect a moving car located at $1150$ m with a radial velocity $\sim 13$ m/s as illustrated in Figure~\ref{fig:18}. We use $N_c=1024$ for the three phase lag compensated PC-FMCW. The range-Doppler profiles of the waveforms are demonstrated in Figure~\ref{fig:19} where the peak location of the target is obtained at $1150$ m for each waveform. The noise level of the range profile is around $\sim-55$ dB for FMCW, and it has $\sim55$ dB dynamic range after windowing (Figure~\ref{fig:19} a). Similar to the stationary target scenario, GMSK provides the best sensing performance between three phase lag compensated PC-FMCW. In particular, the range profile of BPSK PC-FMCW has increased sidelobe level due to limited ADC sampling, and it provides a dynamic range around $\sim30$ dB (Figure~\ref{fig:19} b), while the sidelobe level of Gaussian PC-FMCW provides $\sim40$ dB dynamic range (Figure~\ref{fig:19} c). However, GMSK PC-FMCW provides $\sim55$ dB dynamic range, and it has a range profile very similar to FMCW as shown in Figure~\ref{fig:19} d. Consequently, GMSK PC-FMCW can provide similar sensing performance that is offered by FMCW, and it can also ensure the ability to distinguish different signals due to coding as discussed in Section~\ref{sec:MutualOrthogonality}.

\begin{figure*}[!t]
\centering
\minipage{0.5\textwidth}
\centering
\includegraphics[width=0.9\textwidth]{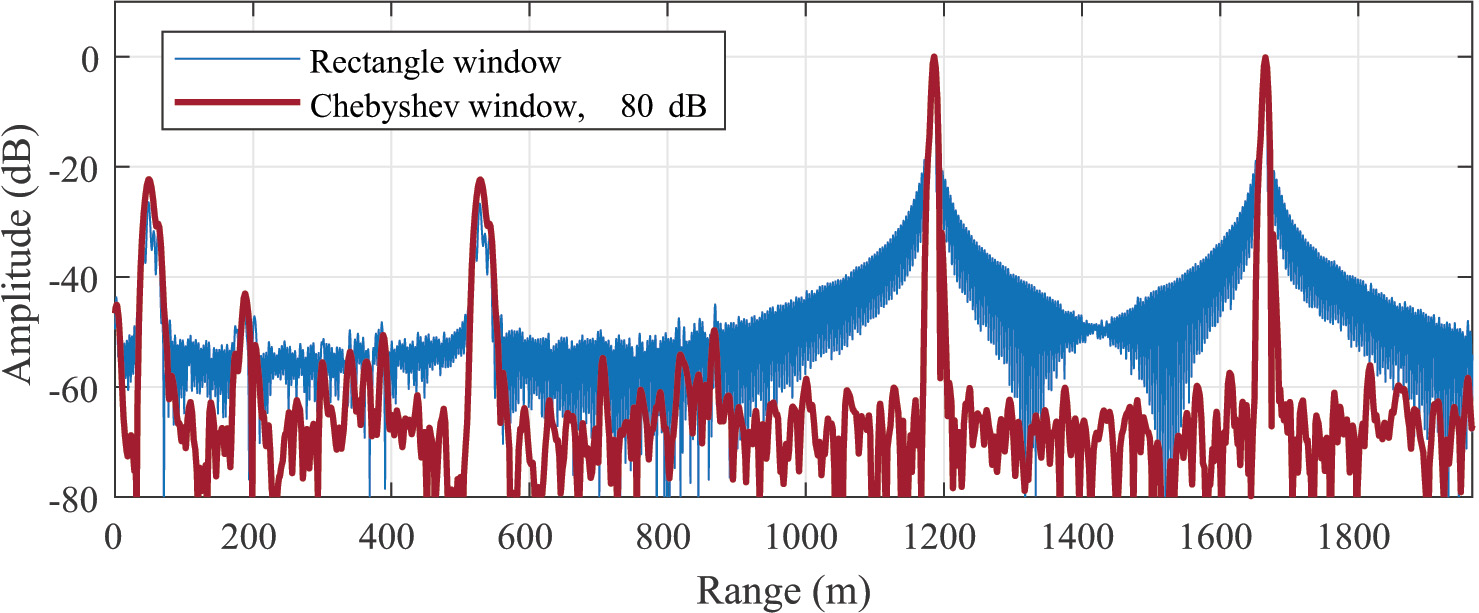}\\
\textit{a)}%
\endminipage  \hfill
\minipage{0.5\textwidth}
\centering
\includegraphics[width=0.9\textwidth]{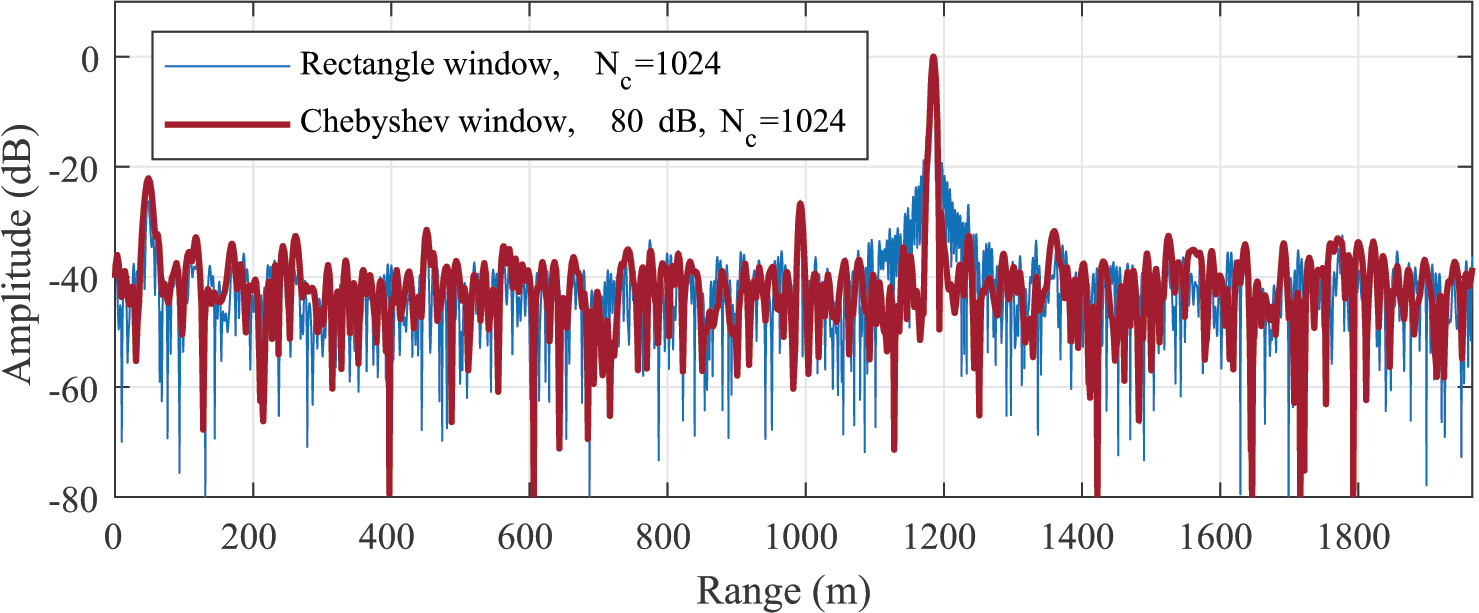}\\
\textit{b)}%
\endminipage  \hfill
\vspace{1mm}
\minipage{0.5\textwidth}
\centering
\includegraphics[width=0.9\textwidth]{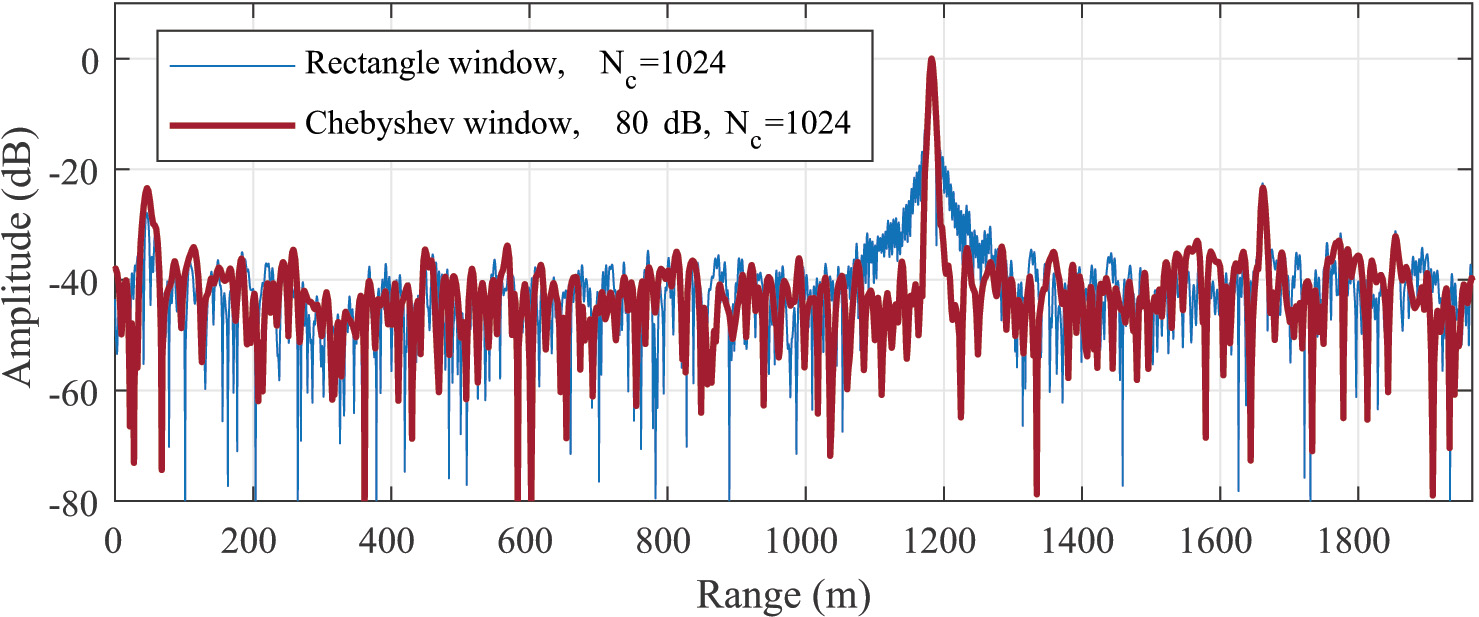}\\
\textit{c)}%
\endminipage  \hfill
\minipage{0.5\textwidth}
\centering
\includegraphics[width=0.9\textwidth]{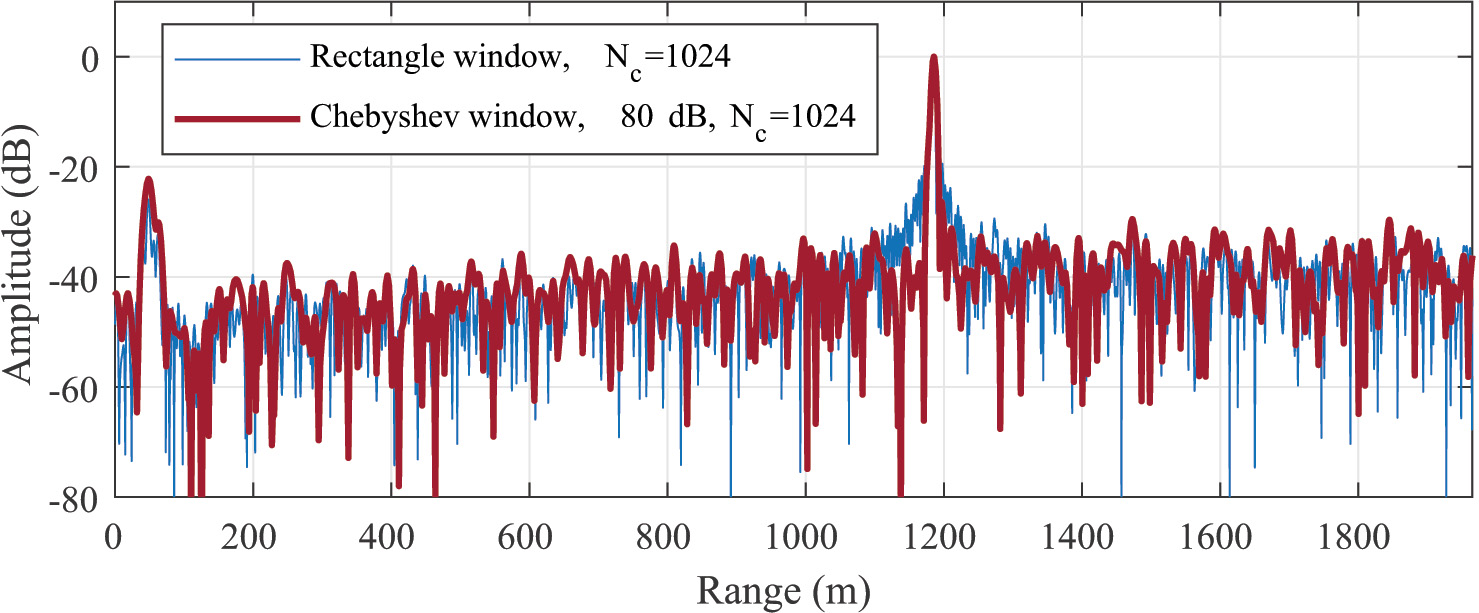}\\
\textit{d)}%
\endminipage  \hfill
\caption{Cross-isolation experiment for a stationary target. Range profiles for phase lag compensated PC-FMCW waveforms: a) FMCW b) BPSK c) Gaussian d) GMSK }
\label{fig:20}
\end{figure*}

\subsection{Cross-isolation Performance Between Two Waveforms}

In this subsection, we transmit two same types of waveforms simultaneously to validate the cross-isolation performance of the waveforms. For the proof of the mutual orthogonality concept and to mimic the worst-case scenario as explained in Section~\ref{sec:MutualOrthogonality}, we apply linear time delay to the second waveform so that it has a range offset compared to the first waveform that corresponds to $480$ m. For each PC-FMCW, the first waveform uses phase lag compensated random code $\hat{s}_1(t)$ with $N_c=1024$, and the second waveform uses phase lag compensated random code $\hat{s}_2(t)$ with $N_c=1024$. Moreover, we use $32$ chirp pulses, and each PC-FMCW chirp pulses use different random code sequences. Consequently, combined with the suppression in both slow-time and fast-time, the theoretical upper-boundary limit of cross-isolation is $10\log_{10}(32)+10\log_{10}(1024)=45$ dB for a perfectly orthogonal code. However, the cross-isolation performance is expected to be degraded due to the loss of orthogonality between codes after applying the phase lag compensation and filtering. The resulting two waveforms are transmitted together and performed in the real scenario to detect the chimney located at $1185$ m away from the radar. Subsequently, the received signal for each PC-FMCW is processed with the proposed transceiver structure and decoded with the reference code $s_1(t)$.

The range profiles of the four different waveforms are demonstrated in Figure~\ref{fig:20}. It can be seen that the second FMCW waveform leads to a beat signal that causes ghost targets at $530$ m and $1665$ m in addition to real targets at $50$ m and $1185$ m. Since there is no mutual orthogonality between two FMCW waveforms, the second waveform can not be distinguished from the first waveform in the traditional FMCW radar. By using PC-FMCW waveforms, only the beat signal associated with the first waveform is decoded with $s_1(t)$ and the beat signal initiated by the second waveform is spread over both fast-time and slow-time as it remains coded. Consequently, the ghost targets created by the second waveform are suppressed using PC-FMCW. Among three phase lag compensated PC-FMCW, GMSK provides the best dynamic range. Particularly, Gaussian PC-FMCW provides weak suppression performance, and the ghost target still appears with $\sim-22$ dB power. BPSK PC-FMCW suppresses the ghost target power around $\sim 34$ dB but only provides $\sim26$ dB dynamic range as it suffers from limited ADC sampling. On the other hand, GMSK suppresses the ghost target's power and provides $\sim40$ dB dynamic range in the vicinity of the chimney. Thus, experimental results verify the advantages of GMSK PC-FMCW over BPSK PC-FMCW and Gaussian PC-FMCW.

\section{Conclusion}\label{sec:Conclusion}

The smoothing of the phase-coded frequency modulated continuous waveform has been introduced as an efficient tool to enhance the coexistence of multiple radars within the same spectrum. The impact of the spectrum widening due to the abrupt phase changes of BPSK is investigated, and the Gaussian filter is proposed to smooth the phase transition of PC-FMCW. We have suggested a receiving strategy with a low sampling requirement and analysed the group delay filter effect on the coded beat signals. In addition, the phase lag compensation is performed on the transmitted phase-coded signal to eliminate the undesired effect of the group delay filter and recover the beat signals properly after the decoding.

The properties of the investigated waveforms for the first time are analysed theoretically and verified experimentally. It is shown that the PSL, PAPR and the cross-isolation between signals increase as the bandwidth of the code raises for the three phase lag compensated PC-FMCW. The simulations and the experimental results demonstrate that the phase lag compensated GMSK PC-FMCW can provide sensing performance similar to that of uncoded FMCW. At the same time, it can provide high mutual orthogonality that can be used to improve cross-isolation between multiple radars.

\appendices

\section{Derivatives of Different Types of Phase Code}
In this proof, we demonstrate the taking derivative of the different types of phase code with respect to time. Recall that the rectangle function can be written as:
\begin{equation}
    {\rm rect}\left(\frac{t-x}{y}\right)=u\left(t-x+\frac{y}{2}\right)-u\left(t-x-\frac{y}{2}\right),
\end{equation}
where $u$ is a unit step function. Similarly, the code term can be written as:
\begin{equation}
\begin{split}
\phi_n {\rm rect}\left( {\frac{{t - (n-\frac{1}{2}){T_c}}}{{{T_c}}}} \right)&= \phi_n u\left(t- (n-\frac{1}{2}){T_c}+\frac{T_c}{2}\right)\\
& -\phi_n u\left(t- (n-\frac{1}{2}){T_c}-\frac{T_c}{2}\right),
\end{split}
\end{equation}
and
\begin{equation}
\begin{split}
\phi_{n+1} {\rm rect}&\left( {\frac{{t - (n+\frac{1}{2}){T_c}}}{{{T_c}}}} \right) \\
&=\quad \phi_{n+1} u\left(t- (n+\frac{1}{2}){T_c}+\frac{T_c}{2}\right)\\
& \qquad -\phi_{n+1} u\left(t- (n+\frac{1}{2}){T_c}-\frac{T_c}{2}\right),
\end{split}
\end{equation}
for the $n^{\text{th}}$ and $(n+1)^{\text{th}}$ elements, respectively. Note that the $\phi_n \in \{ 0,\pi\}$ denotes the phase corresponding to the $n^{\textrm{th}}$ bit of the  $N_c$ bits sequence. To take the summation of unit step functions, we have to consider a junction point in which the adjacent elements are linked. Thus, the relevant junction point includes the right part of the $n^{\text{th}}$ and left part of the $(n+1)^{\text{th}}$ elements and the phase of the BPSK code can be represented as:
\begin{equation}
    \phi_{\text{bpsk}}(t)=\sum\limits_{n = 1}^{{N_c}} (\phi_{n+1} -\phi_n ) \, u(t-nT_c).
\end{equation}
where the amplitude of the unit step function varies between $\pi$ and $-\pi$ depending on the value $(\phi_{n+1} -\phi_n)$, and the summation of the unit step functions gives the phase of the BPSK code sequence $\phi_{\text{bpsk}}(t)\in\{ 0,\pi\}$. In the following subsections, we derive the instantaneous frequency of the different types of phase code.

\subsection{BPSK}
Taking the derivative of the $\phi_{\text{bpsk}}(t)$ gives:
\begin{equation}
    \frac{1}{2\pi}\frac{d}{dt} \phi_{\text{bpsk}}(t)= \frac{1}{2\pi}\sum\limits_{n = 1}^{{N_c}} (\phi_{n+1} -\phi_n ) \, \delta(t-nT_c).
\end{equation}
where $\delta$ is the Dirac delta function. Same result can be seen in \cite{UtkuGeneralized}.

\subsection{Gaussian}
The convolution of the unit step function with filter $h_0(t)=e^{-t^2}$ where $t\geq0$ can be represented as:
\begin{equation}\label{conv_unit}
\begin{split}
    u(t)\otimes h_0(t)&=\int_{-\infty}^{\infty} h_0(\tau) u(t-\tau)d\tau=\int_{0}^{t} h_0(\tau)d\tau \\
    &=\int_{0}^{t} e^{-{\tau}^2}d\tau=\frac{\sqrt{\pi}}{2} {\rm erf} (t),
    \end{split}
\end{equation}
where $\rm{erf}(t)$ represents the error function as:
\begin{equation}
    \rm{erf}(t)=\frac{2}{\sqrt{\pi}} \int_{0}^{t} e^{-t^2}\, dt \, .
\end{equation}
Subsequently, the convolution of the unit step function and the Gaussian filter $h(t)=\frac{\eta}{\sqrt{\pi}} \, e^{- {\eta}^2 t^2}$ can be written as:
\begin{equation}
    u(t)\otimes h(t)=\int_{0}^{t} h(\tau)d\tau=\int_{0}^{t} \frac{\eta}{\sqrt{\pi}} e^{-{\eta}^2 {\tau}^2}d\tau .
\end{equation}
Replacing $\gamma=\eta\tau$ and $d\gamma=\eta d\tau$, the equation becomes:
\begin{equation}\label{conv_erf}
\begin{split}
    u(t)\otimes h(t)&=\frac{1}{\sqrt{\pi}}\int_{0}^{\eta t} e^{-{\gamma}^2}d\gamma \\
    &=\frac{1}{2} {\rm erf} (\eta t).
\end{split}
\end{equation}
Consequently, the phase of the Gaussian binary code can be written as:
\begin{equation}\label{gaussianPhase} 
\begin{split}
   \phi_{\text{gauss}}(t)&= \phi_{\text{bpsk}}(t)\otimes h(t) \\
    &=\frac{1}{2}\sum\limits_{n = 1}^{{N_c}} (\phi_{n+1} -\phi_n ) \, {\rm erf} \left(\eta (t-nT_c)\right).
\end{split}
\end{equation}
The derivative of the error function can be obtained as:
\begin{equation}
    \frac{d}{dt}\left(\rm{erf}(t)\right)=\frac{2}{\sqrt{\pi}} e^{-t^2}.
\end{equation}
Subsequently, taking the derivative of \eqref{gaussianPhase} with respect to time gives:
\begin{equation}
    \frac{1}{2\pi}\frac{d}{dt}\phi_{\text{gauss}}(t)=\frac{\eta}{2\pi\sqrt{\pi}} \sum\limits_{n = 1}^{{N_c}} (\phi_{n+1} -\phi_n ) e^{-{\eta}^2(t-n{T_c})^2}.
\end{equation}

\subsection{GMSK}
The phase of the GMSK can be represented as:
\begin{equation}
\phi_{\text{gmsk}}(t)=\int_{-\infty}^{\infty}\phi_{\text{gauss}}(t)dt=\int_{-\infty}^{\infty} (\phi_{\text{bpsk}}(t)\otimes h(t)) dt.
\end{equation}
Taking the derivative of the $\phi_{\text{gmsk}}(t)$ gives:
\begin{equation}
    \begin{split}
    \frac{1}{2\pi}\frac{d}{dt}\phi_{\text{gmsk}}(t)&=\frac{1}{2\pi} \phi_{\text{gauss}}(t) \\
    &=\frac{1}{4\pi}\sum\limits_{n = 1}^{{N_c}} (\phi_{n+1} -\phi_n ) \, {\rm erf} \left(\eta (t-nT_c)\right).
\end{split}
\end{equation}

\section{Convolution with Phase Lag Compensation}
In this proof, we demonstrate the result of the convolution with phase lag compensation. Let $\beta=\sqrt{\frac{\pi k}{j}}$, then the \eqref{phaseLagEq} becomes:
\begin{equation}
    h_{\text{lag}}(t)=\frac{\beta}{\sqrt{\pi}} e^{-{\beta}^2 t^2}
\end{equation}
Following the steps between \eqref{conv_unit} and \eqref{conv_erf} given in Appendix A, and replacing $\eta=\beta$, the convolution of the unit step function and phase lag compensation filter can be found as:
\begin{equation}
    u(t)\otimes h_{\text{lag}}(t)=\frac{1}{2}{\rm erf} \left(\sqrt{\frac{\pi k}{j}}t\right).
\end{equation}
Subsequently, the result of the convolution for the BPSK code sequence becomes:
\begin{equation}
\begin{split}
    \hat{s}(t)&=c(t) \otimes h_{\text{lag}}(t)\\
    &=\frac{1}{T} \frac{1}{T_c} \frac{1}{2}\sum\limits_{n = 1}^{{N_c}}{e^{j({\phi_{n+1}}-\phi_n)}}{\rm erf}\left(\sqrt{\frac{\pi k}{j}}\left(t-n T_c\right)\right).
\end{split}
\end{equation}

\section{Group Delay Filter Phase Response}
In this proof, we demonstrate the relationship between phase response and group delay. To ease of mathematical manipulations, let:
\begin{equation}
\begin{split}
    d&=\theta(f_b)-f_b\,\frac{d\theta(f)}{df}\big|_{f=f_b} \\
    p&=\frac{d\theta(f)}{df}\big|_{f=f_b}\\
    \varphi(f)&=\sum\limits_{m = 2}^{{\infty}} \frac{1}{m!} \frac{d^m \theta(f)}{df^m}\big|_{f=f_b} (f-f_b)^m.
\end{split}
\end{equation}
Subsequently, multiplying the group delay filter with the mixer output in frequency domain gives:
\begin{equation}
\begin{split}
     Z_{\text{o}}(f)&=X_{\text{M}}(f) H_{\text{g}}(f)\\
  &=S(f-f_b)e^{-j\left(\frac{2\pi f_b }{k}(f-f_b)\right)} e^{j(d+pf)} e^{j\varphi(f)}.
\end{split}
\end{equation}
Note that $\varphi(f)$ term is small compared to first two terms due to the Taylor series expansion. Taking the inverse Fourier transform of the group delay filter output gives:
\begin{equation}
\begin{split}
     z_{\text{o}}(t)=&\mathcal{F}^{-1}\left\{S(f-f_b)e^{-j\left(\frac{2\pi f_b }{k}(f-f_b)\right)}  e^{j(d+pf)} \right\}\\
     &\qquad \otimes \mathcal{F}^{-1}\left\{e^{j\varphi(f)}\right\}\\
     =&z_{1}(t) \otimes z_{2}(t)
\end{split}
\end{equation}
The resulting signal can be considered as the convolution of two signal as $ z_{\text{o}}(t)=z_{1}(t) \otimes z_{2}(t)$. The right part of the convolution $z_{2}(t)=\mathcal{F}^{-1}\left\{e^{j\varphi(f)}\right\}$ comes from the higher order terms in Taylor series expansion and leads to so-called dispersion effect. The left part of the convolution $z_{1}(t)$ causes the group delay that we are interested in and can be obtained as:
\begin{equation}
\begin{split}
     z_{1}(t)=&\mathcal{F}^{-1}\left\{S(f-f_b)e^{-j\left(\frac{2\pi f_b }{k}(f-f_b)\right)}  e^{j(d+pf)} \right\}\\
     =&\int_{-\infty}^{\infty} S(f-f_b)e^{-j\left(\frac{2\pi f_b }{k}(f-f_b)\right)}  e^{j(d+pf+2\pi ft)} \, df \\
     =&\int_{-\infty}^{\infty} S(f_1)e^{-j\left(\frac{2\pi f_b }{k}(f_1)\right)}  e^{j(d+p(f_1+f_b)+2\pi (f_1+f_b) t)} \, df_1\\
     =&\int_{-\infty}^{\infty} S(f_1)e^{j\left(2\pi f_1 (t- \frac{f_b}{k}+\frac{p}{2\pi})\right)} \, df_1 \, \,  e^{j(d+f_b(2\pi t+p))}\\
     =&s\left(t-\frac{f_b}{k}+\frac{p}{2\pi}\right)\,  e^{j(d+f_b(2\pi t+p))}\\
     =&s\left(t-\tau_0+\frac{1}{2\pi}\frac{d\theta(f)}{df}\big|_{f=f_b}\right)\,  e^{j(2\pi f_b t)} e^{j(\theta(f_b))}\\
     =&s\left(t-\tau_0-\tau_g(f)\right)\,  e^{j(2\pi f_b t)} e^{j(\theta(f_b))}.
\end{split}
\end{equation}
As a result, the filter leads to the group delay $\tau_g(f)$, which shifts the envelope of the signal.

\section*{Acknowledgment}
Part of this research activity was performed within the TU Delft Industry Partnership Program (TIPP), which is funded by NXP Semiconductors N.V. and Holland High Tech Systems and Materials (TKIHTSM/18.0136) under the project `Coded Radar for Interference Suppression in Super-Dense Environments’ (CRUISE).

The authors would like to thank Fred van der Zwan and Yun Lu for their support during the experiments and Tworit Dash for his helpful suggestions. Finally, the authors would like to thank the anonymous reviewers for their constructive comments that improved the quality of the paper significantly.

\ifCLASSOPTIONcaptionsoff
  \newpage
\fi

\bibliography{GMSKJournal}
\bibliographystyle{IEEEtran}

\begin{IEEEbiography}[{\includegraphics[width=1in,height=1.25in,clip,keepaspectratio]{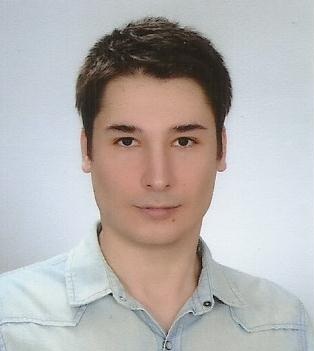}}]{Utku Kumbul} received his B.Sc. degrees with honors in the Department of Electrical and Electronics Engineering in 2014 and in the Department of Computer Engineering (Double Major Program) in 2015, both from the Atilim University, Ankara, Turkey. In 2017, he received the M.Sc. degree in Electrical and Electronics Engineering from the TOBB University of Economics and Technology, Ankara, Turkey. Between 2017 and 2018, he worked as a hardware engineer at Havelsan Ehsim, Air Electronic Warfare Systems Eng. Inc., in Ankara, where he was part of the passive radar and decoy projects. In March 2019, he joined to Microwave Sensing, Signals and Systems (MS3) section of the Faculty of Electrical Engineering, Mathematics, and Computer Science (EEMCS) at Delft University of Technology as a Ph.D. candidate. His current research interests include waveform design, radar signal processing, interference mitigation, MIMO systems and automotive radars. He has been a Reviewer of the IEEE TRANSACTIONS ON AEROSPACE AND ELECTRONIC SYSTEMS and the IET RADAR, SONAR \& NAVIGATION
\end{IEEEbiography}

\begin{IEEEbiography}[{\includegraphics[width=1in,height=1.25in,clip,keepaspectratio]{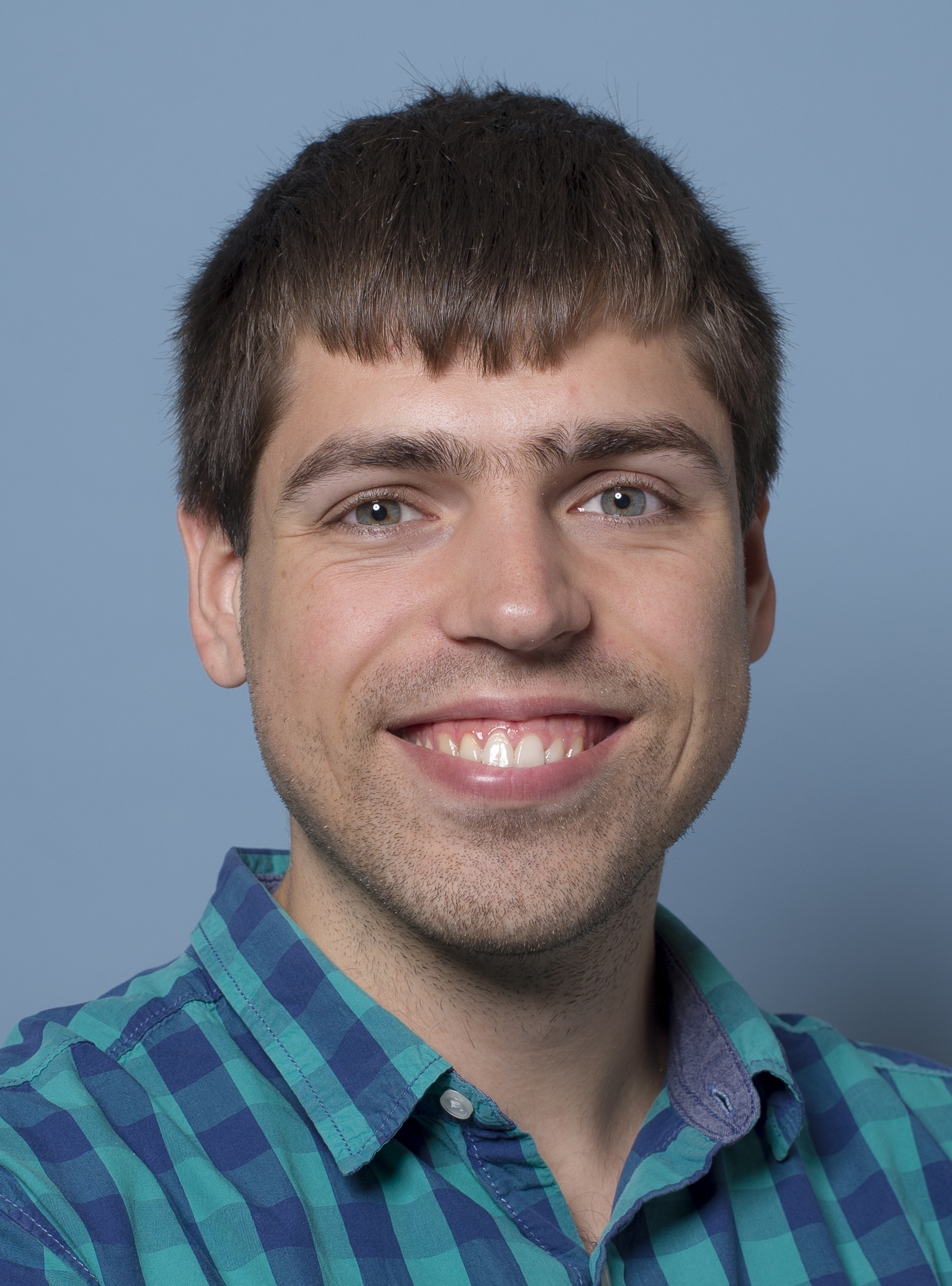}}]{Nikita Petrov}
received the Engineering degree in radio-electronic control systems from Baltic State Technical University “VOENMEH” D.F. Ustinov, Saint Petersburg, Russia, in 2012 and the Ph.D. degree in radar signal processing from the Delft University of Technology, Delft, The Netherlands, in 2019. Since then, he has been a Postdoctoral Researcher with the Microwave Sensing, Signals and Systems Section, Faculty of Electrical Engineering, Mathematics, and Computer Science, Delft University of Technology. Since 2022 he is with NXP Semiconductors, Eindhoven, The Netherlands. His research interests include modern radar technologies, radar signal processing, multichannel and multiband signals and systems, high resolution, and automotive radars. He is currently a Reviewer of the IEEE TRANSACTIONS ON AEROSPACE AND ELECTRONIC SYSTEMS and the IEEE
TRANSACTIONS ON GEOSCIENCE AND REMOTE SENSING
\end{IEEEbiography}


\begin{IEEEbiography}[{\includegraphics[width=1in,height=1.25in,clip,keepaspectratio]{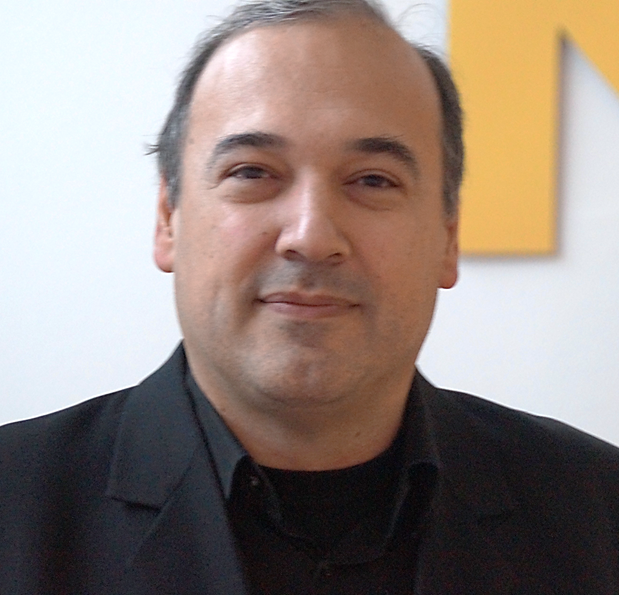}}]{Cicero S. Vaucher} received the Ph.D. degree in Electrical Engineering from the University of Twente in 2001. He was with Philips Research Laboratories Eindhoven from 1990 to 2006, when he joined NXP Semiconductors. He is presently Automotive Radar Product Architect with the ADAS Product Line. He is also a part-time professor at TU Delft, working on mmWave Front-ends. His research activities and interests include micro-wave and mm-Wave transceiver architectures, radar system implementation and signal processing, and implementation of circuit building blocks. He is the author of Architectures for RF Frequency Synthesizers (Boston, MA: Kluwer, 2002) and is a co-author of Circuit Design for RF Transceivers (Boston, MA: Kluwer, 2001). He is (co-) inventor of 28 unique patent families. Dr. Vaucher was a member of the Technical Programme Committee of the IEEE Custom Integrated Circuits Conference (CICC) from 2005 to 2013, acting in the wireless subcommittee. Presently, he is acting on the IEEE-MTT Connected and Autonomous Systems Technical Committee 27. He is a Senior Member of the IEEE, and an NXP Technical Fellow.
\end{IEEEbiography}


\begin{IEEEbiography}[{\includegraphics[width=1in,height=1.25in,clip,keepaspectratio]{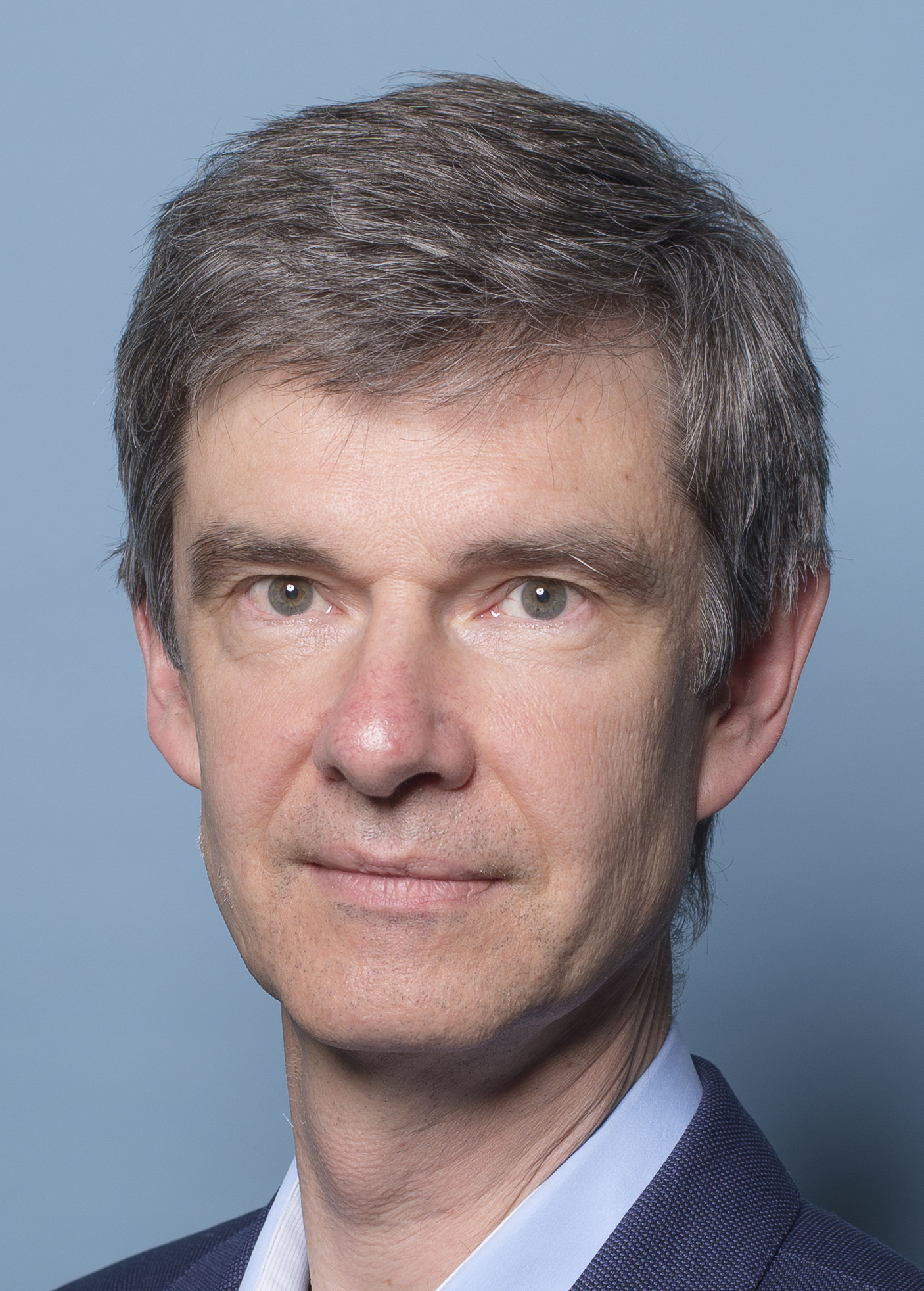}}]{Alexander Yarovoy}
(FIEEE’ 2015) graduated from the Kharkov State University, Ukraine, in 1984 with the Diploma with honor in radiophysics and electronics. He received the Candidate Phys. \& Math. Sci. and Doctor Phys. \& Math. Sci. degrees in radiophysics in 1987 and 1994, respectively. In 1987 he joined the Department of Radiophysics at the Kharkov State University as a Researcher and became a Full Professor there in 1997. From September 1994 through 1996 he was with Technical University of Ilmenau, Germany as a Visiting Researcher. Since 1999 he is with the Delft University of Technology, the Netherlands. Since 2009 he leads there a chair of Microwave Sensing, Systems and Signals. His main research interests are in high-resolution radar, microwave imaging and applied electromagnetics (in particular, UWB antennas). He has authored and co-authored more than 500 scientific or technical papers, seven patents and fourteen book chapters. He is the recipient of the European Microwave Week Radar Award for the paper that best advances the state-of-the-art in radar technology in 2001 (together with L.P. Ligthart and P. van Genderen) and in 2012 (together with T. Savelyev). In 2010 together with D. Caratelli Prof. Yarovoy got the best paper award of the Applied Computational Electromagnetic Society (ACES).  Prof. Yarovoy served as the General TPC chair of the 2020 European Microwave Week (EuMW’20), as the Chair and TPC chair of the 5th European Radar Conference (EuRAD’08), as well as the Secretary of the 1st European Radar Conference (EuRAD’04). He served also as the co-chair and TPC chair of the Xth International Conference on GPR (GPR2004). He served as an Associated Editor of the International Journal of Microwave and Wireless Technologies from 2011 till 2018 and as a Guest Editor of five special issues of the IEEE Transactions and other journals. In the period 2008-2017 Prof. Yarovoy served as Director of the European Microwave Association (EuMA). \end{IEEEbiography}




\end{document}